\documentclass[epj]{svjour}
\usepackage{amsmath,amssymb}
\usepackage{graphicx}
\usepackage{psfrag}
\usepackage{times}
\DeclareGraphicsExtensions{.eps}

\begin{document}
\title{\boldmath A Coherent Interpretation of the Form Factors of 
the Nucleon in Terms of a Pion Cloud and Constituent Quarks} 
\titlerunning{Form Factors of the Nucleon}
\author{J.~Friedrich and Th.~Walcher}

\institute{Institut f\"ur Kernphysik, Universit\"at Mainz, D-55099 Mainz}

\date{Received: date / Revised version: date}

\abstract{The recent measurements of the electric form factor of 
the neutron suggest that its shape may be interpreted as a smooth broad
distribution with a bump at $Q^2 \approx$ 0.3~(GeV/c)$^2$
superimposed. As a consequence the corresponding charge distribution 
in the Breit frame shows a negative charge extending as far out as
2~fm. It is natural to identify this charge with the pion cloud. This
realisation is then used to reanalyse all old and new data of the
electric and magnetic from factors of the proton and the neutron by a
phenomenological fit and by a fit based on the constituent quark model. It is
shown that it is possible to fit all form factors coherently with both
ansaetzen and that they all show the signal of the pion cloud.         
}

\maketitle


\section{Introduction}
\label{sec:introduction}
Form factors encode unique information about the internal structure of
a scatterer provided they are determined with sufficient
precision over a sufficiently large range of momentum transfer. 
Depending on the interaction, the Fourier transformation of the form factors 
gives the spatial distribution of e.g. mass, charge or magnetisation,
which provides insight into several 
aspects of the internal structure of the scatterer:
\begin{itemize}
\item the constituents present in the system,
\item their interaction,
\item and their wave functions.
\end{itemize}
Therefore, form factors represent very significant tests of any model of 
the scatterer.

The nucleon is realized in nature in two species, the proton with one 
charge unit and the neutron with no net charge. While the proton
should dominantly be describable by the s-state wave functions of the 
two up and the one down constituent quark,
these contributions cancel to first order in the neutron
and its electric form factor should be zero in this approximation. 
In this simple picture it is assumed that the quarks are dressed by
gluons and sea quarks forming ``constituent quarks'' which represent
effective Fermions with equal masses of about one third of the nucleon
mass. However, already before the realization of the quark-gluon 
structure of the nucleon the perception of a pion cloud around the 
nucleon was used in order to account for the Yukawa 
interaction between the nucleons. After early pure quark models of 
the nucleon, like the MIT bag, the necessity of a pion cloud was soon 
realized in order to preserve chiral symmetry at the nucleon surface, 
and the ``little bag'' and ``cloudy bag'' models were invented. Based 
on the fundamental chiral symmetry of the QCD Lagrangian 
``chiral dynamics'' was developed identifying the pion as the almost 
Goldstone Boson of the strong interaction. The ``chiral perturbation 
theory'' based on it has shown through many experimental tests that 
indeed the pion is a decisive constituent of the nucleon besides the 
elementary quarks and gluons. It is the purpose of this paper to check
whether and how the pion cloud is reflected in the nucleon form factors.

Because of its zero charge,
the contribution of the bare neutron $n^0$ to the electric form factor
of the physical neutron $n$ is small, and the dissociation of a nucleon 
into its counterpart (here: the proton), and a charged pion 
(here: a negative pion) should
emerge most clearly in the neutron's electric form factor $G_{En}$.
We therefore start out in section~\ref{sec:gen} with a discussion of 
$G_{En}$ for which now data exist from polarisation measurements
\cite{Ed94,Os99,He99,Pa99,Ro99,Be99,Ber01,Gol01,Zhu01,Sei02,Gla02,Mad02,Rei02,Day02}
having a smaller model dependence than previous determinations.
In section~\ref{sec:data} we give an overview of the existing relevant
form factor measurements and of our data selection for the present
investigation. 
In section~\ref{sec:phenomen} we show that, at the percentage level,
the peculiar structure observed in $G_{En}$,
namely a kind of bump around $Q^2 \approx 0.2-0.3$~(GeV/c)$^2$,
is also present in the other form factors $G_{Ep}$, $G_{Mp}$, and $G_{Mn}$.
While this is discussed in section~\ref{sec:phenomen} in terms of a
purely phenomenological
ansatz for the form factors, we show in section~\ref{sec:model} how
this can be viewed in the light of the decomposition of the nucleon states
into a constituent quark core and a polarisation term reflecting the
contribution of the pion cloud. The findings 
are discussed in the concluding section~\ref{sec:conclusion}.


\section{\boldmath The triggering conjecture: 
The electric form factor of the neutron $G_{En}$}
\label{sec:gen}

Previous efforts to determine the electric form factor of the neutron,
e. g. from elastic electron scattering on the deuteron, were hampered 
by severe model-dependencies of the results, which therefore
were uncertain to about 50\%. The emerging results
were describable by the so called Galster 
parameterisation, which started out from the usual dipole fit,
which reproduced $G_{Ep}$, $G_{Mp}$, and $G_{Mn}$ reasonably well
and which was multiplied by some appropriate function
in order to account for the condition $G_{En}(Q^2 = 0) = 0$ required
by the vanishing charge of the neutron.
This Galster form is given by 
\begin{equation}
G_{En}(Q^2) = 
\frac{a_G~\tau}{(1+ b_G~\tau)} \cdot \frac{1}{(1+ Q^2/m_D^2)^2} \, ,
\label{eq:Galster}
\end{equation}
where $\tau = Q^2/(2 m_n)^2$ and $m_n = 0.939~$GeV/c$^2$ is the neutron
mass. The parameter $m^{2}_D$ was taken as the standard dipole value
$m_D^2 = 0.71~$(GeV/c)$^2$ and $a_G$ = 1.73 in order to reproduce the 
measured root mean square radius of the neutron of 
$\langle r^2 \rangle=-6~d\,G_{En}(Q^2)/d\,Q^2|_{Q^2=0}=-0.115~$fm$^2$
as determined from the scattering of thermal neutrons~\cite{Kop97}.
Thus the only parameter free to be fitted to the data was $b_G$, and
it was determined to $b_G=4.59$. The Galster form has no particular 
theoretical justification and may rather hide the essential physics.

\begin{figure}[ht] 
\begin{center}
\psfrag{p1}[t][][1.]{$Q^2/$(GeV/c)$^2$}
\psfrag{p2}[b][][1.]{$G_{En}$}
\includegraphics[width=0.47\textwidth]{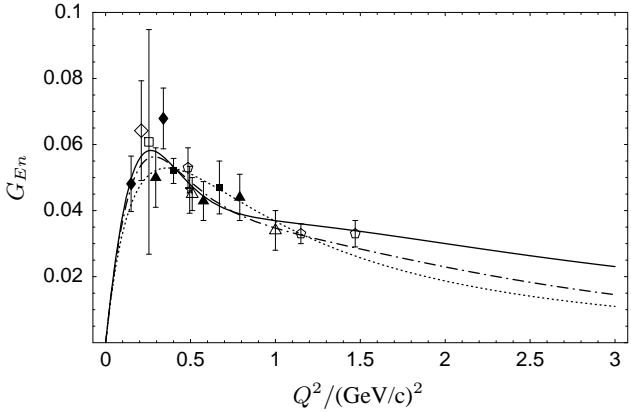} 
\end{center}
\caption{The $G_{En}$ data from polarisation
  measurements. The coding for reactions with the deuteron as a
  neutron target is: open square \cite{Ed94}, filled diamonds
  \cite{Os99,He99}, open diamond \cite{Pa99}, open star \cite{Zhu01}, 
  open triangle \cite{Day02}, open pentagon \cite{Mad02,Rei02}, 
  and filled triangle \cite{Sei02,Gla02}, the measurements with $^3$He are
  shown as filled squares \cite{Ro99,Be99,Ber01,Gol01}. The full curve
  depicts the fit of the parameters of eq.~(\ref{eq:GE}) to the data,
  the dashed dotted curve is a variant with slightly changed
  parameters as explained in the text, while the dotted curve is a fit
  using the Galster form, i.e. eq.~(\ref{eq:Galster}).}
\label{fig:1}
\end{figure}

The collected data for $G_{En}$ determined recently from polarisation
measurements are depicted in fig.~\ref{fig:1}. 
These 15 data points, which are not hampered by model assumptions,
have been taken with 8 very different experimental setups, 
and the data points taken with the same setups were taken over periods 
separated by long time intervals. Also, the setups had very different 
systematic errors and corrections due to nuclear binding effects. 
Therefore, it is justified to consider the data  as statistically independent. 
Since the corrections are less certain for the measurements on $^3$He
than for the loosely bound deuterium, 
the measurements on the two targets are distinguished in fig.~\ref{fig:1}
by markedly different symbols. 
It is not the aim of this paper, however, to discuss critically these 
experiments but just to take this data set seriously and to 
investigate its essential features.

It is evident from fig.~\ref{fig:1} that the data can be as well
regarded as a broad distribution and a peak around 
$Q^2 \approx 0.3 $~(GeV/c)$^2$ not present in the smoother Galster fit.
 
In order to get some insight into the consequences of this alternative
form we have added a term to the form of eq.~(\ref{eq:Galster}) which 
is able to describe an additional peak with reasonable boundary conditions.
\begin{equation}
\begin{split} 
G_{En}(Q^2) = 
\frac{a~Q^2}{(1+ b~Q^2 + c~Q^{4})^5} 
            + \\ + \frac{d~Q^2}{(1 + e~Q^2)(1 + f~Q^2)^2} \, .
\end{split}
\label{eq:GE}
\end{equation}
The rms radius is now given by the sum of $a$ and $d$, 
constrained to $(a+d)(2 m_n)^2 = a_G$ = 1.73, and we fixed
$a$ and $d$ to $a=0.37~$(GeV/c)$^{-2}$ and $d=0.12~$(GeV/c)$^{-2}$. 
The parameters $e$ and $f$ were kept fixed at $0.5~$(GeV/c)$^{-2}$. 
Minimising $\chi^2$ yielded $b=0.39~$(GeV/c)$^{-2}$ and 
$c=1.68~$(GeV/c)$^{-4}$. Here we only want to have a parametrisation
which reproduces the data within the experimental error bars
without associating any particular physical meaning to the single
parameters. In fact, as seen in fig.~\ref{fig:1}, this form 
reproduces the data well. 
It is not meaningful to go into any detail of an error analysis, 
instead we only show by the example of the dashed-dotted curve that
with above parametrisation the ``peak-region'' and
the tail to higher momentum transfers are essentially 
described independently from each other. For completeness we just
mention that the $\chi^2$ of the Galster form is by  
$\Delta \chi^2 = 4.8$ bigger than that of the two others.

As is well known \cite{HaM90}, though sometimes questioned (for a
discussion of this problem see ref.~\cite{Isg98}), the
Fourier transform of the electric and magnetic Sachs form factors 
$G_{E}(Q^2)$ and $G_{M}(Q^2)$ represent the charge and magnetic
density distribution in the Breit frame, where 
the energy transfer $\omega = 0$ and the 
three-momentum transfer $|\vec{q}_{Breit}| = Q$;
we denote these distributions by $\rho (r)$, which thus is 
given by
\begin{equation}
\rho(r) =
\frac{4\pi}{(2\pi)^3}\int_0^{\infty}G(Q)\frac{sin(Qr)}{Qr}Q^2dQ \, .
\label{eq:Fourier}
\end{equation}
Refinements to this relation are discussed in detail in 
ref.~\cite{Kel02} where it is also pointed out that corrections 
cannot be defined without model assumptions.
Since we are interested in the gross features of the
measured form factors and the spatial distributions, 
we base our further discussion on 
eq.~(\ref{eq:Fourier}). A more refined approach may result
in some compression of the resulting distributions
in r-space, which should not alter their salient features
and which are therefore left out of consideration in this paper.
 
\begin{figure}[ht] 
\begin{center}
\psfrag{p1}[t][][1.]{$r$/fm}
\psfrag{p2}[b][][1.]{$r^2 \rho_n(r)\cdot$fm}
\includegraphics[width=0.47\textwidth]{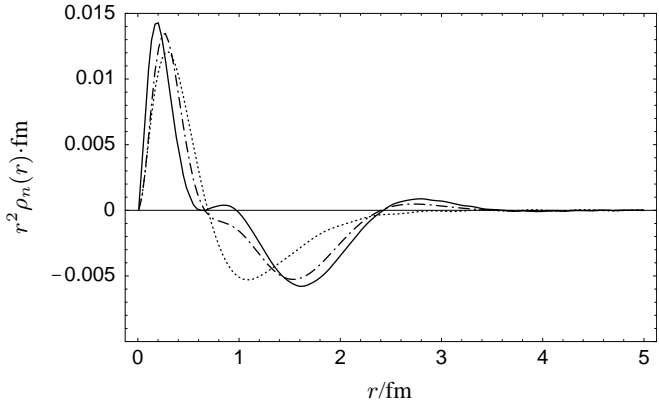} 
\end{center}
\caption{The differential radial charge distribution of the neutron in
  the Breit frame as derived by a Fourier transform. The coding of the 
  lines is that of fig.~\ref{fig:1}.}
\label{fig:2}
\end{figure}

Fig.~\ref{fig:2} shows the charge distribution in the neutron,
$\rho_{En}$, calculated via eq.~(\ref{eq:Fourier}) with above 
given fits to $G_{En}$.
We have plotted $r^2\rho_{En}(r)$ which represents the 
charge in a spherical shell at radius $r$.
The charge distribution of the Galster fit shows the well known 
``aperiodic'' shape  with a positive
bump in the interior and a negative bump at the outside of the neutron.
This characteristic feature also results from an ansatz for the form
factor with the superposition of two appropriate dipole forms, to which
the Galster parameterisation is a good approximation.

The fit with eq.~(\ref{eq:GE}), however, which accounts for the bump
in $G_{En}$ at $Q^2 \approx 0.3~$(GeV/c)$^2$, results in an oscillatory 
behaviour of $\rho_{En}(r)$ (see fig.~\ref{fig:2}). Though the
oscillatory behaviour depends on the particular fitting form we shall
show that it is the bump which shifts more charge to the outside than 
does the Galster fit.
Since this outer region should be dominated by the pion cloud,
the corresponding contribution should show up as a general feature
also in the other form factors, where, however, 
a form factor bump of the same order of magnitude can only be
expected to be a few-percent contribution.

With this in mind, we reconsider all four nucleon form factors in the following.


\section{The data base}
\label{sec:data}

Table~\ref{tab:ff_data} gives an overview of the data which we have
taken into consideration together with the $Q^2$-ranges which they cover.

\begin{table}
\begin{center}
\tabcolsep6.8pt
\renewcommand{\arraystretch}{1.3}
\begin{tabular}{||l|l|l||}
\hline
\hline
Measurement  & $Q^2$-range     & reference                      \\
\hline
$G_{Ep}$     &                 &                                 \\
$p(e,e')$    & ~0.01 - ~0.05   & Simon et al.~\cite{Sim80}       \\
             & ~0.04 - ~1.75   & Price et al.~\cite{Pri71}       \\
             & ~0.39 - ~1.95   & Berger et al.~\cite{Ber71}      \\
             & (1.75 - ~8.83)  & Andivahis et al.~\cite{And94}   \\
$d(e,e'p)$   & ~0.27 - ~1.76   & Hanson et al~\cite{Han73}       \\
$p(\stackrel{\rightarrow}{e},e'\stackrel{\rightarrow}{p})$ 
             & ~0.37 - ~0.44   & Pospischil et al.~\cite{Pos01}  \\
             & ~0.38 - ~0.50   & Milbrath et al.~\cite{Mil00}    \\
             & ~0.40           & Dieterich et al.~\cite{Die01}   \\
             & ~0.49 - ~3.47   & Jones et al.~\cite{Jon00}       \\
             & ~3.50 - ~5.54   & Gayou et al.~\cite{Gay02}       \\
\hline
$G_{Mp}$     &                 &                                 \\
$p(e,e')$    &  ~0.02 - ~0.15  & Hoehler et al.~\cite{Hoe76}     \\
             &  ~0.16 - ~0.86  & Janssens et al.~\cite{Jan66}    \\
             &  ~0.39 - ~1.75  & Berger et al.~\cite{Ber71}      \\
             &  ~0.67 - ~3.00  & Bartel et al.~\cite{Bar73}      \\
             &  ~1.00 - ~3.00  & Walker et al.~\cite{Wal94}      \\
             &  ~1.50 - ~3.75  & Litt et al.~\cite{Lit70}        \\
             &  ~1.75 - ~7.00  & Andivahis et al.~\cite{And94}   \\
             &  ~2.86 - 31.2   & Sill et al.~\cite{Sil93}        \\
$d(e,e'p)$   &  ~0.27 - ~1.76  & Hanson et al~\cite{Han73}       \\
\hline
$G_{En}$     &                &                                  \\
$d(\stackrel{\rightarrow}{e},e'\stackrel{\rightarrow}{n})p$ 
             & 0.15           & Herberg et al.~\cite{He99}       \\
             & 0.26           & Eden et al.~\cite{Ed94}          \\
             & 0.30,~~0.58    & Seimetz et al.~\cite{Sei02}      \\
             & 0.34           & Ostrick et al.~\cite{Os99}       \\
             & 0.49 - ~1.47   & Madey et al.\cite{Mad02,Rei02}   \\
             & 0.76           & Glazier et al.~\cite{Gla02}      \\             
             & 1.00           & Day et al.\cite{Day02}           \\
$\stackrel{\rightarrow}{d}(\stackrel{\rightarrow}{e},e'n)p$ 
             & 0.21           & Passchier et al.~\cite{Pa99}      \\
             & 0.50           & Zhu et al.~\cite{Zhu01}          \\
$\overrightarrow{^3He}(\stackrel{\rightarrow}{e},e'n)$ 
             & 0.40           & Becker et al.~\cite{Be99,Ber01,Gol01}\\
             & 0.67           & Rohe et al.~\cite{Ro99,Gol01}    \\
$d(e,e')$    & (0.27 - 1.76)  & Hanson et al~\cite{Han73}        \\
             & (1.75 - 4.00)  & Lung et al.~\cite{Lun93}         \\
\hline
$G_{Mn}$     &                &                                  \\
$d(e,e'n)p$  & ~0.07 - ~0.89  & Kubon et al.~\cite{Kub02}        \\
             & ~0.10 - ~0.20  & Xu et al.~\cite{Xu00}            \\
             & ~0.11          & Anklin et al.~\cite{Ank94}       \\ 
             & (0.11 - ~0.26) & Markowitz et al.~\cite{Mar93}    \\
             & (0.13 - ~0.61) & Bruins et al.~\cite{Bru95}       \\
             & ~0.24 - ~0.78  & Anklin et al.~\cite{Ank98}       \\ 
$d(e,e'p)$   & (0.27 - ~1.76) & Hanson et al~\cite{Han73}        \\
$d(e,e')$    & ~1.75 - ~4.00  & Lung et al.~\cite{Lun93}         \\
             & ~2.50 - 10.0   & Rock et al.~\cite{Roc82}         \\
\hline
\hline
\end{tabular}
\end{center}
\caption{Overview of data taken into consideration ($Q^2$ in (GeV/c)$^2$). 
The data left out in the final analysis are put into parentheses. The
reactions are as indicated; $d(e,e')$ refers to quasi elastic scattering.}
\label{tab:ff_data}
\end{table}

For $G_{Ep}$ we have omitted in the final analysis the data by Andivahis 
et al.~\cite{And94}. In the $Q^2$-range of these data, $G_{Ep} \ll G_{Mp}$, 
thus its determination via a Rosenbluth separation is quite uncertain.
In fact these data are clearly incompatible with the new results
from polarisation measurements in which not the sum of $G^2_{Ep} + 
\tau \cdot G^2_{Mp}$ is measured but the ratio $G_{Ep}/G_{Mp}$. 
It is straightforward to determine $G_{Ep}$ from this ratio if one 
takes the prevailing $G_{Mp}$ as known from measurement.

For $G_{Mp}$ we took into account the same data as Kelly~\cite{Kel02}, i. e. 
the data by Hoehler et al.~\cite{Hoe76}
up to $Q^2$ = 0.15~(GeV/c)$^2$ and those revised and compiled recently by
Brash et al.~\cite{Bra02}. 
In addition we also used the data by Hanson et al.~\cite{Han73}.

For $G_{En}$ we have only taken into account the data from polarisation
measurements. The measurement in~\cite{Lun93,Han73} give only
$G_{En}^2$, thus the sign of $G_{En}$ remains undetermined, and the
errors are so large that the 
data can essentially be regarded as upper limits only;
we did not take them into account in the fits. 
Other determinations of $G_{En}$ were
very uncertain due to the model dependency of the extraction of
$G_{En}$ from the measured cross sections,  and we did not take them into
consideration. 

For $G_{Mn}$ the data by Markowitz et al.~\cite{Mar93} and by Bruins
et al.~\cite{Bru95} were omitted in the analysis as was already done 
(and argued) in Kubon et al.~\cite{Kub02}. Also,
the data by Hanson et al.~\cite{Han73} are omitted 
since they deviate substantially from the trend of the more recent
data measured with the detection of the neutron in coincidence.
\footnote{After these investigations were finished a new measurement of
$G_{Mn}$ was published \cite{Xu03} and came to our attention. Since 
the $G_{Mn}$ values were derived from the reaction 
$\overrightarrow{^3He}(\stackrel{\rightarrow}{e},e'n)$ 
they carry a large systematic error due to the correction of the binding
of the neutron in $^3He$. They lie even below the values of 
ref.~\cite{Ank98} and support the same trend. We repeated our fits with
these values taking the error into account and found that the results 
presented in the following are not changed.}

\section{A phenomenological description of the nucleon form factors}
\label{sec:phenomen}

\subsection{The phenomenological ansatz}
\label{sec:phenomen_ansatz}

What we are particularly interested in is the
existence of a small bump on top of a large ``smooth main part''.
This raises the question after what is ``smooth main part'' and what 
is ``bump''. After the investigation of several parameterisations we 
decided to keep close to what one is used to in the description of 
the nucleon form factors, namely the dipole form.
In fact it is fascinating that the three form factors $G_{Ep}$,
$G_{Mp}$, and $G_{Mn}$ are describable to quite a precision
by the dipole form with the one parameter $m_D$ given above
as it has entered into the text books, e. g.~\cite{HaM90,Per87}.
Such good description could make one believe that there is some
physical meaning in the parameter $m_D$. In fact there is none.

Looking at the form factors more closely, however, the precision
with which the measurements are reproduced by this simple
parameterisation is limited. While this has been realized 
on the percentage level already in the high-precision 
measurement of $G_{Ep}$ at low $Q^2$ by Simon et al.~\cite{Sim80},
this became completely obvious at high $Q^2$ by the polarisation 
measurements by Jones et al.~\cite{Jon00}. 

Purely phenomenologically, we describe the smooth part of the form 
factors, $G_s(Q^2)$, by the dipole form. In order to be
somewhat flexible, however, we took into account the superposition 
of two dipoles: 
\begin{equation}
G_s(Q^2) = \frac{a_{10}}{(1+Q^2/a_{11})^2} 
         + \frac{a_{20}}{(1+Q^2/a_{21})^2} \, .
\label{eq:ff_dipol}
\end{equation}
To account for a possible bump on top of the smooth
form factor we now take a parameterisation which is easy to 
handle and the  parameters of which give direct
insight into the characteristic features of such a bump,
namely its amplitude, position  and width, $a_b$, $Q_b$, 
and $\sigma_b$, respectively.
A quite natural choice would be a Gaussian positioned at
$Q_b$. For $Q_b \neq 0$, however, such Gaussian contains uneven 
powers in $Q$, which is not allowed for a function
representing a form factor. This shortcoming can be healed 
by superimposing two Gaussians as introduced in r-space by Sick 
for his ansatz for a model-independent analysis of nuclear
charge distributions~\cite{Sic74}. We thus parameterise
the bump as
\begin{equation}
G_b(Q^2) = e^{\textstyle -\frac{1}{2}(\frac{Q - Q_b}{\sigma_b})^2 } 
         + e^{\textstyle -\frac{1}{2}(\frac{Q + Q_b}{\sigma_b})^2 } \, . 
\label{eq:ff_gauss}
\end{equation}

In order to keep as close as possible to the accustomed
description of the form factors, we attribute the full
normalisation to the dominating smooth part,
i. e. $G_N(Q^2 = 0) = G_s(Q^2 = 0)$, i. e. in this
ansatz the smooth part accounts for the full charge or magnetic
moment, respectively. To make sure, that, independent
of the fitted values for its parameters,
$G_b$ does not interfere with this normalisation, we
multiply it by $Q^2$.
We thus parameterise the nucleon form factors by the ansatz
\begin{equation}
G_N(Q^2) = G_s(Q^2) + a_b \cdot Q^2 G_b(Q^2) \, ,
\label{eq:ff_nucleon}
\end{equation}
where $a_b$ is essentially the amplitude of the bump.

\subsection{Fit of the form factors with the phenomenological ansatz}
\label{sec:results_phenomen_ansatz}

The parameters from the fits of above phenomenological ansatz 
to the form factors $G_{Ep}$, $G_{Mp}$, $G_{En}$, and $G_{Mn}$
are compiled in table~\ref{tab:par_phen}.
The given errors are the standard errors from the
fit procedure, which also account for correlations, therefore
the parameters cannot just be varied independently within
these margins.
 
The main purpose of these fits is to allow a coherent view
on the measured form factors in order to reveal certain common
features. Therefore, we do not go into details of these fits. 
However, a large number of fits with other analytical forms were 
also tried. They all point to the same feature of a structure with 
a width of $\approx 0.2~$(GeV/c)$^2$.

The final fitting results given here have been performed with fixed
normalisation which was guaranteed by setting $a_{20} = 1 - a_{10}$
and fitting only $a_{10}$ to the data ($a_{20} = -a_{10}$ for $G_{En}$). 
Fits with both $a_{10}$ and $a_{20}$ as free parameters did not improve
the fits by more than $\Delta \chi^2 = -2$;
this means that there is no hint for normalisation problems
in the data.

\begin{figure}
\begin{center}
\vspace*{-0.0cm} \rotatebox{-90} {\resizebox{6.0cm}{!}
{\includegraphics[width=1.0\textwidth]{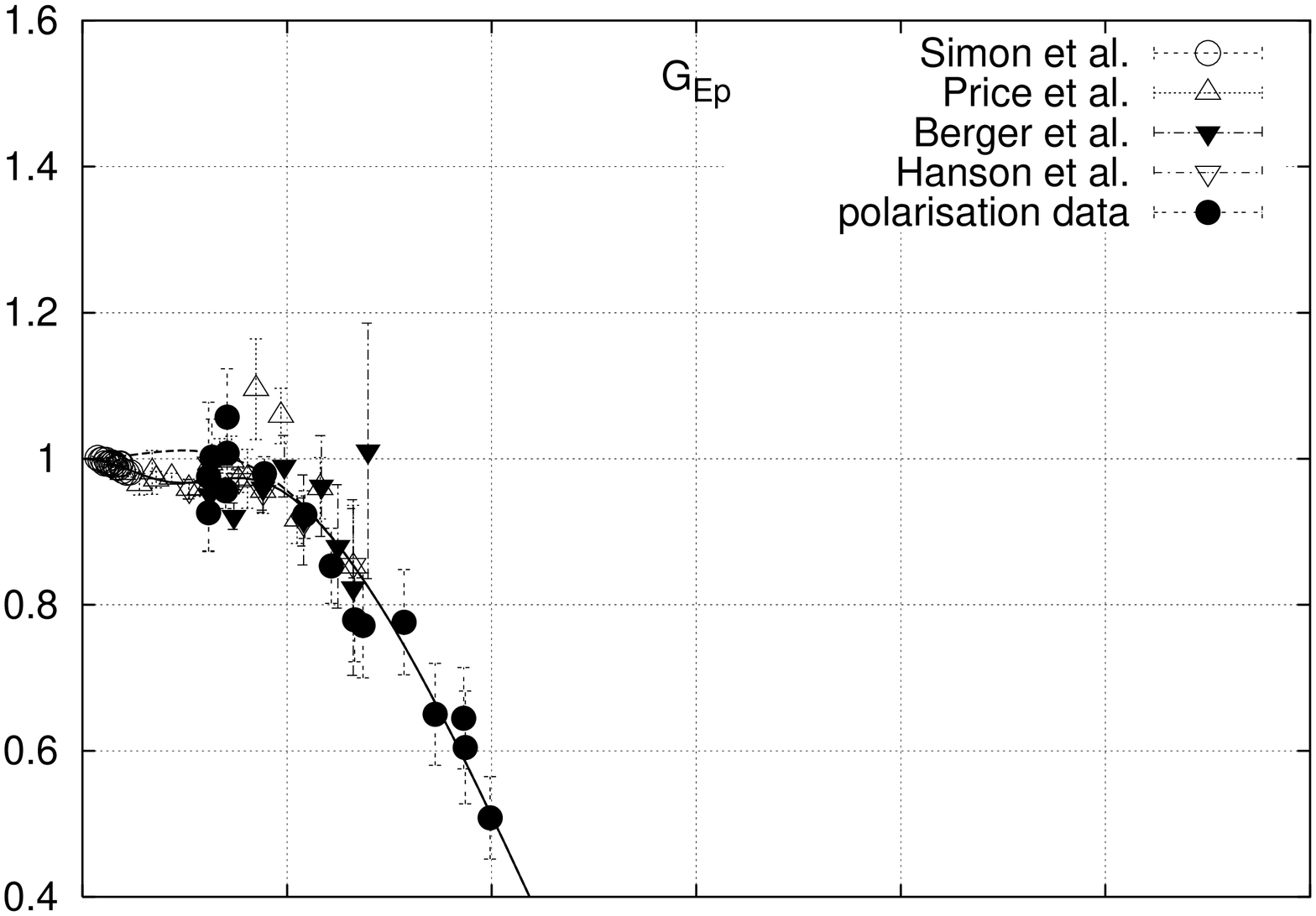}} } \\
\vspace*{-0.5cm} \rotatebox{-90} {\resizebox{6.0cm}{!}
{\includegraphics[width=1.0\textwidth]{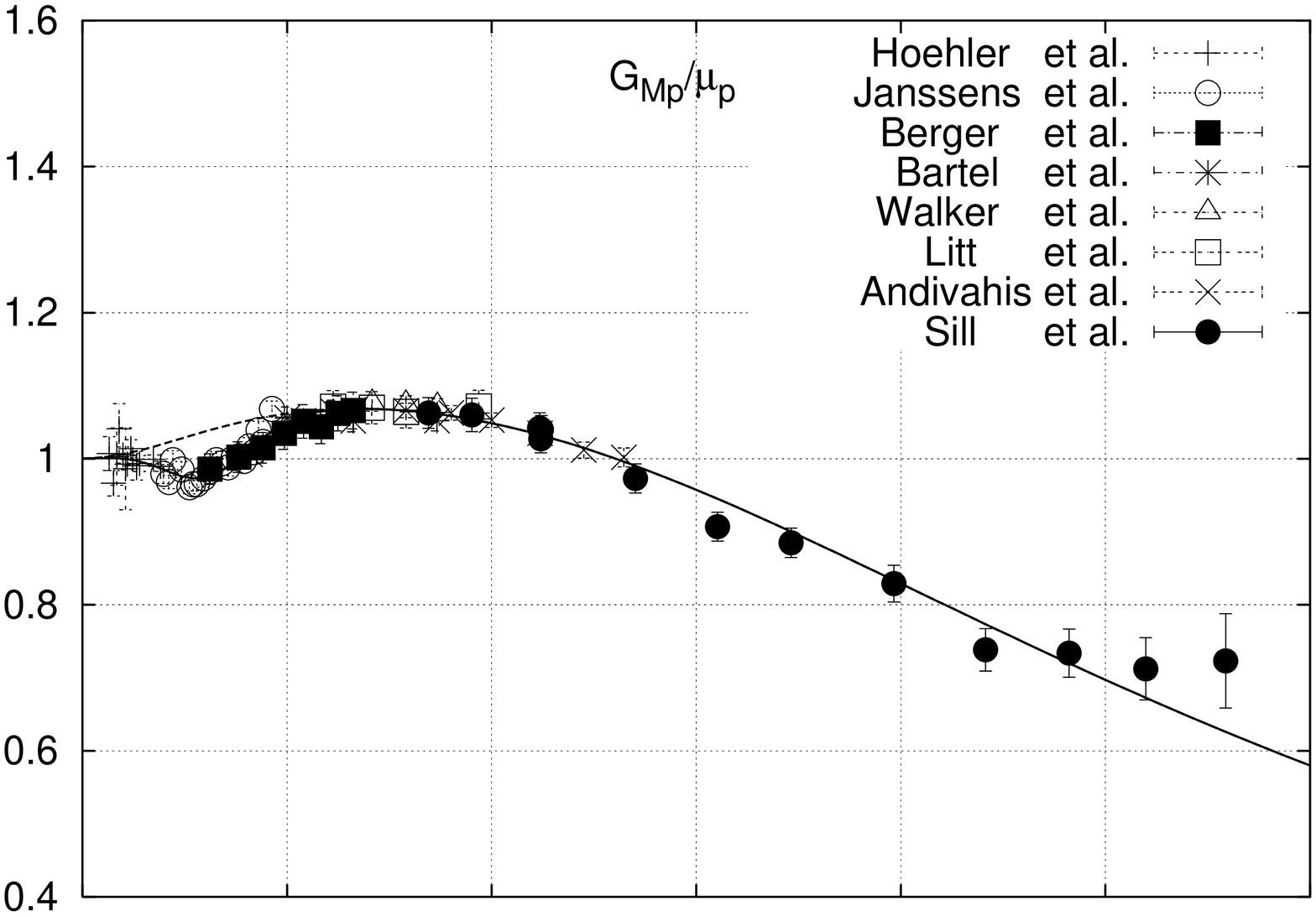}} } \\
\vspace*{-0.5cm} \rotatebox{-90} {\resizebox{6.0cm}{!}
{\includegraphics[width=1.0\textwidth]{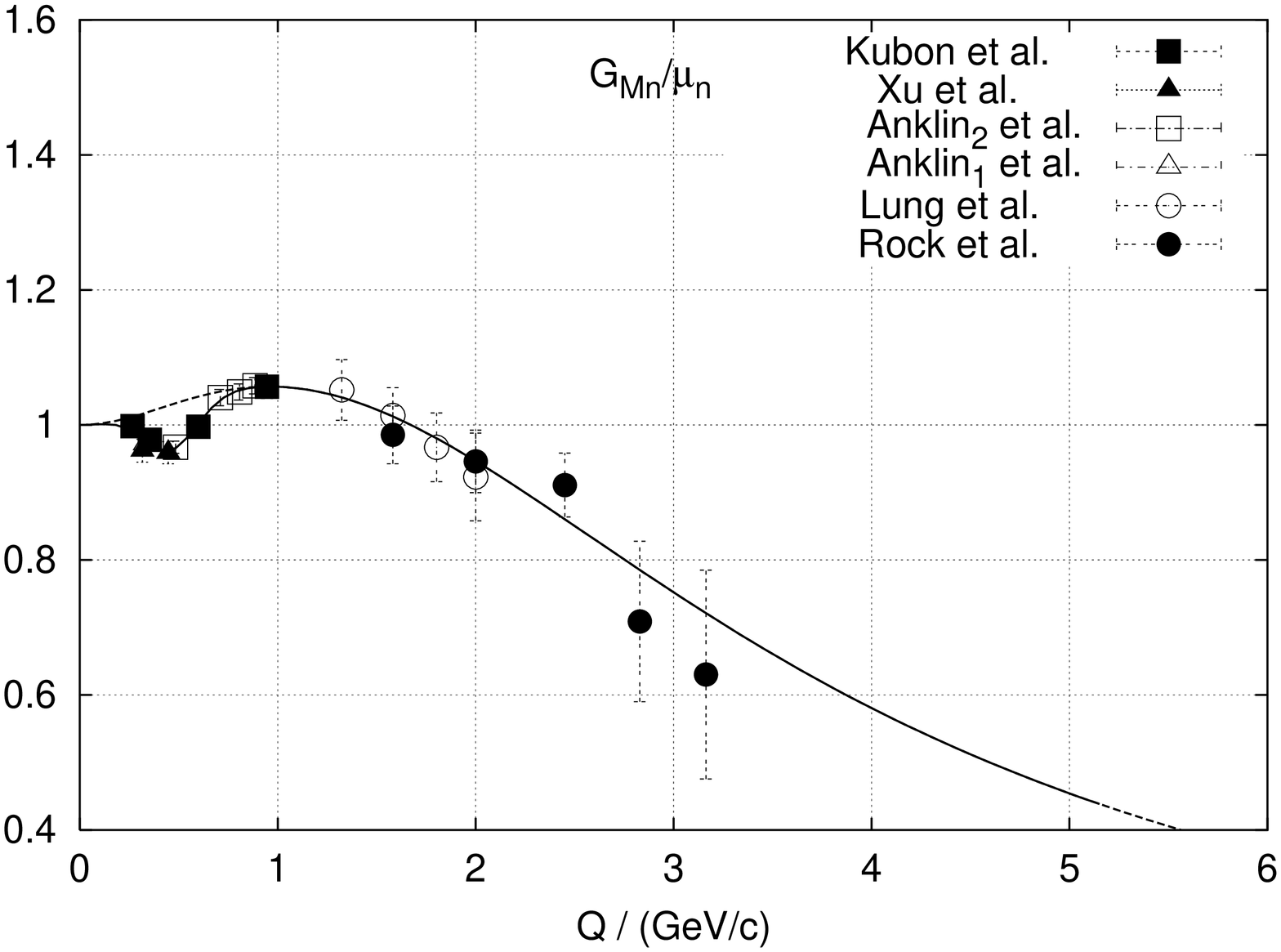}} } \\ [0.3cm]
\end{center}
\caption{The measured nucleon form factors and their
 phenomenological description divided by the standard dipole
 form factor. The full line represents the full fit, while the
 broken line is only the ``smooth main part'', i. e. the sum of the
 two dipoles.}
\label{fig:ff_durch_dipol}
\end{figure}

Let us first look at the ``standard form factors'' $G_{Ep}$, $G_{Mp}$, 
and $G_{Mn}$, which we refer to as $G_N^{std}$.
For these,
one dipole (the first) accounts for the overwhelming part of the 
strength at low q, i. e. it carries most of the charge or magnetisation, 
respectively. The slope constants of all three $G_N^{std}$ essentially
agree within the errors. In fact, the mean value 0.76~(GeV/c)$^{-2}$
is near to the value $m^2_D$ = 0.71~(GeV/c)$^{-2}$ of the standard 
dipole fit, however the deviation is significant.
In fact one cannot expect equal parameters since the standard
dipole fit deviates systematically from the data 
(c.f. fig.~\ref{fig:ff_durch_dipol}) 
and thus also from our fits which reproduce the data within the
experimental errors.

It is interesting to note, that also the slope parameters
$a_{21}$ of the second dipole form are very similar for all three 
$G_N^{std}$ (cf. fit 1 for $G_{Mp}$ in table \ref{tab:par_phen}).
This term (with negative amplitude) accounts for the fact that the
measurements fall below the dipole fit at larger $Q^2$.
While this became obvious for $G_{Ep}$ at larger $Q^2$ from
polarisation measurements, a systematic deviation from the
dipole fit was already observed by Simon et al.~\cite{Sim80} at low $Q^2$, 
though only at the percent level.
It may now be somewhat surprising that the slope parameters $a_{21}$
are so similar for all three $G_N^{std}$. In fact, if one accounts in $G_{Mp}$
also for the high $Q$ region (fit 2), this is no longer the case.

\begin{table*}
\begin{center}
\begin{tabular}{||l||l|l|l|l||l|l|l||r|c|r||}
\hline
\hline
              &           \multicolumn{4}{c||}{     }      & \multicolumn{3}{c||}{     }      & \multicolumn{3}{c||}{     }       \\
              &           \multicolumn{4}{c||}{$G_s$}      & \multicolumn{3}{c||}{$G_b$}      & \multicolumn{3}{c||}{     }       \\
              &           \multicolumn{4}{c||}{     }      & \multicolumn{3}{c||}{     }      & \multicolumn{3}{c||}{     }       \\
\hline
              &           &          &           &           &           &          &             &            &                  & \\
form factor   & $a_{10}$  & $a_{11}$ & $a_{20}$  & $a_{21}$  & $a_b$     & $Q_b$    & $\sigma _b$ & $N_{d.f.}$ & $\chi ^2_{d.f.}$ & $\chi^2_{total}$ \\ 
              &     & (GeV/c)$^{2}$ &     & (GeV/c)$^{2}$  & (GeV/c)$^{-2}$ &  (GeV/c) & (GeV/c)   &            &                  &                  \\
              &           &          &           &           &           &          &             &            &                  &                  \\
\hline
              &           &          &           &           &           &          &             &            &                  &                  \\
$G_{Ep}$      & 1.041(40) & 0.765(66)& -0.041(-) & 6.2(5.0)  & -0.009(7) & 0.07(88) & 0.27(29)    & 64         & 0.933            & 59.71            \\
              &           &          &           &           &           &          &             &            &                  &                  \\
$G_{Mp}/\mu _p$ &         &          &           &           &           &          &             &            &                  &                  \\
1             & 1.002(7)  & 0.749(6) & -0.002(-) & 6.0(3.4)  & -0.005(1) & 0.35(7)  & 0.21(3)     & 60         & 0.861            & 51.66            \\
2             & 1.003(7)  & 0.753(2) & -0.003(-) & 16.9(6)   & -0.006(2) & 0.33(7)  & 0.23(3)     & 75         & 0.876            & 65.7             \\
\hline
              &           &          &           &           &           &          &             &            &                  &                  \\
$G_{En}$      & 1.04(10.7)& 1.73(-)  & -1.04(-)  & 1.54(1.94)& ~0.009(6) & 0.29(17) & 0.20(9)     & 10         & 0.861            &  8.61            \\
              &           &          &           &           &           &          &             &            &                  &                  \\
$G_{Mn}/\mu _n$ & 1.012(6)& 0.770(10) & -0.012(-) & 6.8(3.0)  & -0.011(1) & 0.33(3)  & 0.14(2)     & 14         & 0.579            &  8.11             \\
              &           &          &           &           &           &          &             &            &                  &                  \\
\hline
\hline
\end{tabular} \\ [0.5cm]
\end{center}
\caption{Parameters from the fit of the phenomenological ansatz eq.~(\ref{eq:ff_nucleon}) to the electric and magnetic nuclear form factors. 
In the usual way the errors on the parameters are given in brackets;
if no decimal point is given they refer to the last given digits of the parameter.
For $G_{Mp}$, fit 1 uses only the data up to 2~GeV/c, thus it is more comparable to the fits to $G_{Ep}$ and $G_{Mn}$ than the 
fit 2 where all data up to 6~GeV/c are taken into account.}
\label{tab:par_phen}
\end{table*}

\begin{figure} 
\begin{center}
\vspace*{-0.0cm} \rotatebox{-90} {\resizebox{5.8cm}{!}
{\includegraphics[width=1.0\textwidth]{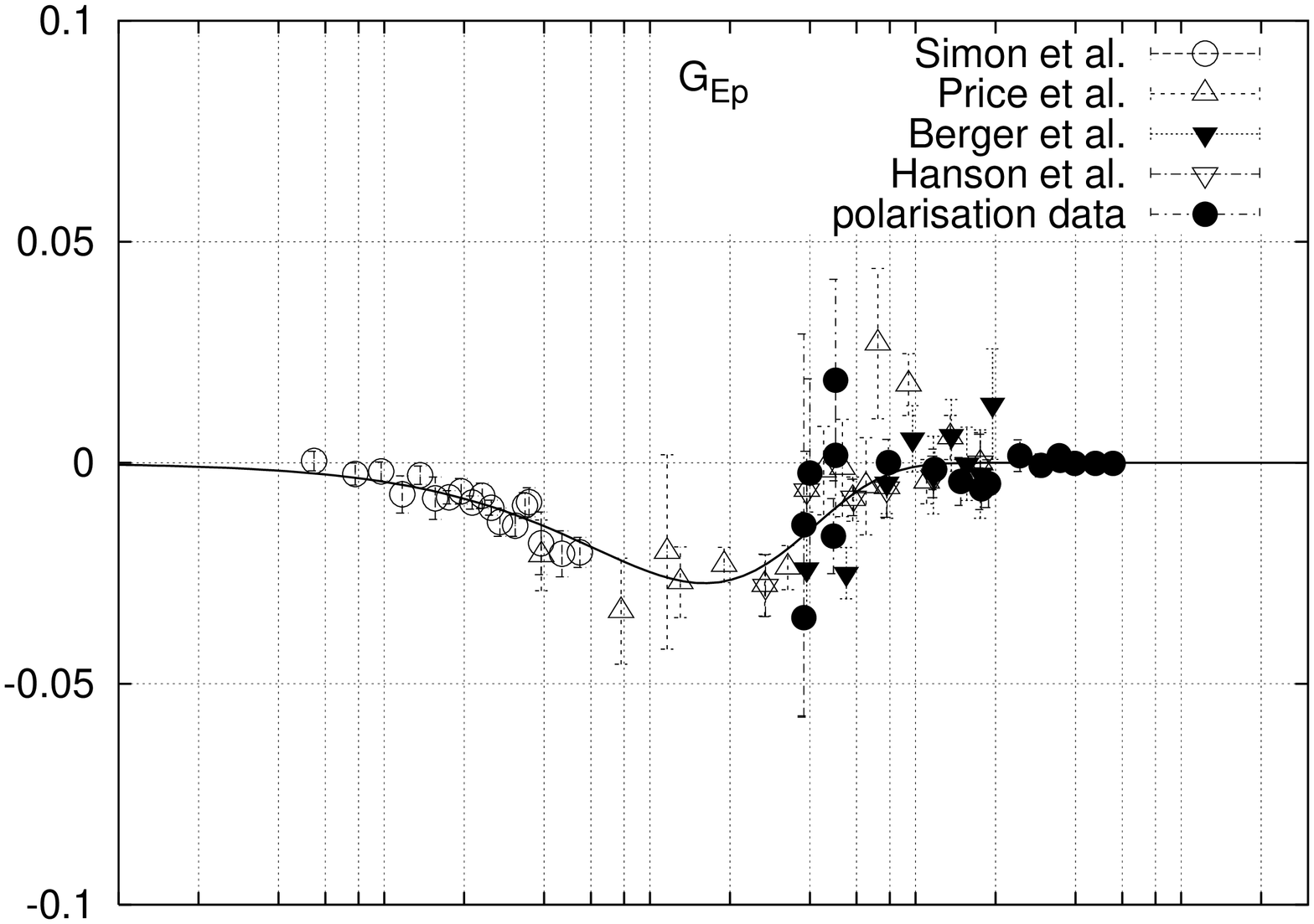}} } \\
\vspace*{-0.5cm} \rotatebox{-90} {\resizebox{5.8cm}{!}
{\includegraphics[width=1.0\textwidth]{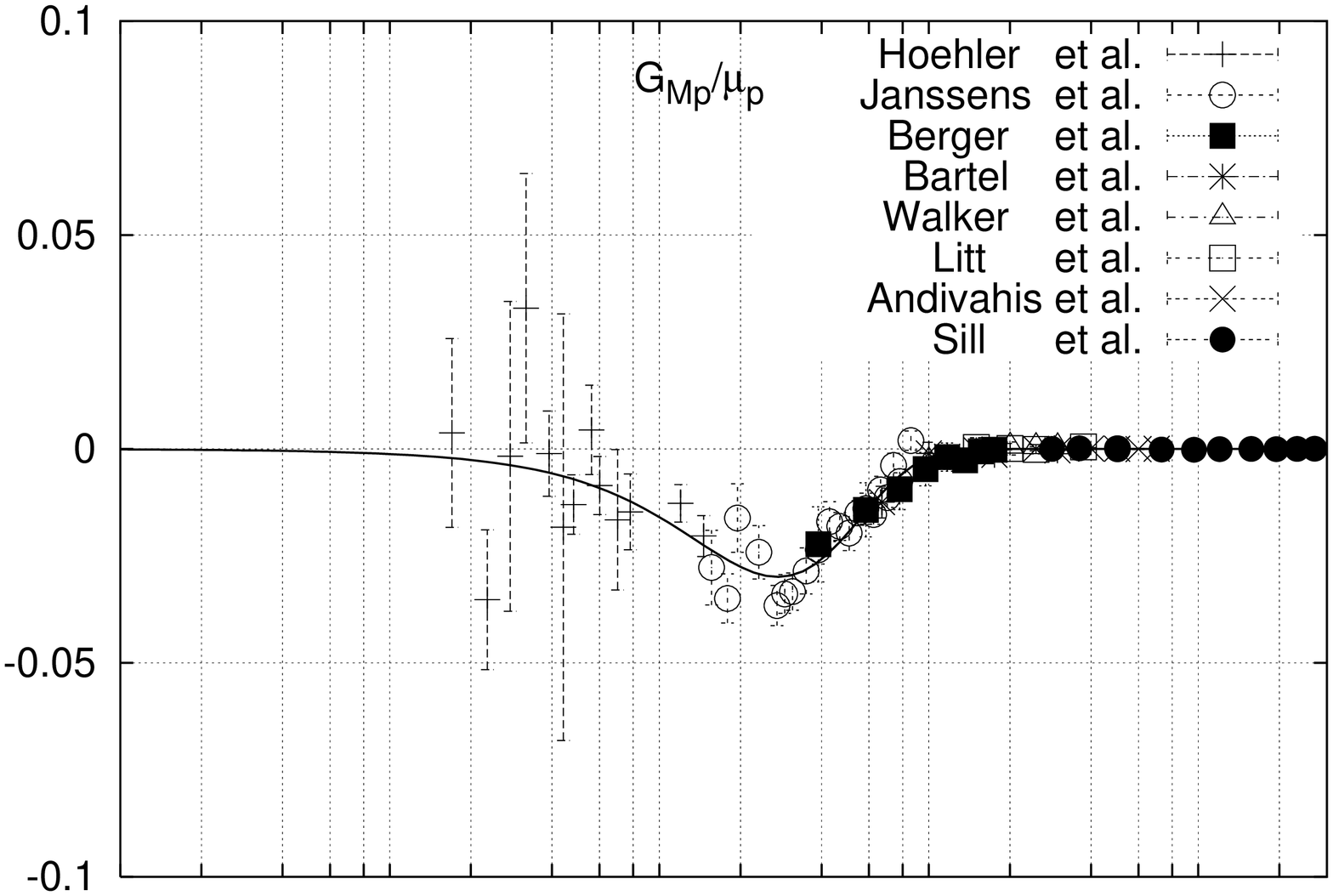}} } \\
\vspace*{-0.5cm} \rotatebox{-90} {\resizebox{5.8cm}{!}
{\includegraphics[width=1.0\textwidth]{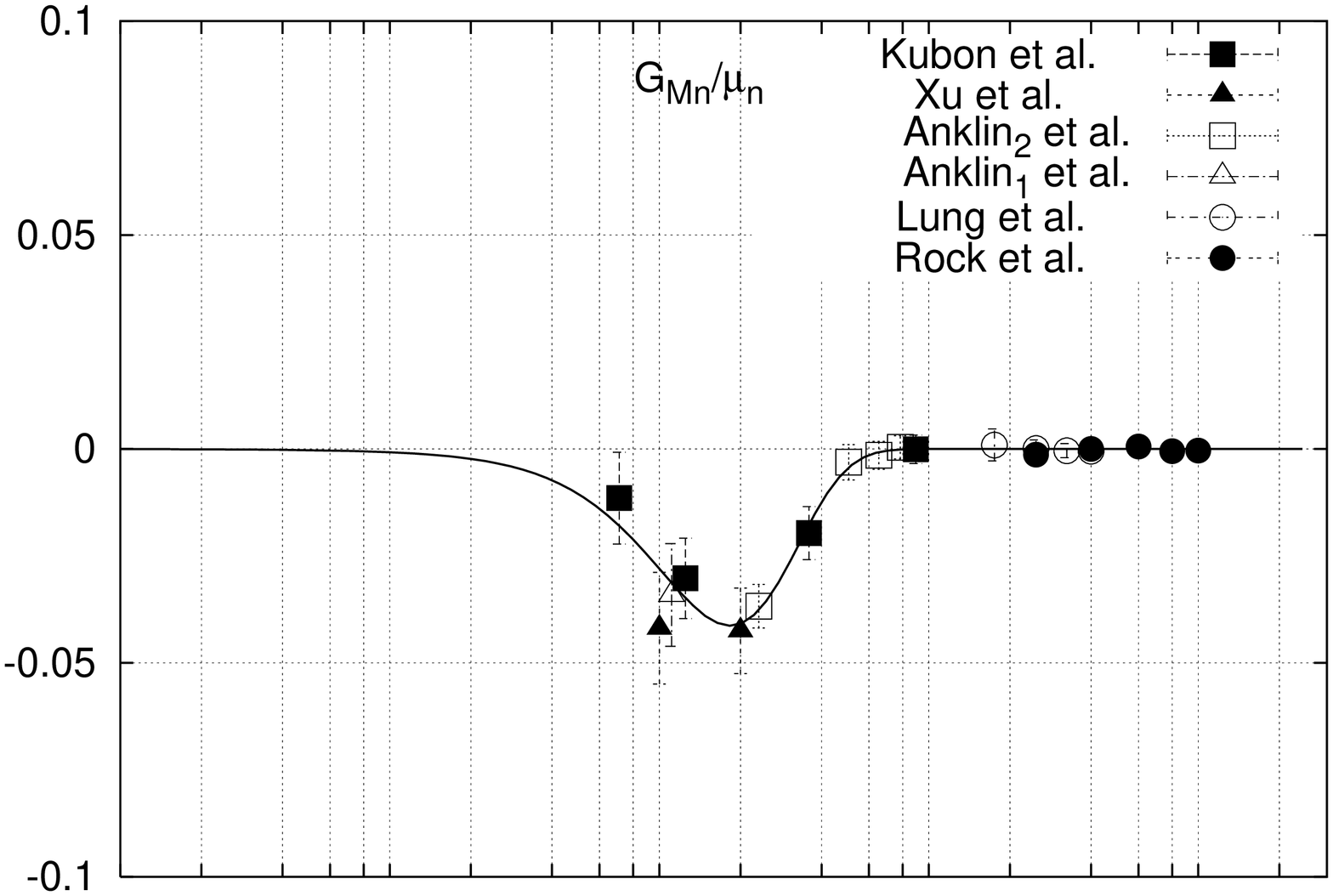}} } \\ [0.3cm]
\vspace*{-0.5cm} \rotatebox{-90} {\resizebox{5.8cm}{!}
{\includegraphics[width=1.0\textwidth]{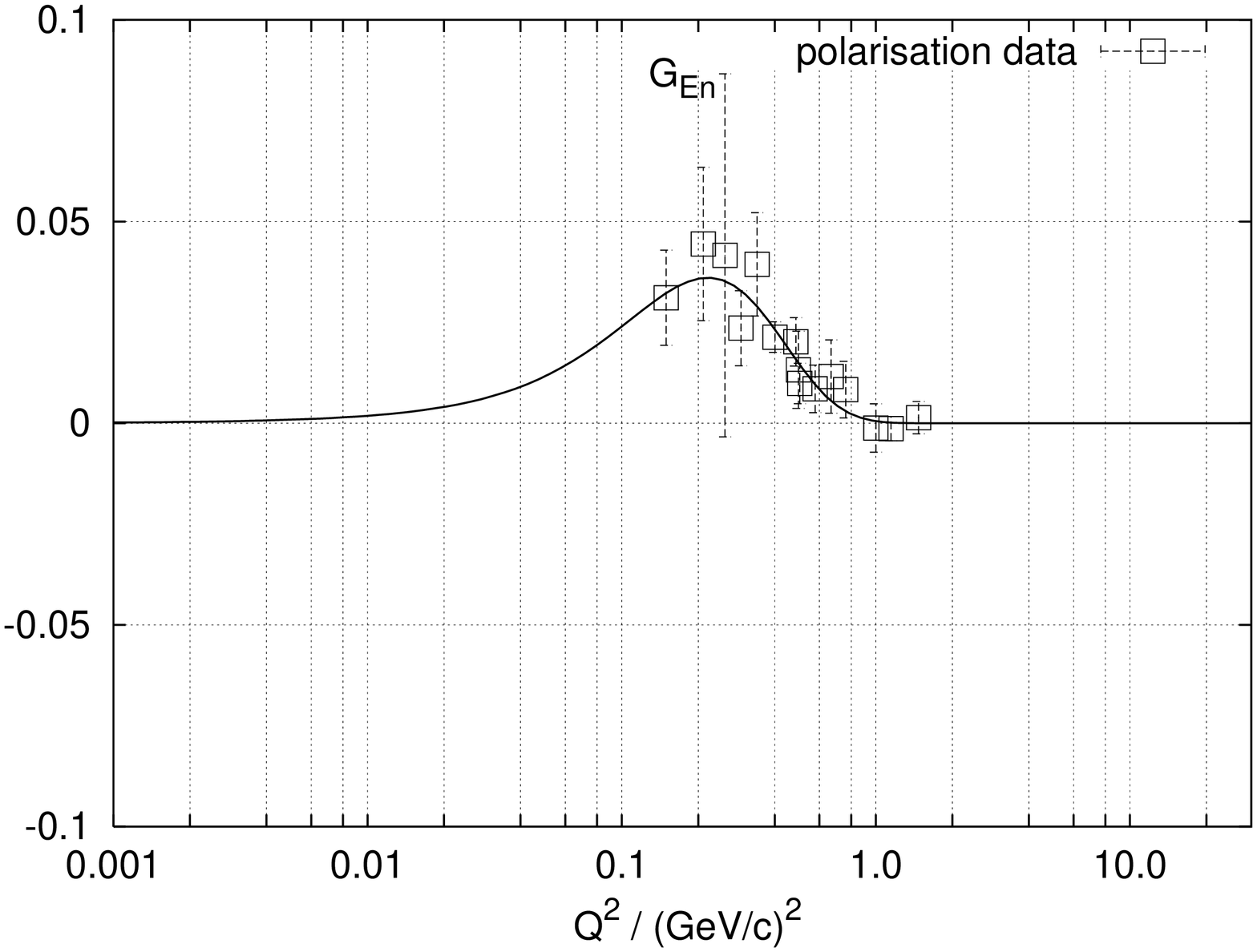}} } \\ [0.3cm]
\end{center}
\caption{The difference between measured nucleon form factors and 
 the smooth part of the phenomenological ansatz with logarithmic
 x-scale for $Q^2/$(GeV/c)$^2$.} 
\label{fig:ff_minus_smooth}
\end{figure}

A direct interpretation of the bump structure in terms of the parameters 
$a_b$, $Q_b$, and $\sigma _b$ is obscured by the multiplication of the 
Gaussian with $Q^2$. Therefore, for a discussion of this structure
we refer to its graphical representation in fig.~\ref{fig:ff_minus_smooth}.

\begin{figure} 
 \vspace*{-0.0cm}\rotatebox{-90} {\resizebox{5.5cm}{!}
 {\includegraphics[width=1.0\textwidth]{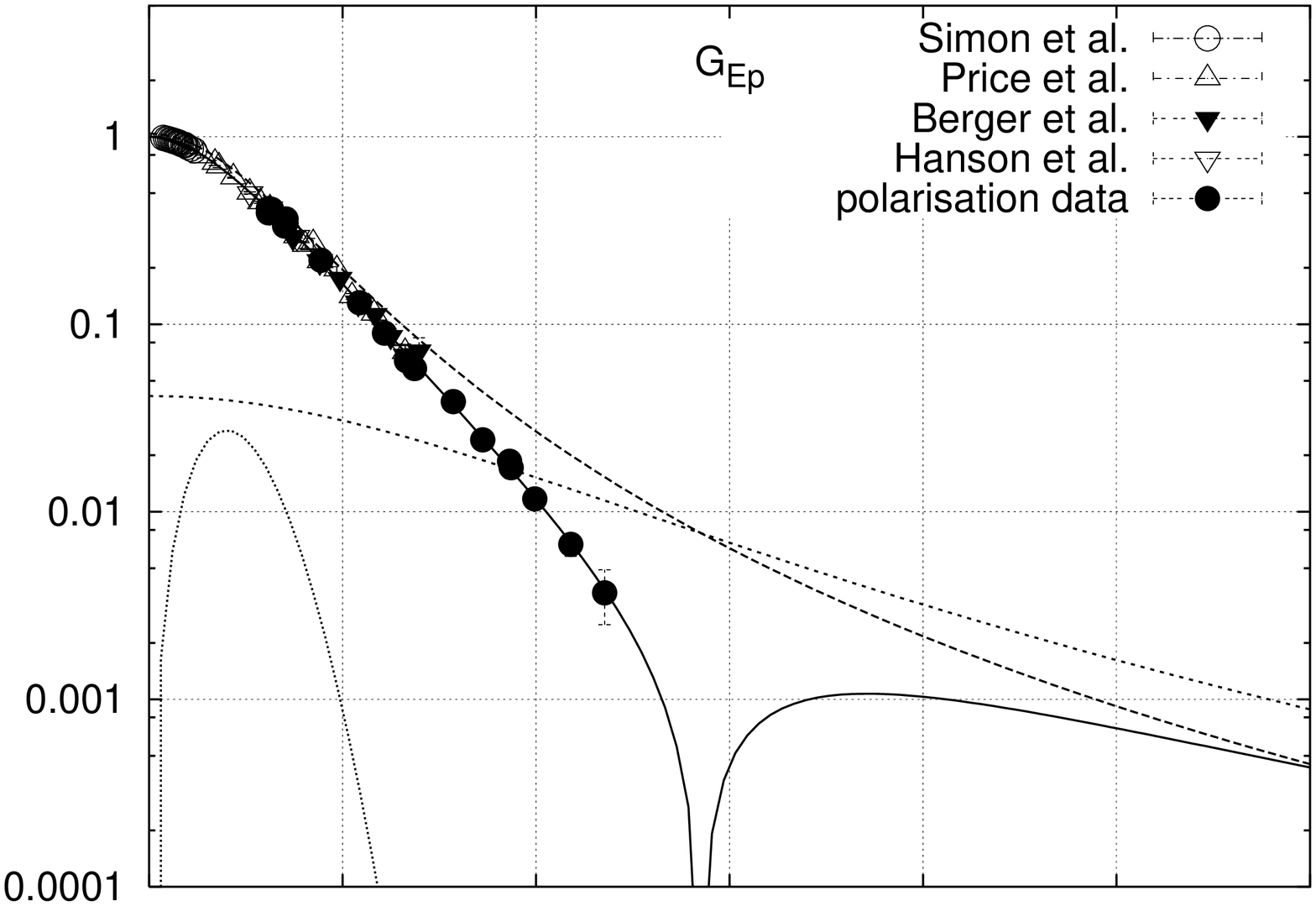}} } \\
 \vspace*{-0.5cm}\rotatebox{-90} {\resizebox{5.5cm}{!}
 {\includegraphics[width=1.0\textwidth]{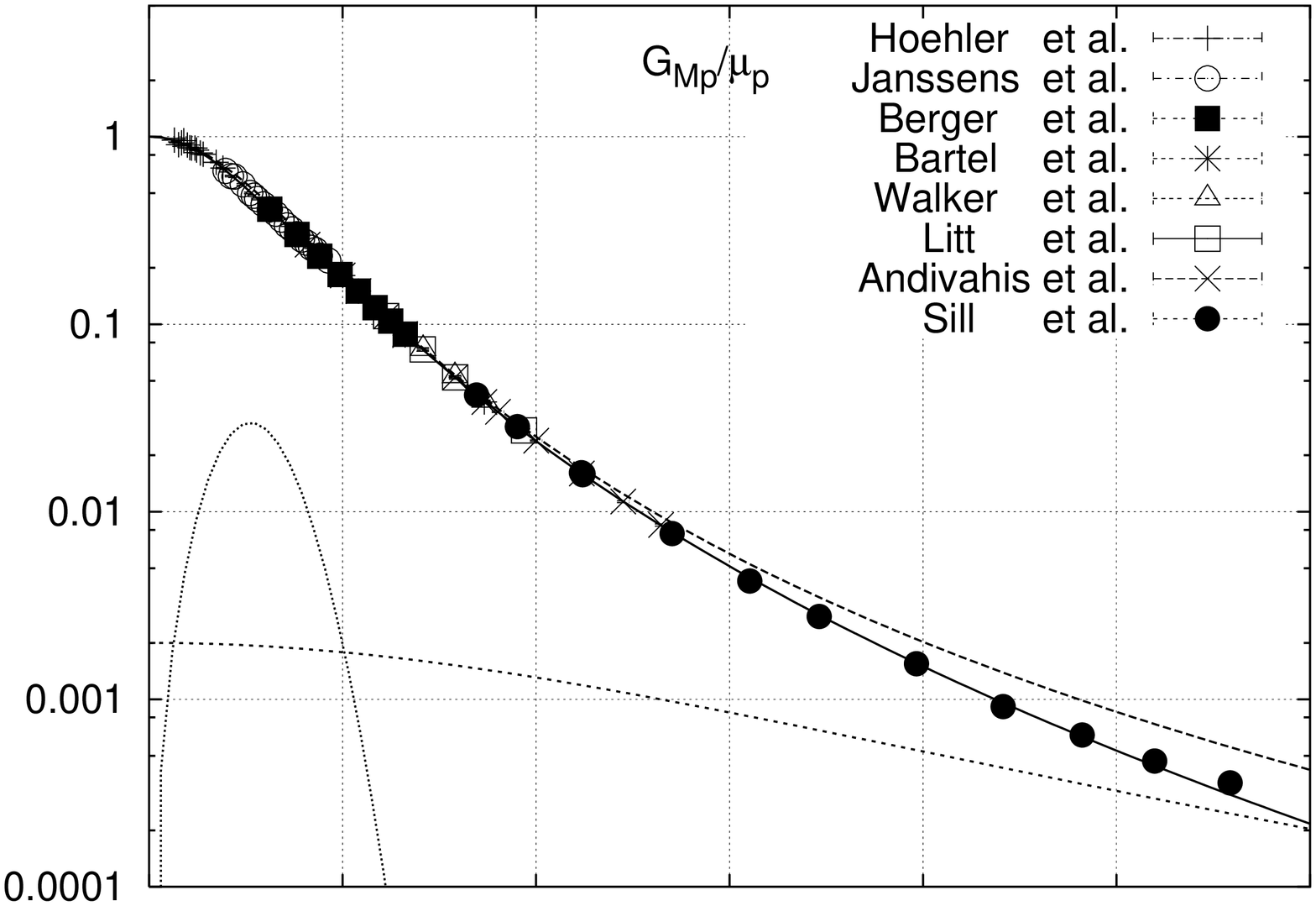}} } \\
 \vspace*{-0.5cm}\rotatebox{-90} {\resizebox{5.5cm}{!}
 {\includegraphics[width=1.0\textwidth]{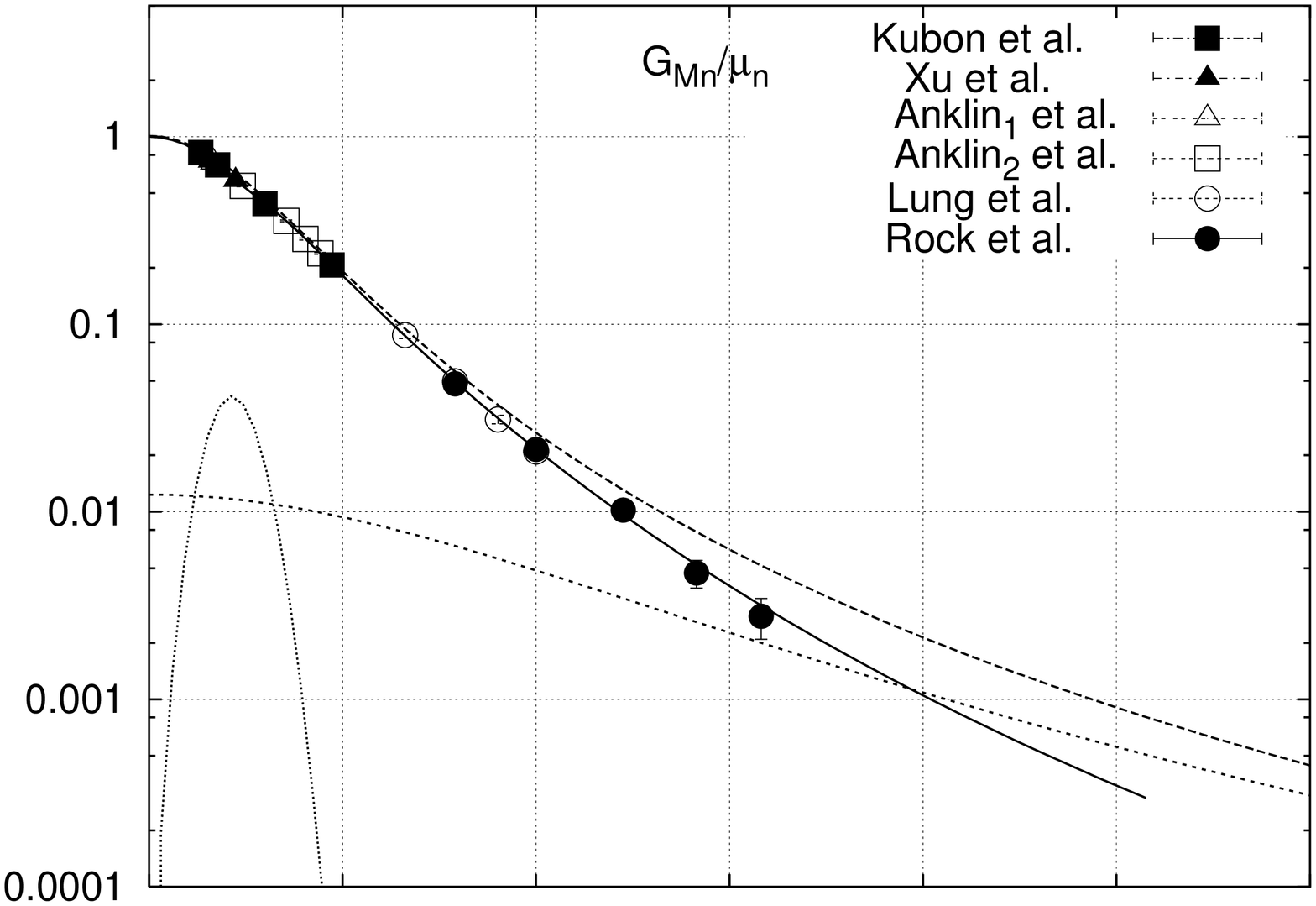}} } \\
 \vspace*{-0.5cm}\rotatebox{-90} {\resizebox{5.5cm}{!}
 {\includegraphics[width=1.0\textwidth]{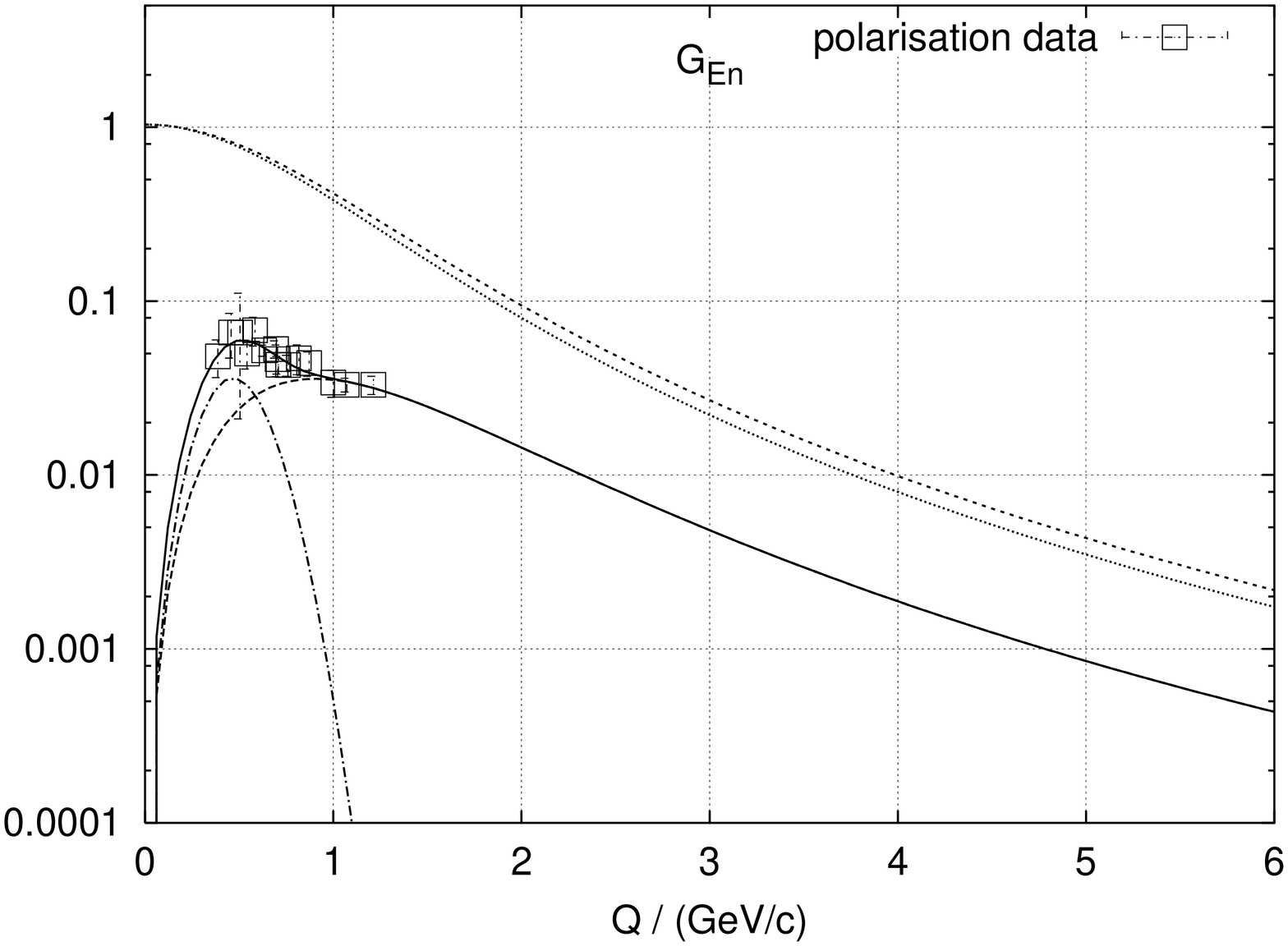}} } \\ [0.3cm]
\caption{The measured nucleon form factors and their description by
the phenomenological fits. The full line represents the 
the sum of the two dipoles and the Gaussian, which are also shown 
separately, the second dipole form being multiplied by -1 in order
to make it positive for this logarithmic plot. For $G_{En}$ we also
show the sum of the two dipoles separately.}
\label{fig:ff_of_q}
\end{figure}

First, fig.~\ref{fig:ff_of_q} shows the overall behaviour of the nucleon
form factors and the quality of the overall agreement with the fits.
For $G_{Ep}$ the relatively large (negative) amplitude $a_{20}$ of the second dipole
results in a zero in the form factor around 3~GeV/c. This makes this
form factor look very differently from $G_{Mp}$ and $G_{Mn}$, though,
in fact, this is only due to the larger amplitude of the dipole with negative sign
and not the form of the single contributions.
At the highest measured momentum transfers above 3~GeV/c the data for 
$G_{Mp}$ are not so well described by this phenomenological fit. Though,
from $\chi ^2$, the overall description of the data is excellent, there
might be some systematic deviation at high $Q^2$ pointing to insufficient
flexibility of the ansatz. 

In order to make the deviation of the measurements (and thus also of our fits)
from the standard dipole fit more obvious, we show in fig.~\ref{fig:ff_durch_dipol}
the ratio of the three standard form factors (data and fit) to the
standard dipole fit (see second factor of eq.~(\ref{eq:Galster})).

In fig.~\ref{fig:ff_minus_smooth} we demonstrate the ``bump''-contribution
to the form factors by the subtraction of the fitted smooth part, i. e.
the two dipoles. In order to emphasise the low-$Q^2$ region where this
phenomenon occurs, we have plotted this difference as function
of log($Q^2$). It is obvious from this graph that the bump
structure around $Q^2 = 0.2 - 0.3$~(GeV/c)$^2$, as discussed in section~\ref{sec:gen}  
for $G_{En}$, is a common feature of all four form factors.
It is striking how similar the contribution from the bump is in all
four form factors. While it is only a small contribution to $G_N^{std}$,
it dominates $G_{en}$ at low momentum transfer where the contribution
from the two dipoles, though both separately are large, cancel.

As also discussed in section~\ref{sec:gen}, such bump structure
in Q-space contributes a certain $\Delta \rho _b$ in r-space, the detailed
structure of which, however, is not easy to foresee.
The net charge in $\Delta \rho _b$ under discussion here is zero 
by construction, therefore it must show some oscillation.
The wavelength $\lambda _\rho $ of this oscillation is given by the position
of the bump in $Q$-space, $Q_b$, as $\lambda _\rho = 2\pi \hbar / Q_b$. 
The damping of the oscillation is related to the relative width
of the bump.
From $Q_b \approx $ 0.45~GeV/c = 2.3~fm$^{-1}$ there thus results the 
wavelength $\lambda _\rho $ = 2.7~fm in agreement with the oscillation
shown in fig.~\ref{fig:2}. We conclude that an oscillation 
with such wavelength results as a common feature from all four form factors 
and is not just a peculiarity of $G_{En}$.

We demonstrate this by showing in the following subsection the Fourier
transforms of the four form factors.

\subsection{The Fourier transform of the fits of the form factors}
\label{sec:charge_phenomen_ansatz}

As mentioned above,
the Fourier transforms of the Sachs form factors can be regarded as
the charge and magnetic distribution, respectively, in the Breit frame.
With this in mind we show in this subsection the Fourier transforms
of the form factors which, for brevity, we all denote by ``charge'' $\rho$.

\begin{figure} 
\begin{center}
\vspace*{-0.0cm} \rotatebox{-90} {\resizebox{5.8cm}{!}
{\includegraphics[width=1.0\textwidth]{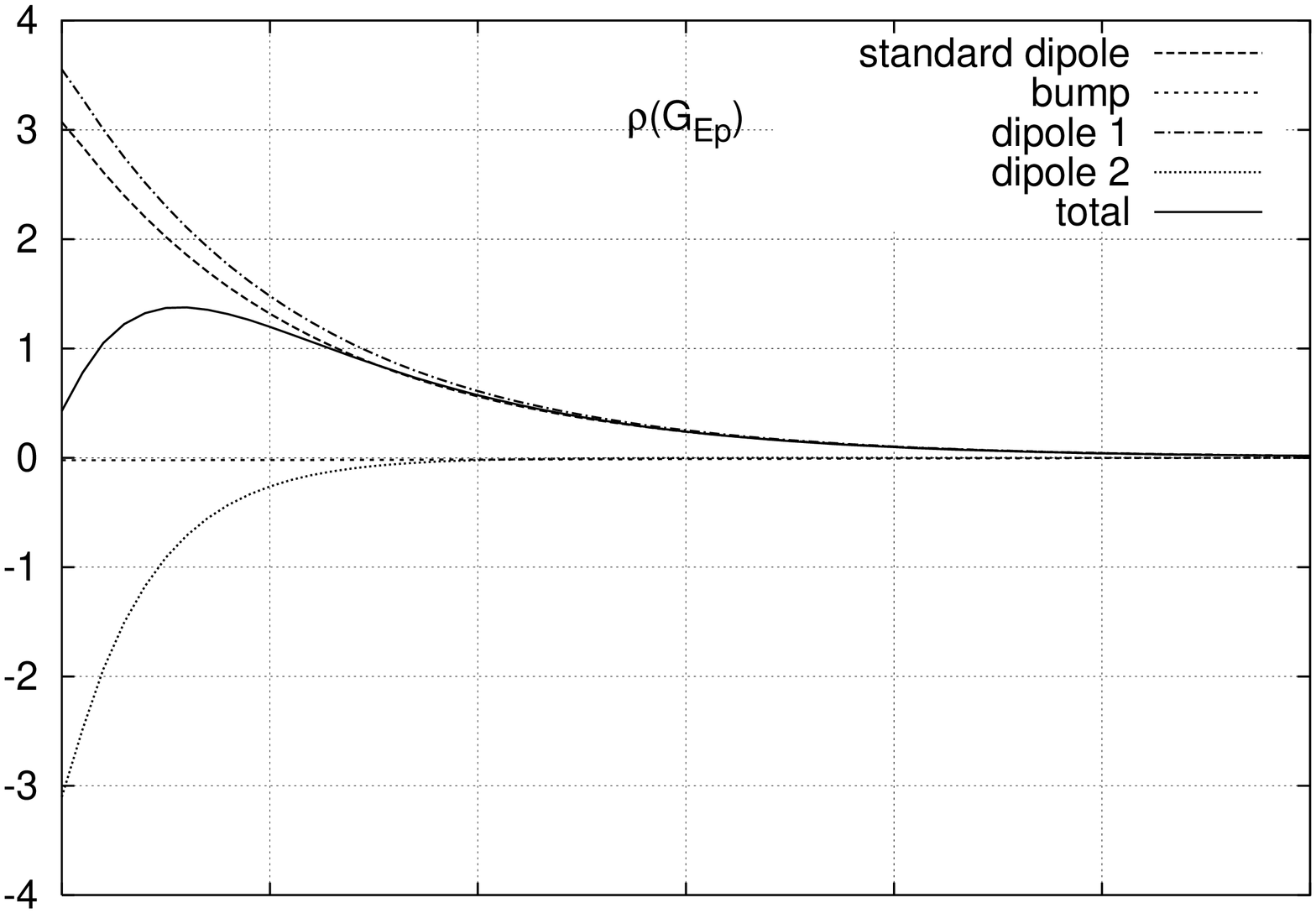}} } \\
\vspace*{-0.5cm} \rotatebox{-90} {\resizebox{5.8cm}{!}
{\includegraphics[width=1.0\textwidth]{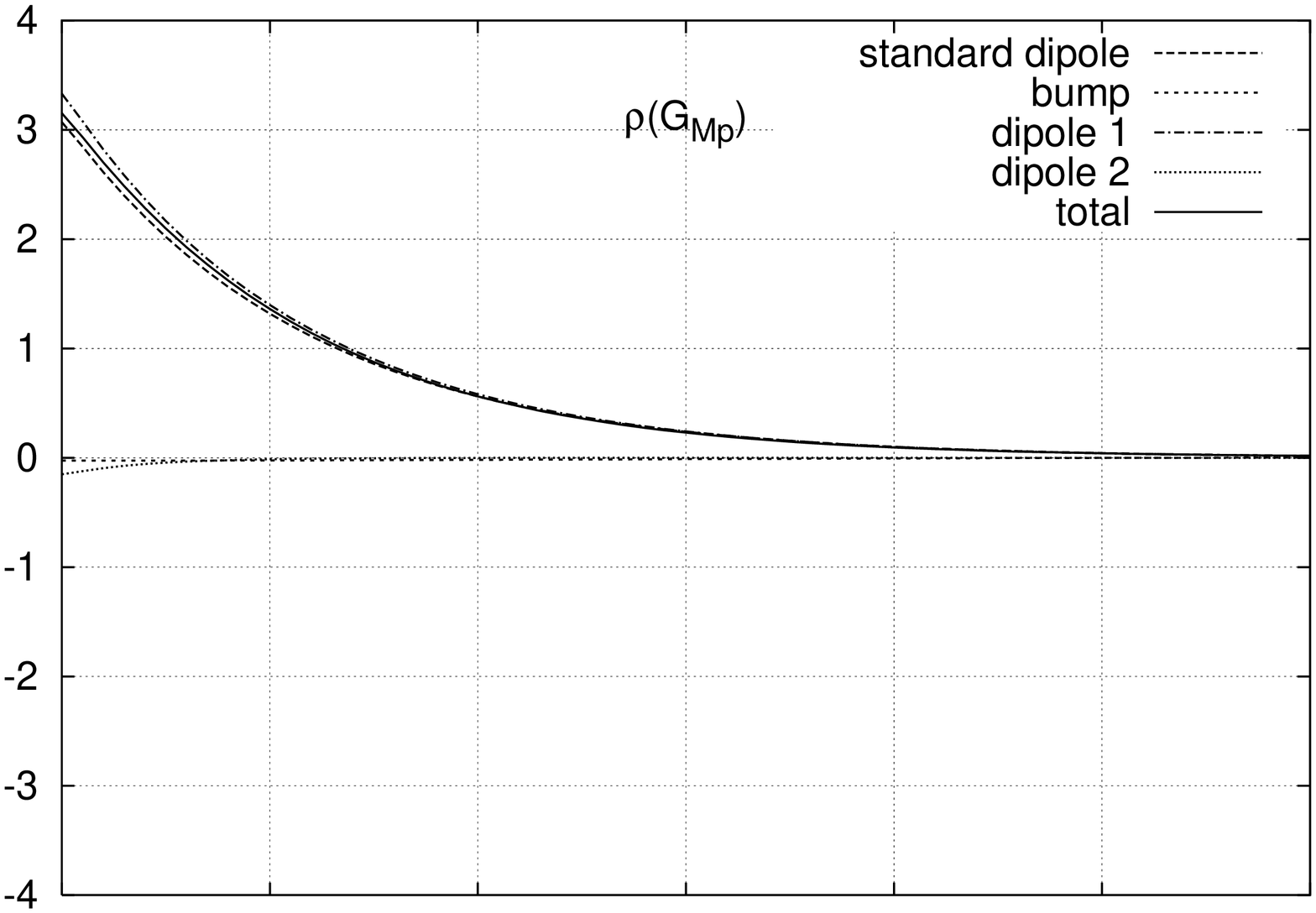}} } \\
\vspace*{-0.5cm} \rotatebox{-90} {\resizebox{5.8cm}{!}
{\includegraphics[width=1.0\textwidth]{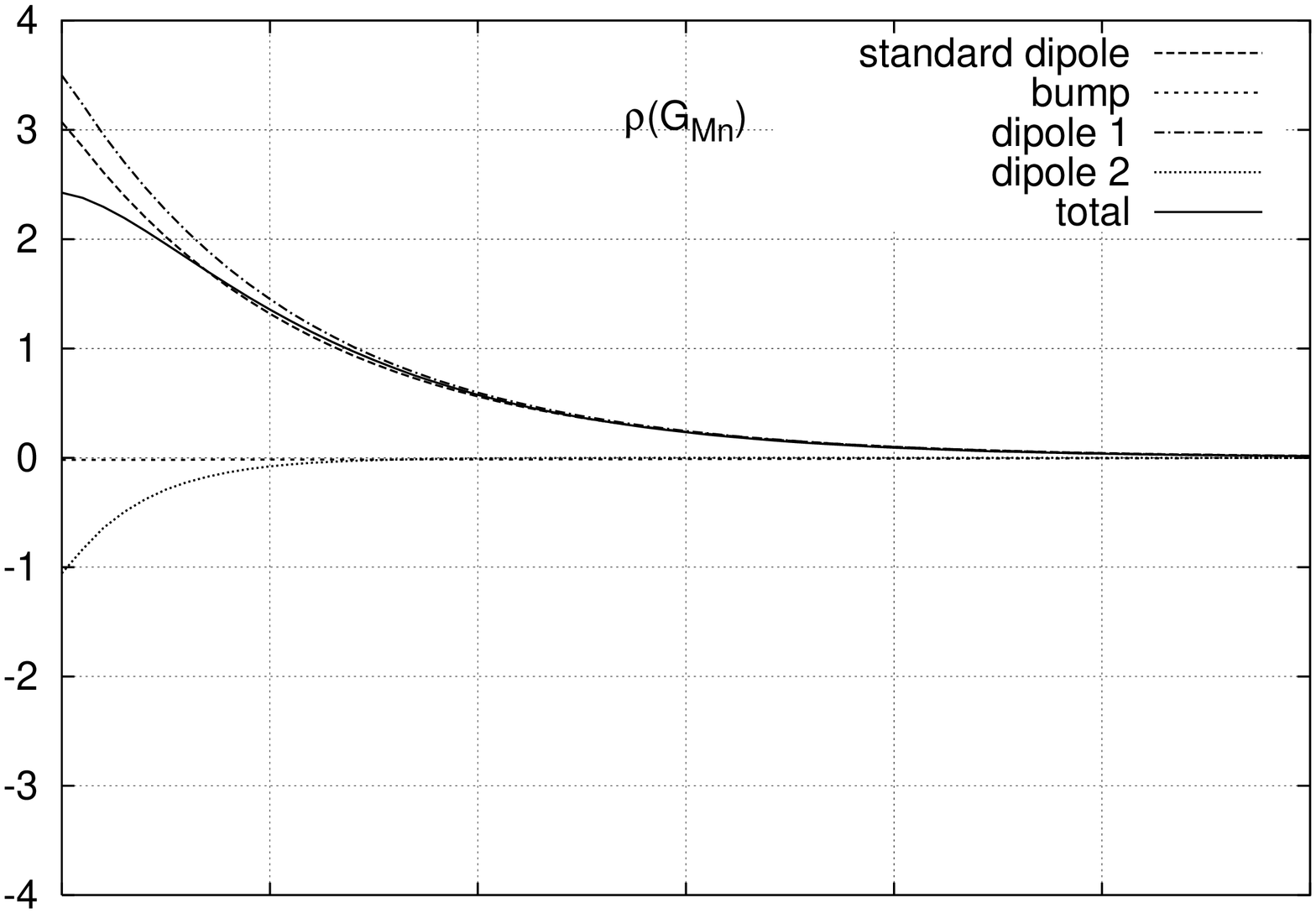}} } \\ [0.3cm]
\vspace*{-0.5cm} \rotatebox{-90} {\resizebox{5.8cm}{!}
{\includegraphics[width=1.0\textwidth]{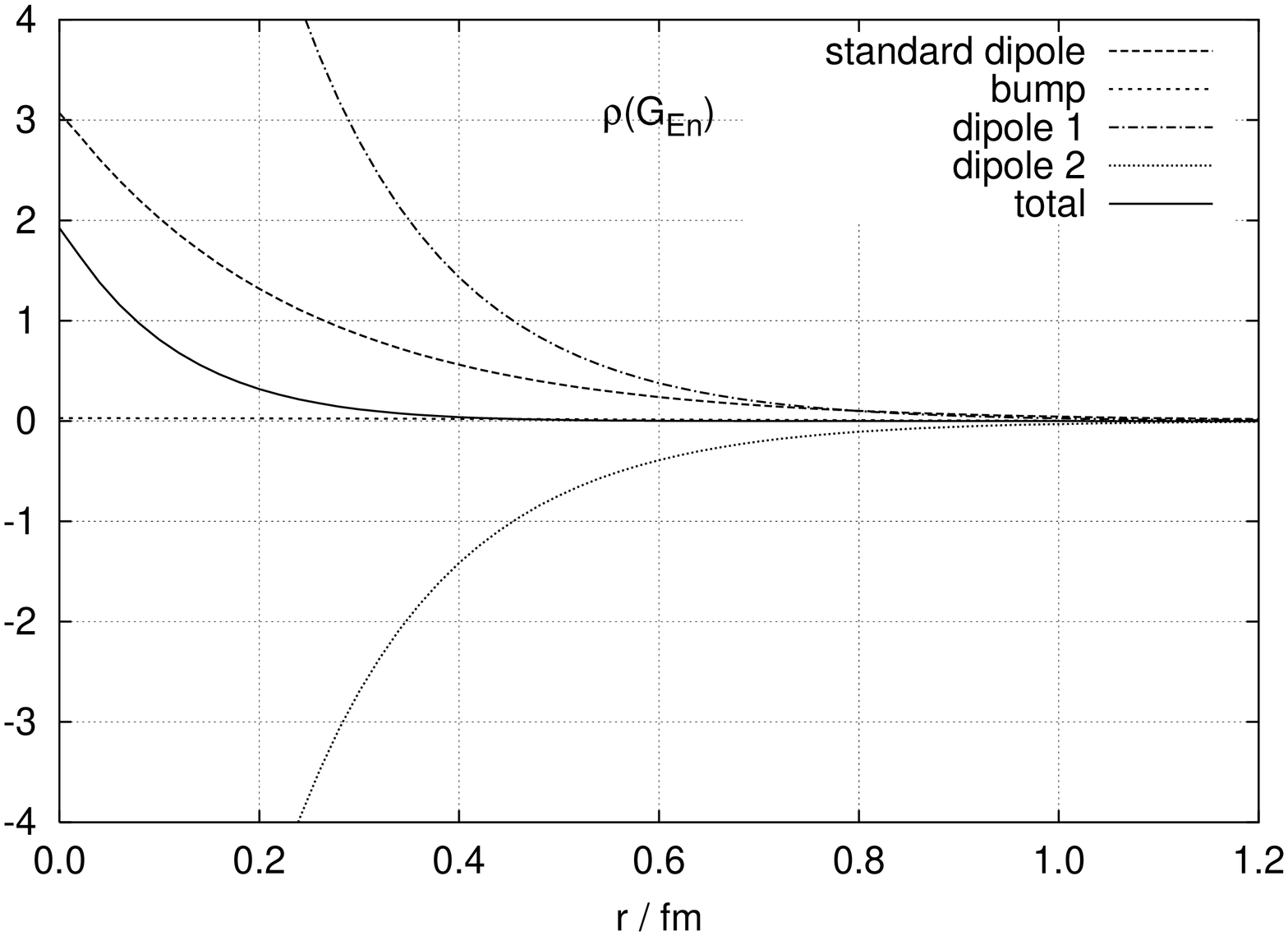}} } \\ [0.3cm]
\end{center}
\caption{$\rho (r)$ of the nucleons in the Breit frame. The units of $\rho
  (r)$ are fm$^{-3}$. The distributions are normalised to 1 for $G^{std}_N$ 
  and to 0 for $G_{En}$.}
\label{fig:rho}
\end{figure}

Fig.~\ref{fig:rho} shows $\rho (r)$ resulting from the two dipole
contributions, the bump and their sum. For reference, the 
transform of the standard dipole is also shown.
In this representation, for all three $G_N^{std}$ $\rho (r)$ is very 
close to that of the standard dipole form for $r >$ 0.2 fm.
This is true for $G_{Mp}$ also down to the centre of the nucleon,
where we see deviations from the dipole form for $G_{Ep}$
and $G_{Mn}$. It has to be admitted, however, that in the
latter cases the measurements do not extend to as high
$Q^2$ as in the former, thus the distribution in $r$-space
is less well fixed at small $r$. Actually, there is only a tiny fraction of
the total charge contained in this inner part. The contribution from
the bump is not visible in this plot.

\begin{figure} 
\begin{center}
\vspace*{-0.0cm} \rotatebox{-90} {\resizebox{5.8cm}{!}
{\includegraphics[width=1.0\textwidth]{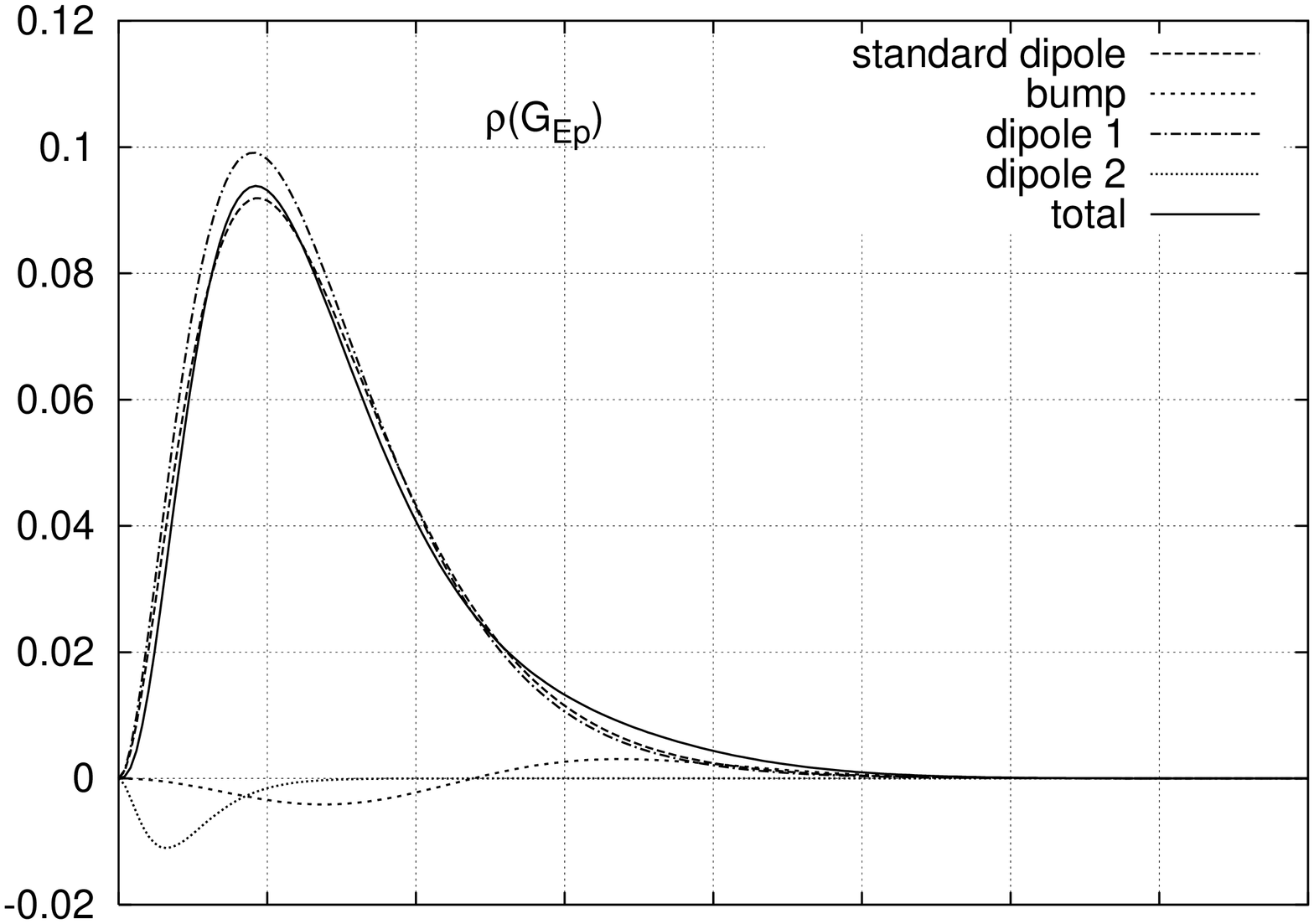}} } \\
\vspace*{-0.5cm} \rotatebox{-90} {\resizebox{5.8cm}{!}
{\includegraphics[width=1.0\textwidth]{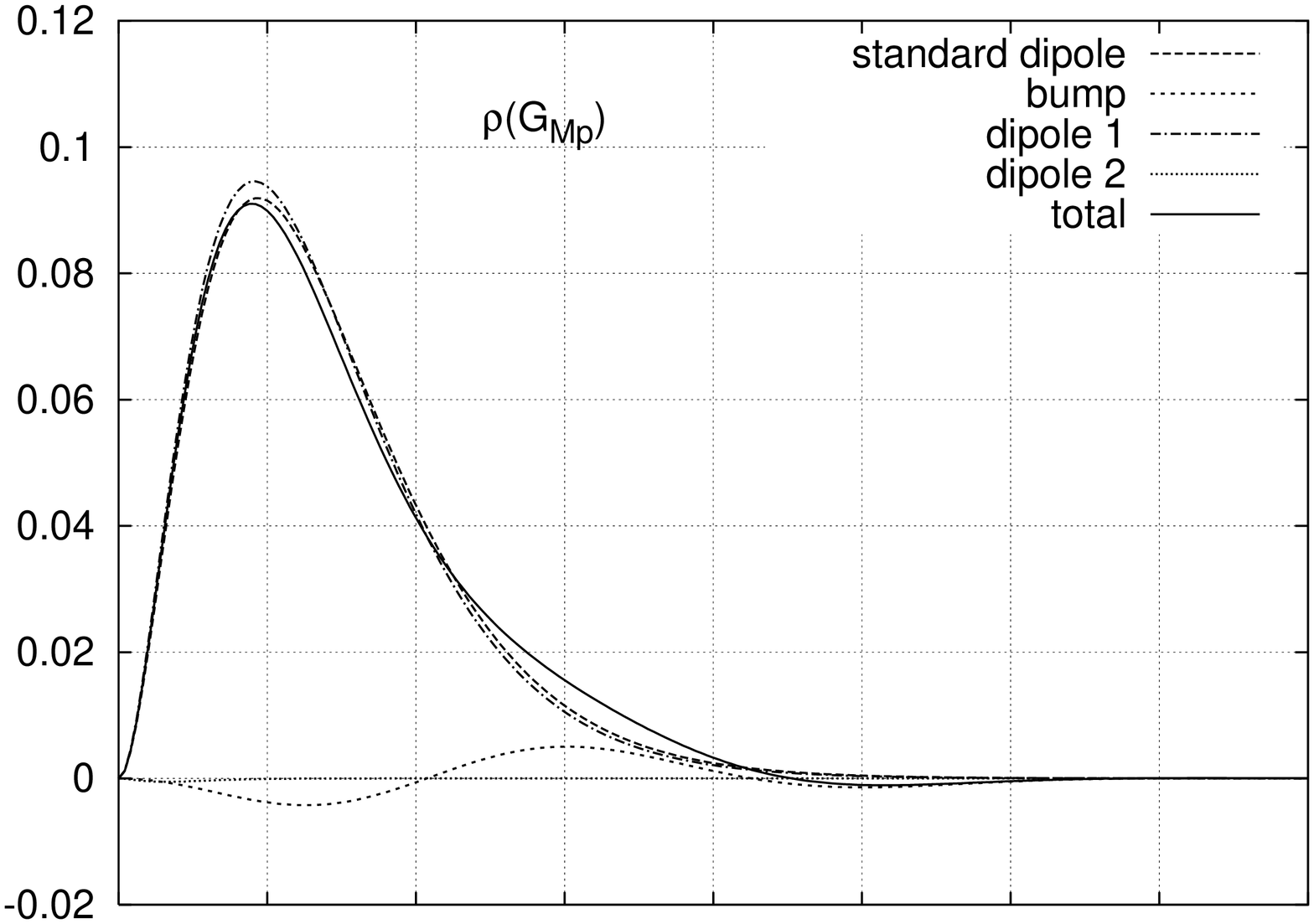}} } \\
\vspace*{-0.5cm} \rotatebox{-90} {\resizebox{5.8cm}{!}
{\includegraphics[width=1.0\textwidth]{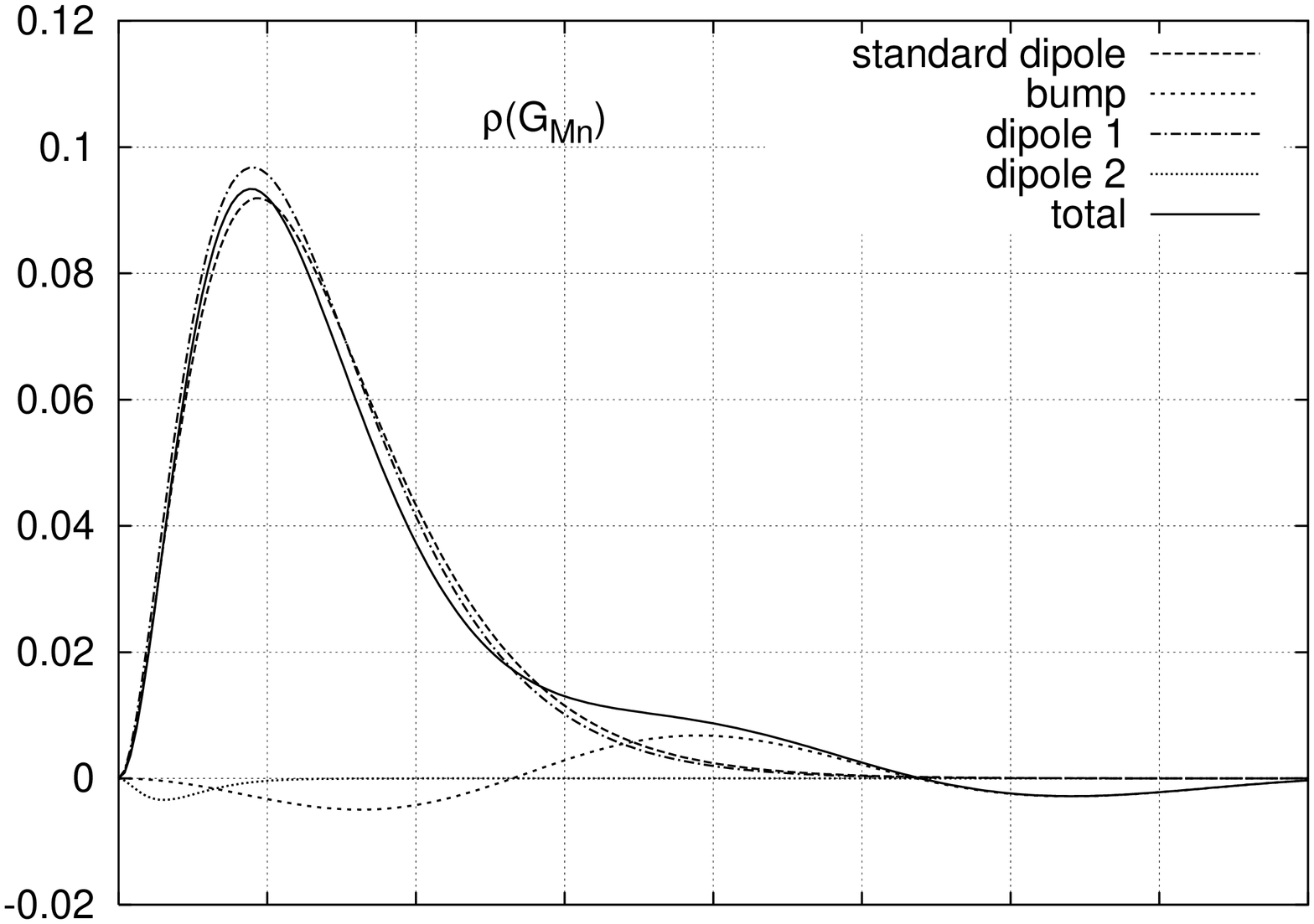}} } \\ [0.3cm]
\vspace*{-0.5cm} \rotatebox{-90} {\resizebox{5.8cm}{!}
{\includegraphics[width=1.0\textwidth]{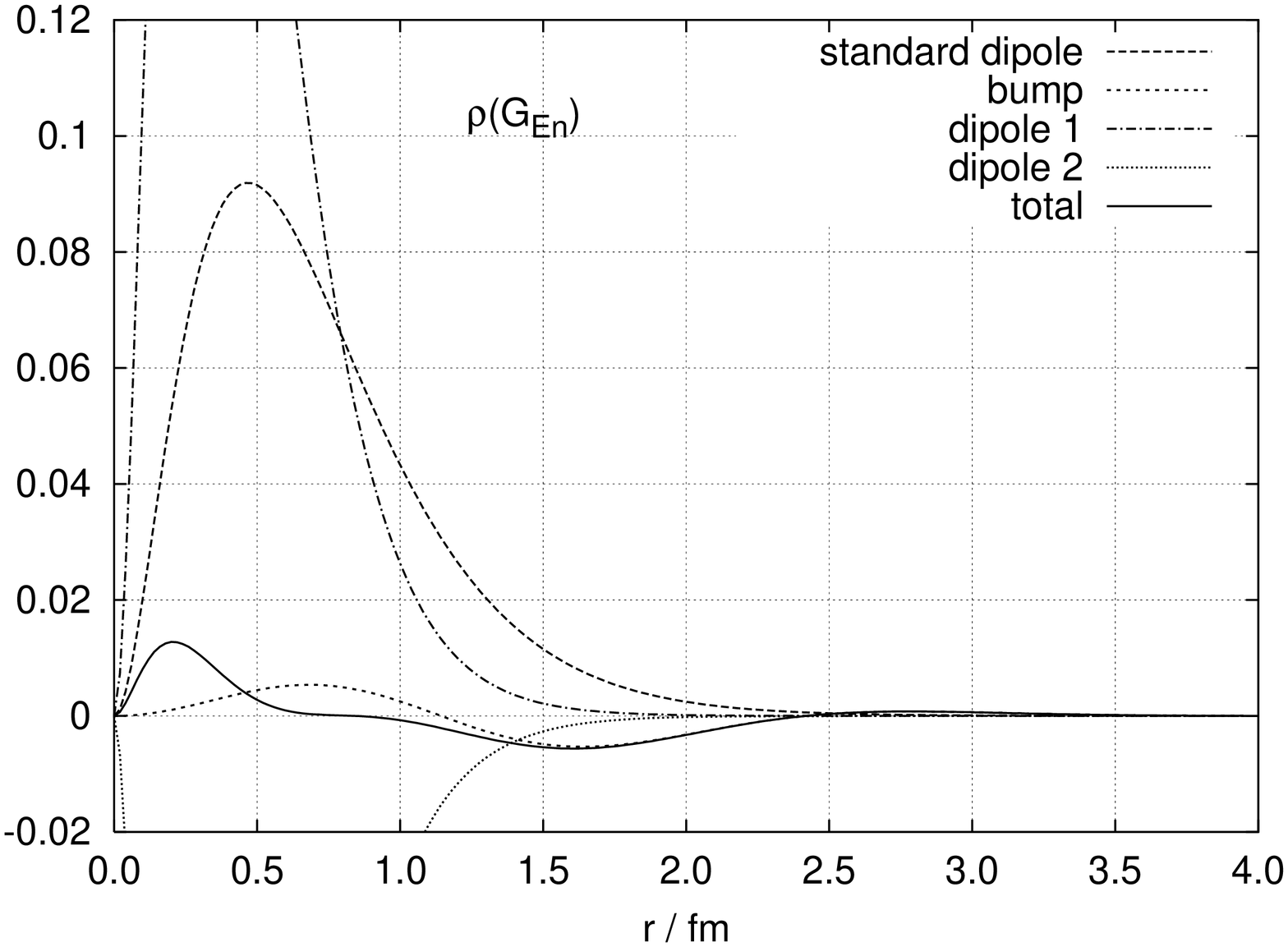}} } \\ [0.3cm]
\end{center}
\caption{$\rho (r) \cdot r^2$ in the Breit frame. }
\label{fig:rho_r2}
\end{figure}

The quantity $\rho (r) \cdot r^2$ gives directly the weight of the charge
contained in a spherical shell at distance $r$, thus the area under the 
curve gives the total charge contained in the respective term. 
This quantity is shown in fig.~\ref{fig:rho_r2}.
Here, the contribution from the bump in the form factor is clearly
visible as oscillation (net charge = 0). 
Its phase in $r$-space is such that it puts additional
strength on the dipole form in the outer region with maxima between 1.5 ($G_{Mp}$)
and 2.0 fm ($G_{Ep}$, $G_{Mn}$). 
The second dipole gives small and tiny contributions in the interior of 
$\rho (G_{Ep})$ and $\rho (G_{Mn})$, respectively, and
is not visible in $\rho (G_{Mp})$.
For $G_{En}$ the oscillation gives the total $\rho (r)$ in the outer
region centred around 1.7 fm, while the inner part is dominated by the difference of the two 
dipoles. In between there is a cancellation between these
two terms, a feature already visible in the analysis in section~\ref{sec:gen}.
We come back to this point below.

\begin{figure} 
\begin{center}
\vspace*{-0.0cm} \rotatebox{-90} {\resizebox{5.8cm}{!}
{\includegraphics[width=1.0\textwidth]{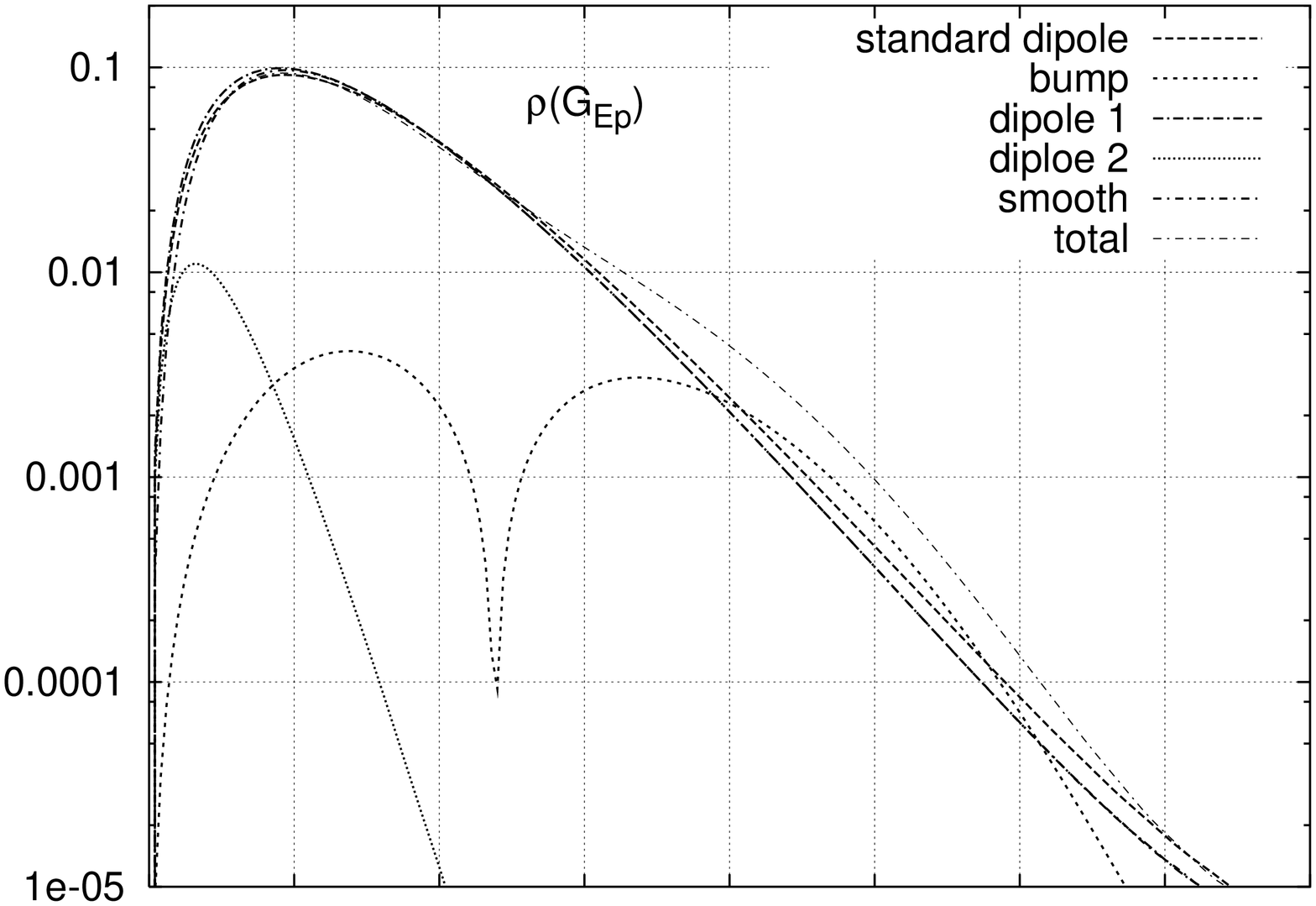}} } \\
\vspace*{-0.5cm} \rotatebox{-90} {\resizebox{5.8cm}{!}
{\includegraphics[width=1.0\textwidth]{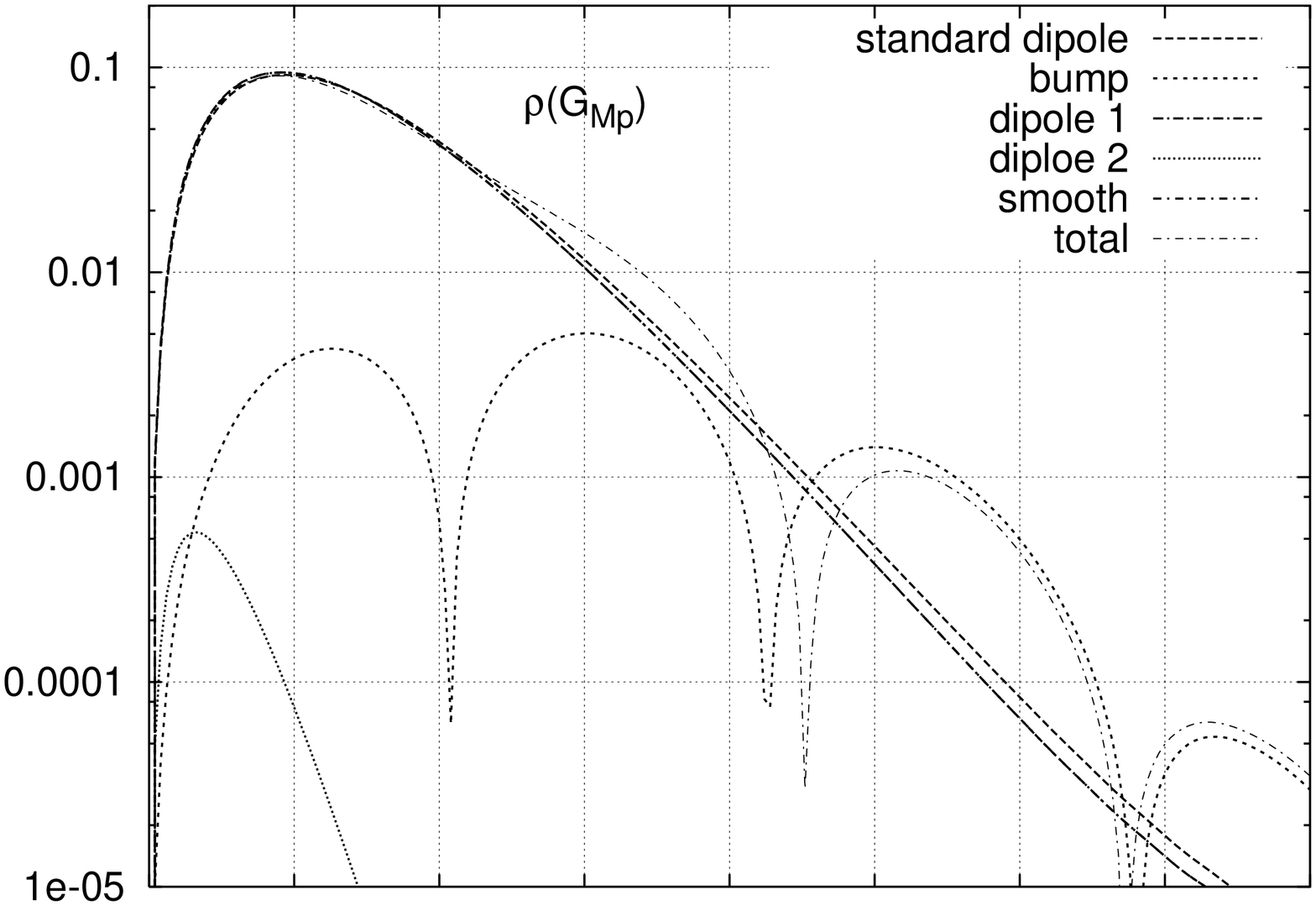}} } \\
\vspace*{-0.5cm} \rotatebox{-90} {\resizebox{5.8cm}{!}
{\includegraphics[width=1.0\textwidth]{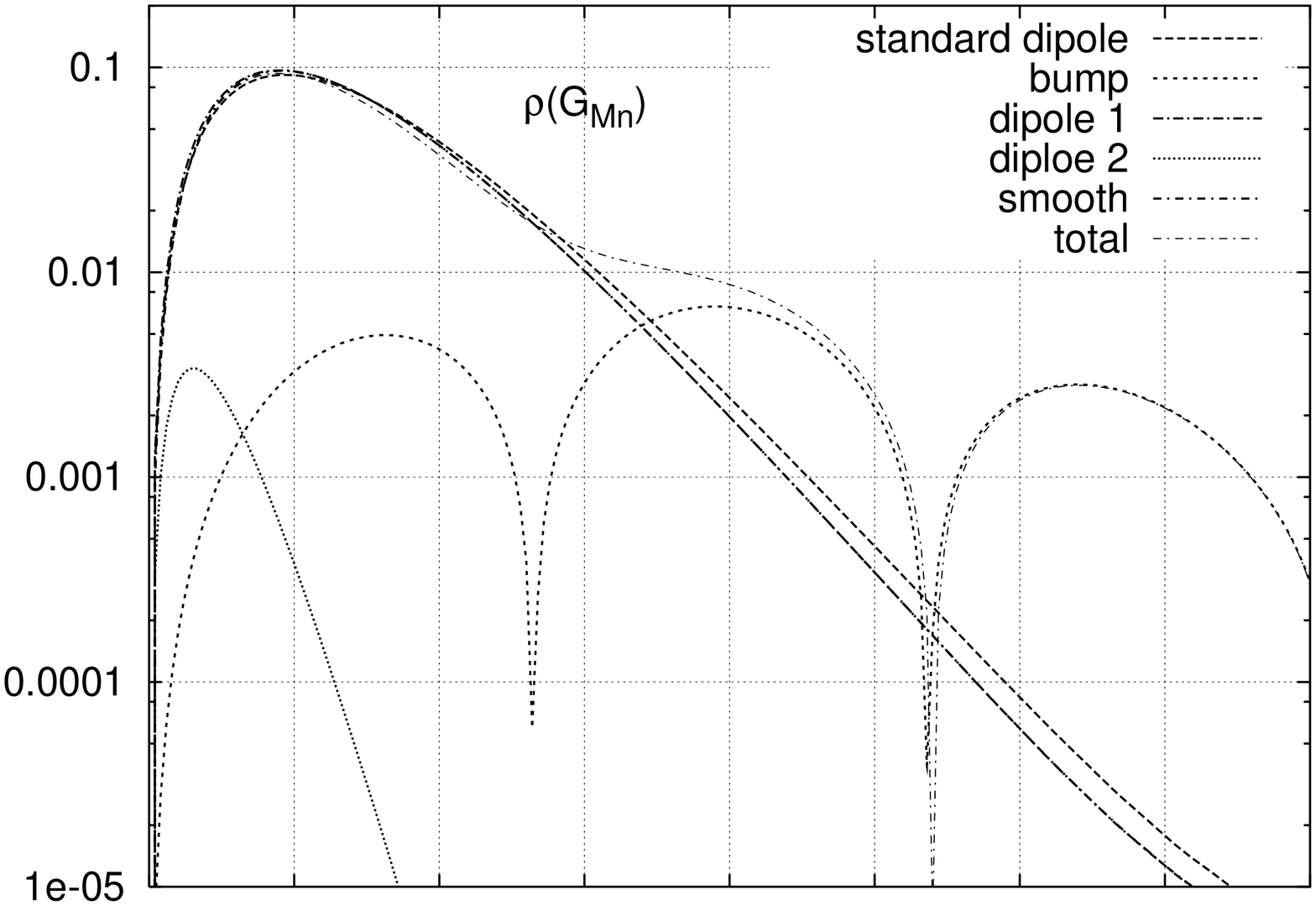}} } \\ [0.3cm]
\vspace*{-0.5cm} \rotatebox{-90} {\resizebox{5.8cm}{!}
{\includegraphics[width=1.0\textwidth]{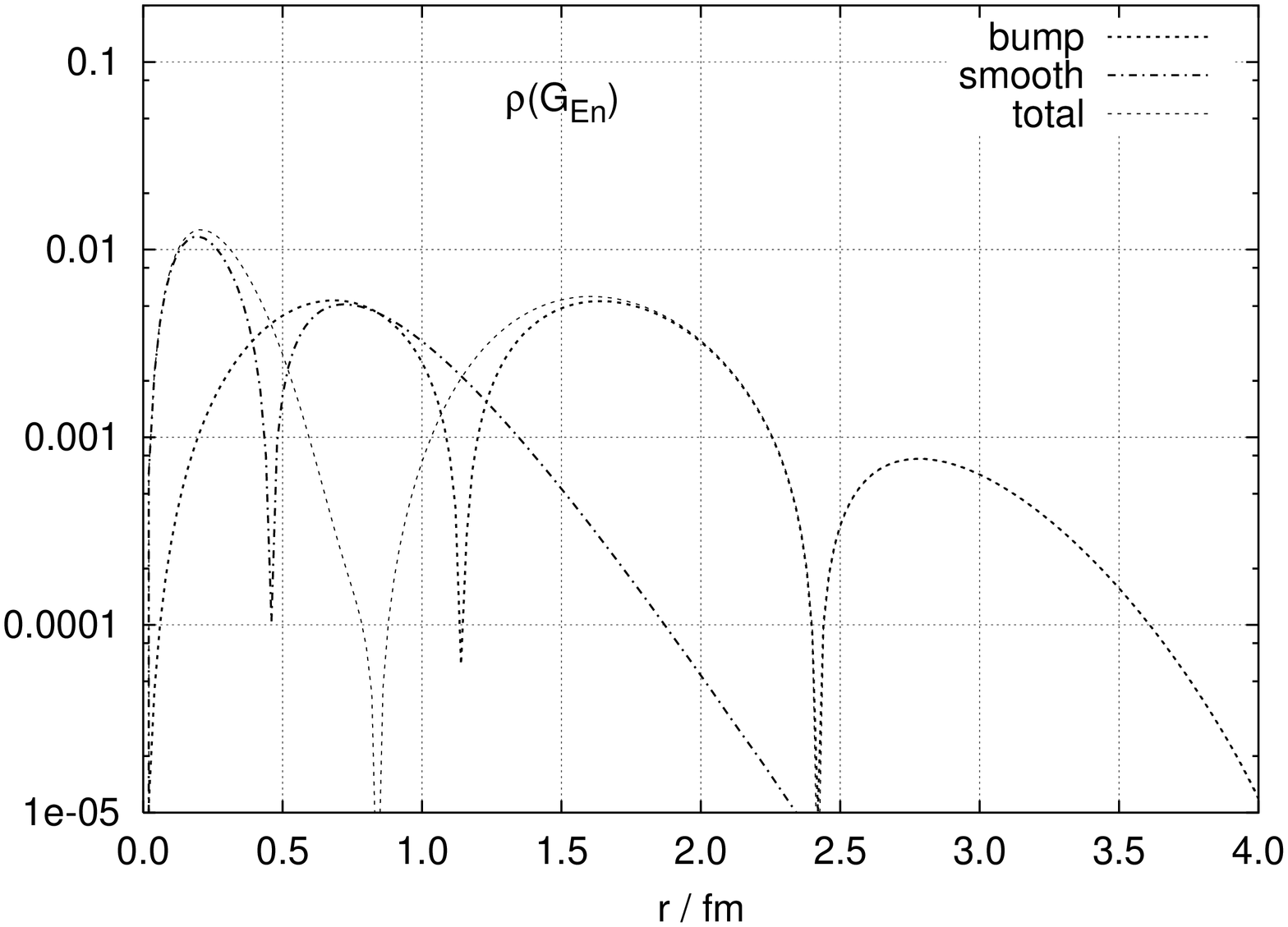}} } \\ [0.3cm]
\end{center}
\caption{$\rho (r) \cdot r^2$ in the Breit frame as in fig.~\ref{fig:rho_r2},
but now in a logarithmic scale.}
\label{fig:rho_r2_log}
\end{figure}

In order to emphasise the outer region of the nucleons even more,
fig.~\ref{fig:rho_r2_log} shows $|\rho (r)| \cdot r^2$ 
but now in a logarithmic scale. In this plot the sign information gets lost
and one has to look at fig.~\ref{fig:rho_r2} to keep track of
the sign. The sign of the first lobe in the contribution
$\Delta \rho_b(r)$ from the bump in the form factor, this means
$a_b Q^2 G_b(Q)$, is that of $\Delta \rho_b(r=0)$. Since
$\Delta \rho_b(r=0) \propto$  $a_b \int G_b(Q) Q^2 dQ$, the sign
of $\Delta \rho_b(r=0)$ and thus the sign of the first lobe
in $\Delta \rho_b(r)$ is given by the sign of the
amplitude $a_b$ of the bump ($G_b(Q) > 0$).
- Here again, we do not want to go into
details. We will see in the next section, that one $\pm$ oscillation on top
of a smooth distribution can be interpreted as dissociation of
the nucleon in its counterpart and a pion cloud. The further
oscillations visible in the logarithmic plot in fig.~\ref{fig:rho_r2_log}
are compatible with the data, but certainly must be regarded as
depending on the special ansatz eq.~(\ref{eq:ff_nucleon}).

\section{A coherent description of the four form factors by a physically 
motivated ansatz}
\label{sec:model}

\subsection{The ansatz}
\label{sec:model_ansatz}

Inspired by the conspicuous graphical representation of the form factors,
which reveal a bump on top of a smooth trend, we make the ansatz
of describing the nucleons by the sum of a bare nucleon plus a
polarisation part according to
\begin{eqnarray}
p & = & a_p \cdot p^0 + b_p \cdot ( n^0  + \pi ^+ ) \nonumber          \\
  & = & ~~~~~~~p^0 + b_p \cdot ( -p^0 + n^0 + \pi ^+)      \, , \label{eq:proton}  \\
n & = & a_n \cdot n^0 + b_n \cdot ( p^0  + \pi ^- ) \nonumber          \\
  & = & ~~~~~~~n^0 + b_n \cdot ( +p^0 - n^0 + \pi ^-)           \label{eq:neutron}
\end{eqnarray}
where we have made use of the normalisation condition $a_N + b_N = 1$ for $N = p, n$.
Neutral pions are not taken into consideration since, to first order, they
do not contribute to elastic electron scattering. 

The form factors can thus be written as
\begin{eqnarray}
G_p & = & G_p^0 + b_p \cdot ( -G_p^0 + G_n^0 + G^{\pi^+} ) = G_p^0 + G_p^{pol} \, , \label{eq:proton_formfactor} \\
G_n & = & G_n^0 + b_n \cdot ( +G_p^0 - G_n^0 + G^{\pi^-} ) = G_n^0 + G_n^{pol} \,   \label{eq:neutron_formfactor}
\end{eqnarray}
where we use the transparent nomenclature of the form factor of the
polarisation
\begin{eqnarray}
G_N^{pol} & = & b_N \cdot ( G_{\bar{N}}^0 - G_N^0 + G^{\pi } ), \label{eq:polarisation_formfactor} 
\end{eqnarray}
where $\bar{N}$ denotes the neutron (proton) and $\pi $ the $\pi ^+$ ($\pi ^-)$ when $N$ is the proton (neutron).

Furthermore, we think of the bare nucleons in terms of their 
constituent quark content, i. e. $ p = (uud)$ and $n = (udd)$.
We denote the form factor of the distribution of quark $q$ in the
nucleon $N$ by $G^{qN}$.
We are thus dealing with the ingredients $G^{up}$, $G^{dp}$, $G^{un}$,
and $G^{dn}$ for which we take the dipole form:
\begin{equation}
G^{qN} = \frac{a^{qN}_0}{(1 + Q^2/a^{qN}_1)^2} \, .
\label{eq:quark_dipole}
\end{equation}

The pion has intrinsic parity -1 which has to be compensated by its spatial
wavefunction; therefore, to lowest order, it should be in an ($l$=1)-state. Taking as simple
ansatz the wave function of a harmonic oscillator, the related form
factor is given by
\begin{equation}
G^{\pi } = a^\pi_0 \cdot ( 1 - \frac{1}{6}(Q/a^{\pi }_1)^2) e^{-\frac{1}{4} (Q/a^{\pi }_1)^2} \, .
\label{eq:pion_ff}
\end{equation}
The form factor of the pion cloud should be the convolution of this
form factor from the wave function with that of the intrinsic distribution of the pion,
the size of which is certainly not negligible compared to that of the nucleon
and thus to the extension of the pion cloud.
Convolution in $r$-space results in a multiplication in $Q$-space.
Assuming a Gaussian for the intrinsic pion distribution, this results in
a multiplication of eq.~(\ref{eq:pion_ff}) by a Gaussian, thus by a change of
the parameter $a^{\pi }_1$ in the exponential. This does not change
the form of $G^{\pi }$, it would only decouple the parameter $a^{\pi }_1$ in
the exponential from that in the brackets. We will, however, not make use
of this additional degree of freedom.

For the electric form factors, we have $a^{\pi ^+}_{0,E} = -a^{\pi ^-}_{0,E} = 1$.
For the magnetic form factor, the situation is not that clear due to the
degrees of freedom of the vector coupling of the magnetic moments.
Furthermore, it is not clear what magnetic moment should be related
with the pion cloud since we are not dealing with a free pion.

Strict isospin invariance would imply 
\begin{eqnarray}
 G^{up}     & \sim &  ~G^{dn}     \, , \nonumber                       \\ 
 G^{dp}     & \sim &  ~G^{un}     \, , \label{eq:isospin_conservation} \\
 G^{\pi ^+} & \sim &  -G^{\pi ^-} \, . \nonumber
\end{eqnarray}
We will check, whether the measured form factors can be described under this condition.

\subsection{The electric form factors}
\label{sec:ff_e}

Since $G_{En}^0(0)=0$ due to the vanishing charge of the neutron,
from eq.~(\ref{eq:proton_formfactor}) we have
\begin{equation}
G_{Ep}(0) = [(1-b_p) \cdot G^0_{Ep}(0) + b_p \cdot G_{\pi ^+}(0)] = 1 \, .
\label{eq:gep0_0}
\end{equation}
We note, that the charge $b_p \cdot 1$ of the pion cloud, which goes at
the expense of the bare proton's charge, contributes to the proton electric form factor 
at $Q^2$ = 0. Therefore, the peak on top of a smooth part of the form factors,
which we have revealed in section~\ref{sec:phenomen}, 
cannot be attributed directly to the pion.
In fact, to the contrary, according to eq.~(\ref{eq:pion_ff})
the contribution from the pion cloud should 
be concentrated around $Q^2 = 0$ since the pion cloud is
expected to extend further out than the bare proton.

According to our model, we evaluate the electric form factor of the proton
with the ansatz
\begin{eqnarray}
G_{Ep} & = & ( G^{up}_E + G^{dp}_E )                          \label{eq:gep_quark_pion}  \\
       & + & b_p \cdot (-(G^{up}_E+G^{dp}_E)\,+\,(G^{un}_E+G^{dn}_E)+G^{\pi ^+}_E)  \nonumber 
\end{eqnarray}
with $G^{qN}$ and $G^{\pi }$ parametrised by eqs.~(\ref{eq:quark_dipole})
and (\ref{eq:pion_ff}), respectively.

The weights $a^{qN}_0$ for the electric quark form factors are given by the
quark charges, i. e. 
$a^{up}_{0,E} = +4/3, a^{un}_{0,E} = +2/3$ and 
$a^{dp}_{0,E} = -1/3, a^{dn}_{0,E} = -2/3$, and that of the pion is $a^{\pi ^+}_0 = +1$. 
This ansatz conserves automatically the normalisation.
We are thus left with the free  parameters
of the quark distributions, $a^{up}_1, a^{dp}_1, a^{un}_1, a^{dn}_1$, the amplitude
$b_p$ of the polarisation term and the oscillator parameter for the pion, $a^{\pi ^+}_1$,
i. e. we have one free parameter less than in the phenomenological model
with two dipoles and the bump discussed in section~\ref{sec:phenomen}.

In a first fit we take these six parameters as free. The resulting values
are given in table~\ref{tab:par_model} as fit 1. 
It is not too surprising that the $n^0$-parameters remain completely undetermined. 
The data are described by this ansatz 
as well as with the phenomenological ansatz with seven parameters
($\chi ^2$ = 59.6 here compared to 59.7 there). 
We note that the pion  parameters in this model
are better determined than the bump parameters in the phenomenological ansatz.
Further, the large values of $a^{qn}_1$ would yield an extremely sharp
localisation of the quarks in the neutron, most likely
an unrealistic scenario. The large error on the neutron parameters, however, 
leave room for applying further model restrictions. 
In fit 2 we subject the quark distributions
to complete isospin invariance, i. e. we demand $a^{un}_1 = a^{dp}_1$ and
$a^{dn}_1 = a^{up}_1$, there are thus only 4 free parameters left.
The proton parameters vary essentially only within their errors,
the same is true for the pion-cloud parameters.
The total $\chi^2$ increases by the omission of the two parameters
by only 1.5 which is an insignificant increase.

\begin{figure}
\begin{center}
\vspace*{-0.0cm} \rotatebox{-90} {\resizebox{6.0cm}{!}
{\includegraphics[width=1.0\textwidth]{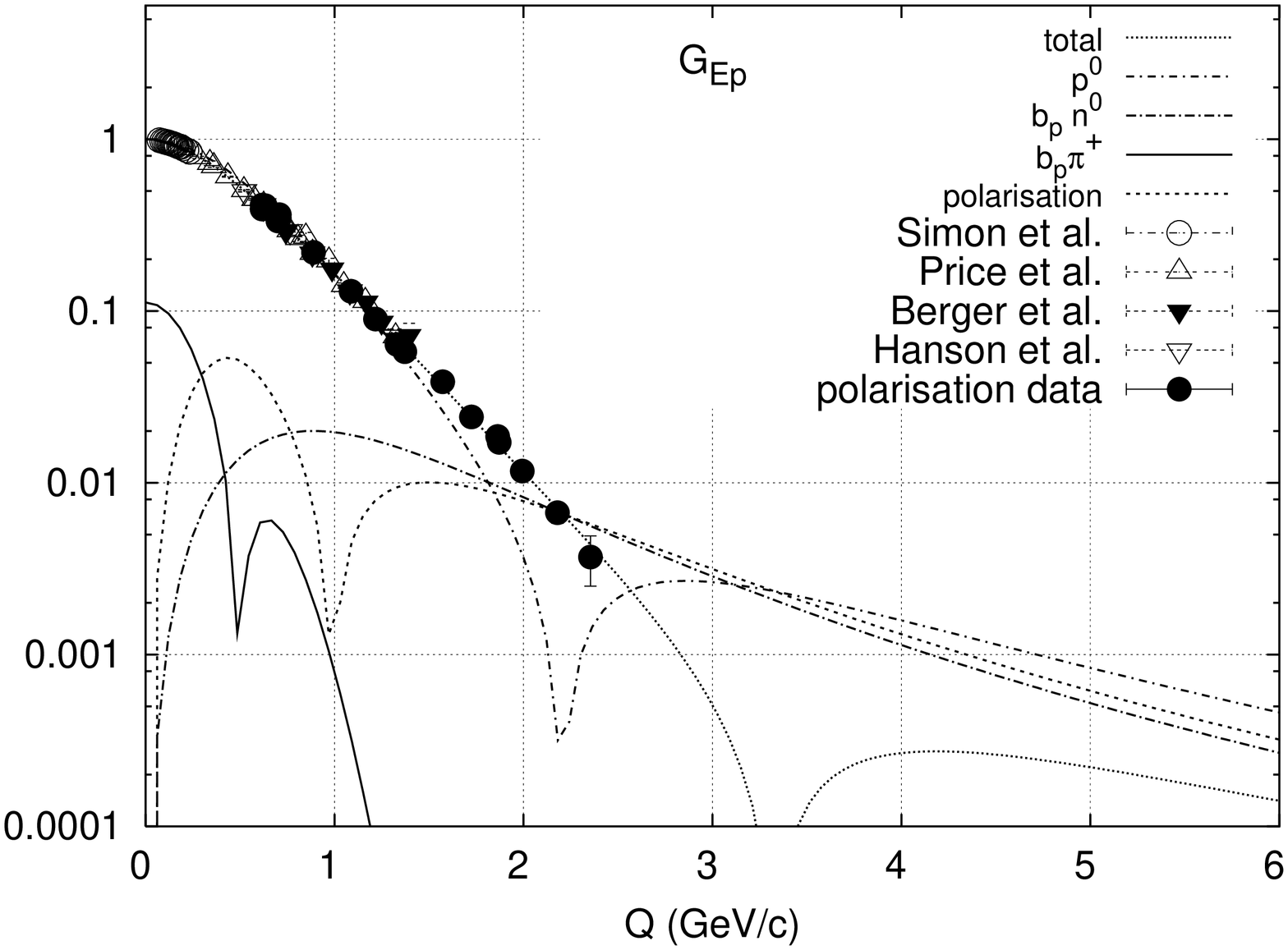}} } \\
\vspace*{-0.0cm} \rotatebox{-90} {\resizebox{6.0cm}{!}
{\includegraphics[width=1.0\textwidth]{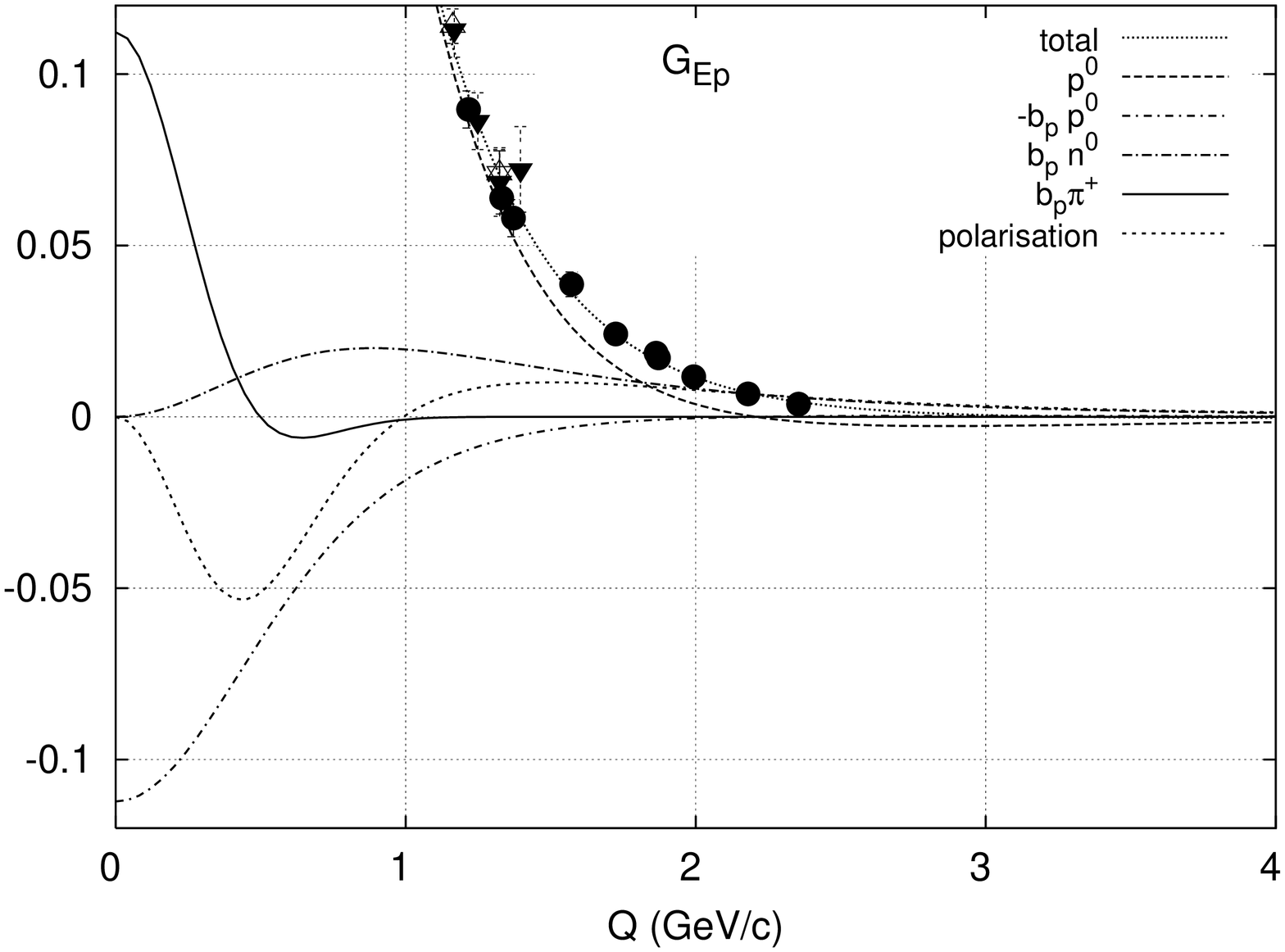}} } \\
\end{center}
\caption{$G_{Ep}$: Fit 2 (cf. table~\ref{tab:par_model}) compared to the measurements.
Upper panel: Full scale comparison. Lower panel: 
The contributions to $G_{Ep}$ in a linear scale
for a discussion of the interplay of the different contributions to $G^{pol}$
and, finally, the making up of $G^{Ep}$ as the sum of $G^0_p$ and $G^{pol}$
(note the extended x-scale in the lower panel). -
The form factors of $p^0$ + $\pi ^+$ are normalised to 1, that of
$n^0$ to 0.}
\label{fig:ff_e_p_model}
\end{figure}
The fit 2 is compared in fig.~\ref{fig:ff_e_p_model} to the data.
Here, we also show the single contributions of the model. 
In the logarithmic plot (upper panel) their signs get lost,
therefore, in order to discuss the interplay between the
single contributions to $G_{Ep}$, we have plotted in the lower panel
these contributions on a linear scale.
The dominating $G^0_p$ from the bare proton $p^0$ has a zero around 2.2~GeV/c
due to the inference between the positive $G^{up}$ and the negative $G^{dp}$, the
latter being suppressed by a factor 4, but extending out much further
($a^{dp}_1 > a^{up}_1$;
these contributions are not shown separately). This minimum, however, is shadowed by the
polarisation term $G_p^{pol}$. 
At $Q^2 = 0$, -$b_p G_p^0$ and $b_p G^{\pi ^+}$ cancel, while $G_n^0$ itself is zero there
(see lower panel of fig.~\ref{fig:ff_e_p_model}).
With increasing $Q$, the negative contribution of -$b_p G_p^0$ prevails, $G_p^{pol}$
thus becomes negative, until it is balanced by the positive contribution from the 
neutron, $b_p G_n^0$. Around $Q \approx $ 0.4~GeV/c $G_p^{pol}$ has a minimum,
around $Q \approx $ 1.0~GeV/c it passes through zero and becomes positive at large $Q$
where the contribution $b_p G_n^0$ prevails. Finally there results a zero
in $G_{Ep}$ around 3.3~GeV/c due to the interference of the positive polarisation
(from $n^0$) and the negative lobe of $G_p^0$ from its $d$-quark contribution. -
The data are certainly not sufficient to fix these numbers precisely;
nevertheless they are useful as a guidance for what might go on physically 
in this $Q^2$-range. The determination of the minimum in $G_{Ep}$ by experiment
is highly desirable.

Comparison with the evaluation in terms of the purely phenomenological
model in section~\ref{sec:phenomen} reveals that the bump structure
there has a different meaning than that of the
polarisation term resulting here from the evaluation in terms of the quark model.
It is clear, that this is due to the definition of what is ``bump'' and what is ``smooth''.
The interpretation of this model, however, makes clear, that the low-$Q^2$ side of the
bump results mainly from the interplay of the form factor from the $\pi ^+$ with
the reduction in $p^0$.

\begin{figure} [ht] 
\begin{center}
\vspace*{-0.0cm} \rotatebox{-90} {\resizebox{6.0cm}{!}
{\includegraphics[width=1.0\textwidth]{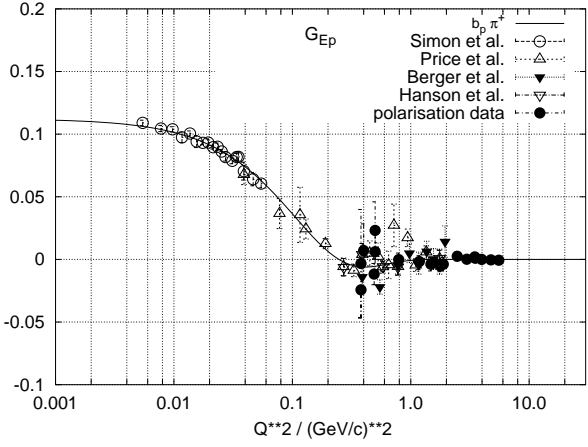}} } \\
\end{center}
\caption{The contribution $b_p G_E^{\pi ^+}$ from the pion cloud to
  $G_{Ep}$ (fit 2). For comparison the data points show the 
  measurements$- [(1-b_p)\cdot G^0_{Ep}+ b_p \cdot G^0_{En}]$. }
\label{fig:ff_e_p_pion_cloud}
\end{figure}

In fig.~\ref{fig:ff_e_p_pion_cloud} we show on a logarithmic $Q^2$-scale 
the contribution of the pion cloud for fit 2. This scale emphasises the
low-$Q^2$ part, where the pion-cloud contribution is concentrated.
It it obvious that the data are not precise enough to fix the zero in $G_E^{\pi ^+}$.
A good determination of this zero, however, would be a prerequisite to distinguish
between the parameter $a_1^{\pi}$ in the exponential and in the brackets, thus
to see the signature of the finite size of the pion.

\begin{figure}[ht] 
\begin{center}
\vspace*{-0.0cm} \rotatebox{-90} {\resizebox{6.0cm}{!}
{\includegraphics[width=1.0\textwidth]{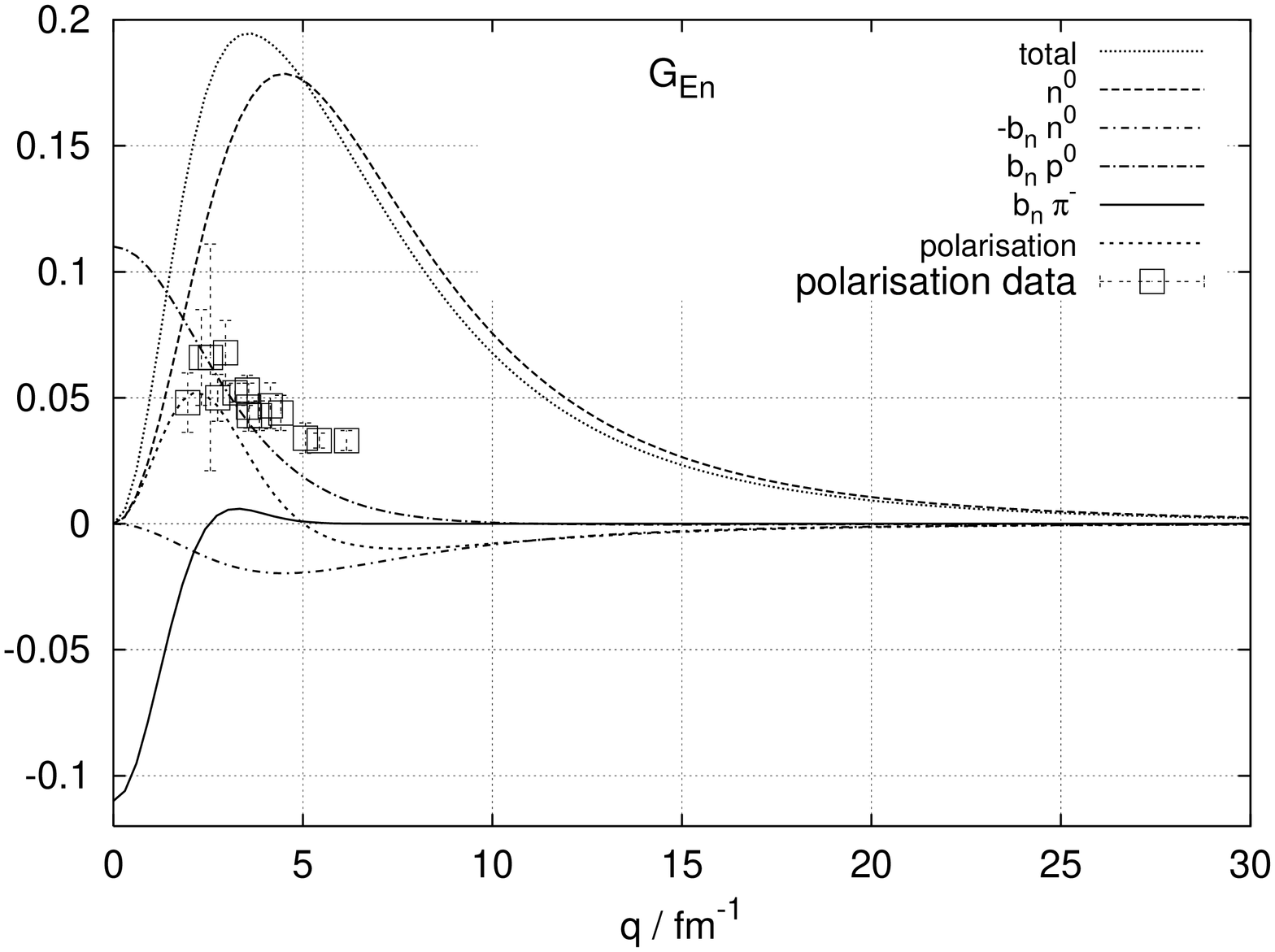}} } \\
\vspace*{-0.5cm} \rotatebox{-90} {\resizebox{6.0cm}{!}
{\includegraphics[width=1.0\textwidth]{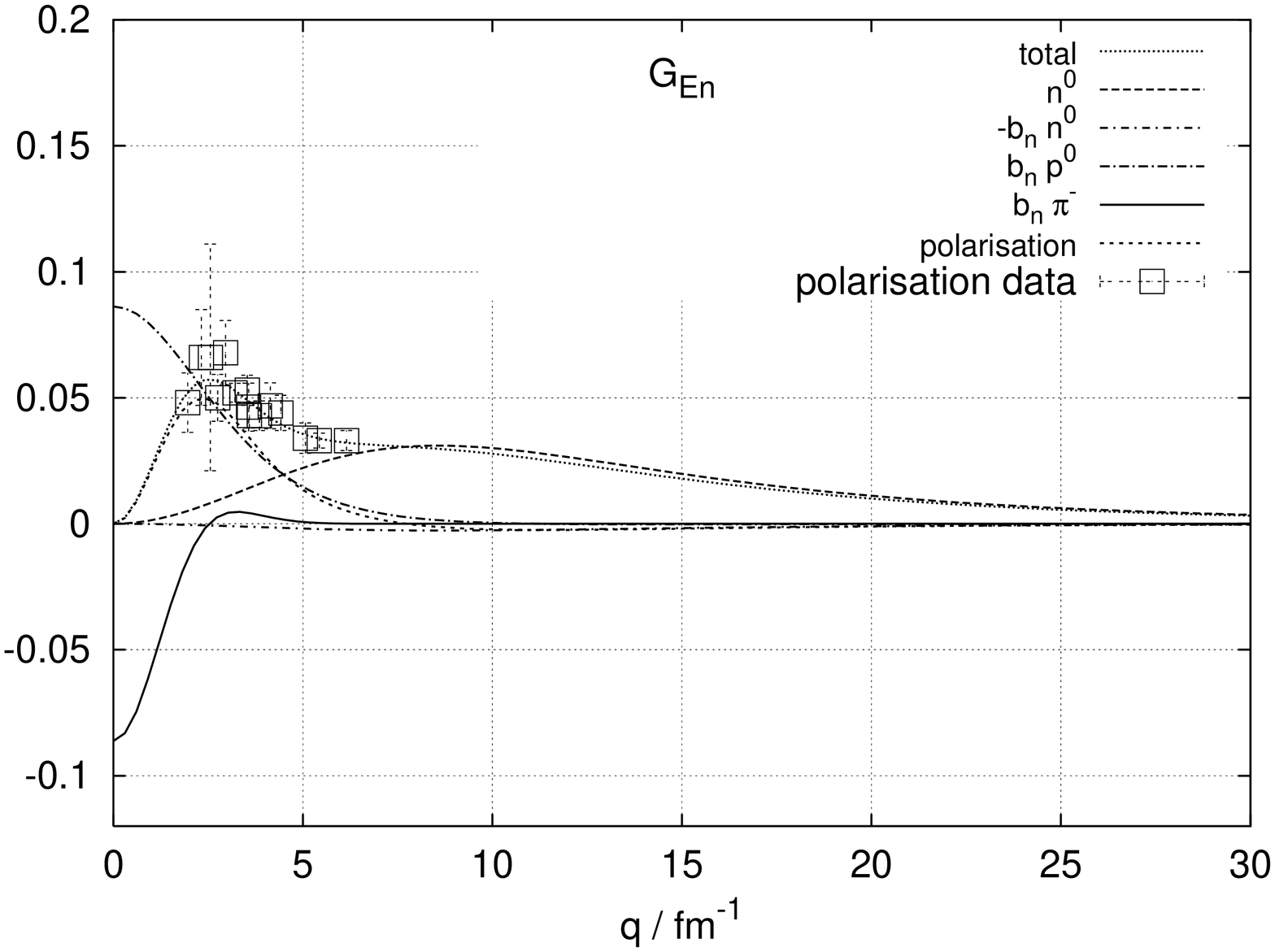}} } \\
\vspace*{-0.5cm} \rotatebox{-90} {\resizebox{6.0cm}{!}
{\includegraphics[width=1.0\textwidth]{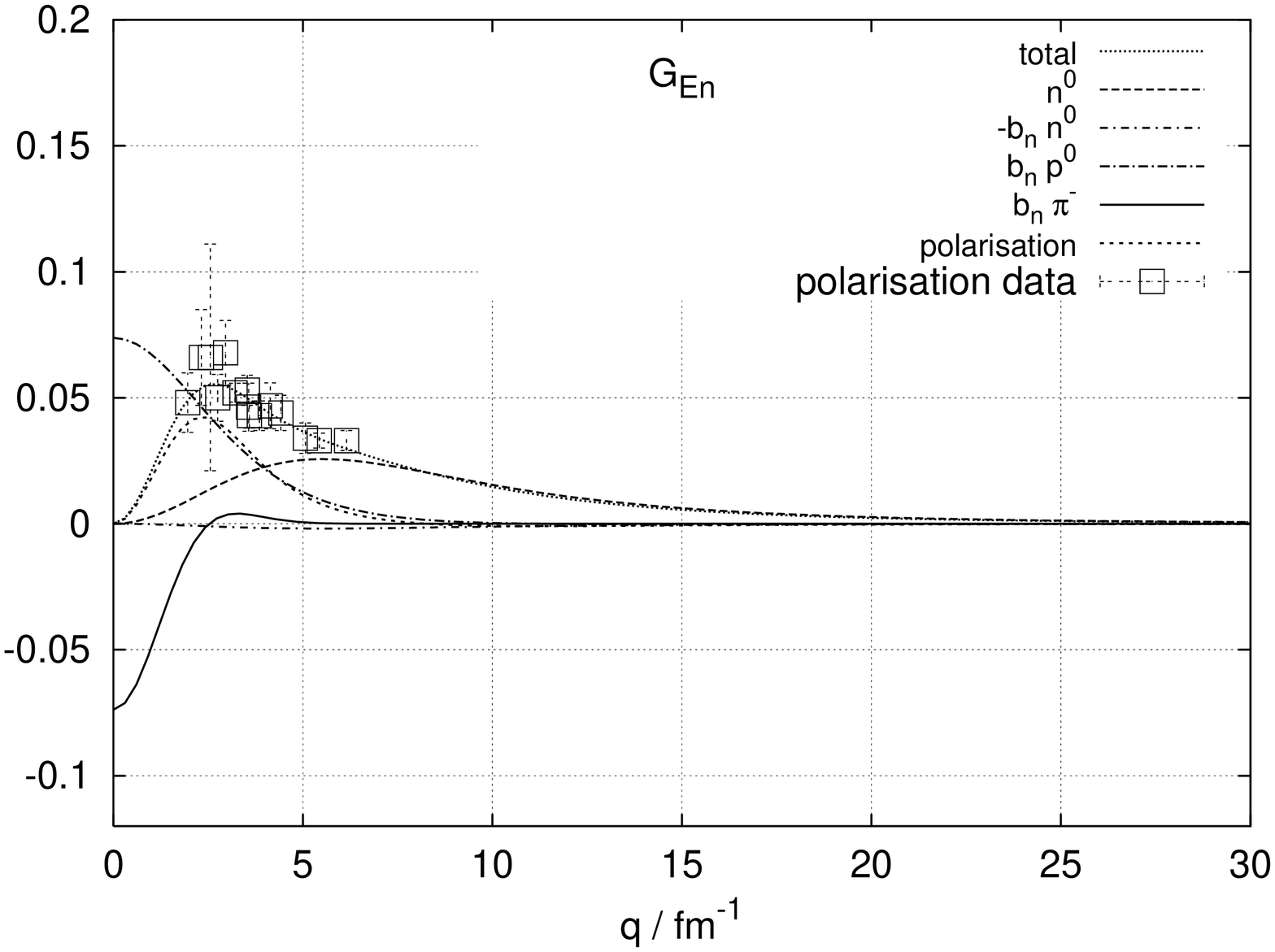}} } \\
\end{center}
\caption{$G_{En}$: Calculation with eq.~(\ref{eq:gen_quark_pion})
compared to the measurements. 
Upper panel: Calculation obeying strict isospin invariance. 
Middle panel: $a^{un}_1$, $a^{dn}_1$, and $b_p$ fitted to the $G_{En}$-data 
(see fit 2 to $G_{En}$ in table~\ref{tab:par_model}).
Lower panel: $a^{dn}_1$ and $b_n$ fitted to the data 
(see fit 3 to $G_{En}$ in table~\ref{tab:par_model}).
}
\label{fig:ff_e_n_model}
\end{figure}

We have seen that in the fits to $G_{Ep}$ the parameters of the contribution from $G^0_n$
remain practically undetermined. We now want to see what we learn about them from the
electric form factor of the neutron, for which the corresponding
expression is
\begin{eqnarray}
G_{En} & = & ( G^{un}_E + G^{dn}_E )                                         \label{eq:gen_quark_pion} \\
       & + & b_n \cdot ((G^{up}_E+G^{dp}_E)\,-\,(G^{un}_E+G^{dn}_E)+G^{\pi ^-}_E) \, .  \nonumber 
\end{eqnarray}
As in the case of the proton, this ansatz conserves automatically the normalisation
(here: to 0). Respecting strict isospin invariance, i. e. 
$a^{un}_1 = a^{dp}_1$, $a^{dn}_1 = a^{up}_1$, and $G_E^{\pi^+} = G_E^{\pi^-}$ 
we can calculate $G_{En}$ from eq.~(\ref{eq:gen_quark_pion}).
In the upper panel of fig.~\ref{fig:ff_e_n_model}
we compare this calculation to the measured data,
using the parameters of fit 2 for $G_{Ep}$.
First, it is remarkable, that and how well the polarisation term, directly calculated with
the parameters from the fit to $G_{Ep}$, reproduces the bump structure of $G_{En}$
at low $Q$! However, second, the contribution from the bare neutron alone,
while being reasonably well positioned in $Q$,
overestimates drastically the total measured $G_{En}$, 

\begin{table*}[ht]     
\begin{center}
\begin{tabular}{||l||l|l||l|l||l|l||r|c|r||}
\hline
\hline
              &\multicolumn{2}{c||}{$G^0_p$}&\multicolumn{2}{c||}{$G^0_n$}& \multicolumn{2}{c||}{$G^{\pi ^+}$}& \multicolumn{3}{c||}{     }                \\
\hline
form factor   &  $a^{up}_1$  &  $a^{dp}_1$  &  $a^{un}_1$  & $a^{dn}_1$   & $b_{p,n}$ & $a^{\pi ^{+/-}}_1$    &$N_{d.f.}$&$\chi ^2_{d.f.}$&$\chi^2_{total}$\\ 
              &  (GeV/c)$^2$ &  (GeV/c)$^2$ &  (GeV/c)$^2$ & (GeV/c)$^2$  &           & GeV/c                 &          &                &                \\
\hline
              &              &              &              &              &           &                       &          &                &                \\
$G_{Ep}$      &              &              &              &              &           &                       &          &                &                \\
1             & 1.000(100)   & 2.03(72)     & 57.(1200.)   & 78.(2500.)   & 0.10(4)   & 0.198(12)             & 64       & 0.932          & 59.6           \\
2             & 1.008(20)    & 2.54(16)     & 2.54(-)      & 1.008(-)     & 0.11(1)   & 0.203(6)              & 66       & 0.926          & 61.1           \\
3             & 1.051(20)    & 2.391(14)    & 2.53(-)      & 2.22(-)      & 0.118(13) & 0.204(6)              & 66       & 0.928          & 61.2           \\
              &              &              &              &              &           &                       &          &                &                \\
$G_{En}$      &              &              &              &              &           &                       &          &                &                \\
1             & 1.008(-)     & 2.54(-)      & 2.54(-)      & 1.008(-)     & 0.11(-)   & 0.203(-)              & -        & -              & -              \\
2             & 1.008(-)     & 2.54(-)      & 6.2(6.4)     & 5.3(5.1)
              & 0.086(10)    & 0.203(-)              & 12       & 0.807          &  9.7           \\
3             & 1.008(-)     & 2.54(-)      & 2.54(-)      & 2.22(2)      & 0.074(5)  & 0.203(-)              & 13       & 0.818          & 10.6           \\
\hline
\hline
\end{tabular} \\ [0.5cm]
\end{center}
\caption{Parameters from the fits of the model ansatz to the electric nucleon form factors. 
For the error convention see caption of table~\ref{tab:par_phen}.
The different fits observe more or less to isospin symmetry (see text).}
\label{tab:par_model}
\end{table*}

\begin{table*}[ht]     
\begin{center}
\begin{tabular}{||l||l|l||l|l||l|l||r|c|r||}
\hline
\hline
              & \multicolumn{2}{c||}{      }& \multicolumn{2}{c||}{      }& \multicolumn{2}{c||}{          }  & \multicolumn{3}{c||}{     }                \\
              & \multicolumn{2}{c||}{outer }& \multicolumn{2}{c||}{inner} & \multicolumn{2}{c||}{$G^{\pi }$}  & \multicolumn{3}{c||}{     }                \\
              & \multicolumn{2}{c||}{distr.}& \multicolumn{2}{c||}{distr.}& \multicolumn{2}{c||}{          }  & \multicolumn{3}{c||}{     }                \\
\hline
              &              &              &              &              &           &                       &          &                &                \\
form factor   & $a^{out}_0$  & $a^{out}_1$  & $a^{in}_0 $  & $a^{in}_1$ &$a^{\pi ^{+/-}}_0$&$a^{\pi ^{+/-}}_1$&$N_{d.f.}$&$\chi ^2_{d.f.}$&$\chi^2_{total}$\\ 
              &              & (Gev/c)$^2$  &              & (Gev/c)$^2$  &           & Gev/c               &          &                &                \\
\hline
              &              &              &              &              &           &                       &          &                &                \\
$G_{Mp}/\mu _p$ &            &              &              &              &           &                       &          &                &                \\
1             & 0.914(5)     & 0.818(8)     & -0.0049(1)   & 9.578(1.2)   & 0.110(7)  & 0.213(7)              & 75       & 0.887          & 66.5           \\
2             & 0.917(6)     & 0.811(16)    & -0.0034(14)  & 13.57(6.0)   & 0.106(8)  & 0.210(8)             & 69       & 0.901          & 62.2           \\
              &              &              &              &              &           &                       &          &                &                \\
$G_{Mn}/\mu _n$ &            &              &              &              &           &                       &          &                &                \\
1            & 1.019(14)     & 0.939(110)   & -0.112(16)   & 2.37(1.1)    & 0.219(47) & 0.152(9)              & 14       & 0.629          & ~8.8           \\
2            & 1.363(3.14)   & 1.173(700)   & -0.511(3.17) & 1.789(2.0)   & 0.140(46) & 0.213(-)              & 15       & 0.946          & 14.9           \\
3            & 1.189(1.34)   & 1.060(460)   & -0.309(1.38) & 1.853(1.8)   & 0.120(40) & 0.189(19)             & 15       & 0.837          & 12.6           \\
             &               &              &              &              &           &                       &          &                &                \\
\hline
\hline
\end{tabular} \\ [0.5cm]
\end{center}
\caption{Parameters for the fits to the magnetic form factors. In all but fit 3 for $G_{Mn}$ the normalisation is free.
$G_{Mp}$ ($Q^2_{max}$ = 31.2 (GeV/c)$^2$) : Fit 1: Fit to all data, all parameters free. 
Fit 2: Fit to the data up to 10 (GeV/c)$^2$ with all parameters free.
$G_{Mn}$ ($Q^2_{max}$ = 10 (GeV/c)$^2$): 
Fit 1: All parameters free. 
Fit 2: $a^{\pi ^-}_1$ kept fixed at 0.213~GeV/c as determined for $G_{Mp}$.
Fit 3: Normali\-sation kept fixed by adding a point with value  1.0000 $ \pm $ 0.0001 at $Q^2$ = 0. }
\label{tab:par_model_gm}
\end{table*}

There are two ways to reduce the amplitude of the superposition of two dipoles with
equal amplitudes of different sign. The first way is to just reduce the common amplitude,
here one would need a reduction of roughly a factor of 6.
The amplitude being given by the charge of the one $up$ and the two $down$ quarks as +2/3
and -2/3, respectively, such reduction would mean to leave the grounds of the
present model, namely the building up of the nucleons by constituent
quarks. Therefore, we prefer the second way which requests letting the two parameters
$a^{un}_1$ and $a^{dn}_1$ approach each other. In the fit we let the
routine search for an appropriate parameter choice by varying only 
$a^{un}_1$, $a^{dn}_1$, and $b_p$. In fact, the program finds a perfect fit
to the measured $G_{En}$ with a $\chi ^2$ per d. f. of 0.81 (fit 2).
While fit 2 reduces the polarisation term by only some 20\,\%, the dipole
parameters are increased by a factor of 2.5 and 5, respectively, which corresponds to
making the distribution in $r$-space very narrow. Furthermore, $a^{un}_1/a^{dn}_1$ = 1.2
whereas $a^{dp}_1/a^{up}_1$ = 2.5, i. e. the $up$- and $down$-quark distributions are much 
more similar in $n^0$ than in $p^0$.
This finding, however, may not be too surprising since at the small distances of
fractions of a $fm$ the difference in the Coulomb interaction in $p^0$ and $n^0$
might make strict isospin symmetry questionable. In other words,
the net positive charge of the two up constituent quarks will repel
them so they reside more outside than the quarks with the net zero
charge in the neutron. One should not mix this up with the opposite
behaviour of the current quark distribution as derived from deep
inelastic scattering. - 
The result of this fit is shown in the middle panel of fig.~\ref{fig:ff_e_n_model}.

In a last step (fit 3) we examine the significance of 
the fit of $a^{un}_1$ and $a^{dn}_1$ 
by setting $a^{un}_1$ equal to $a^{dp}_1$ from fit 2 of $G_{Ep}$ and keeping this fixed.
We thus allow only $a^{dn}_1$ and $b_n$ to vary. While the resulting $b_n$ differs by 
less than one standard deviation from its value in fit 2, $a^{dn}_1$ just follows $a^{un}_1$
in order to keep the difference small, which, as said above, is necessary to keep
the $G_{En}$ small in the high-$Q$ region.
As expected from the large uncertainties in $a^{un}_1$ and $a^{dn}_1$ in fit 2,
$\chi^2$ only varies by 0.9 with this quite drastic variation in $a^{un}_1$.
The result of this fit is shown in the lower panel of fig.~\ref{fig:ff_e_n_model}.

Data at higher momentum transfers are needed to further constrain
the low-distance behaviour of the neutron form factor.

We have checked the significance of the bump-structure in $G_{En}$ by fitting
the data with only a smooth ansatz consisting of two dipoles
with equal but opposite amplitudes, which is equivalent  to the ansatz 
eq.~(\ref{eq:Galster}). With this parameterisation we get $\chi ^2_{total}$ = 11.1
($d.f.$ = 12), i. e. an increase by 1.4 compared to fit 2 in table~\ref{tab:par_model}.
Thus, for a significant determination of the bump more precise data at low $Q^2$
are needed as well as data extending to higher $Q^2$. 

Finally, we check whether the fit of the proton's electric form 
factor is deteriorated when the neutron parameters 
are kept fixed to the values determined now from the fit to the 
neutron data. The resulting parameters are shown as fit 3 of $G_{Ep}$ 
in table~\ref{tab:par_model}. In fact the change in the parameters
and thus also in the graphical representation of the 
form factor are so small that we need not go into any detail here.

Summarising up for the electric form factors, $G_{Ep}$ and $G_{En}$ 
can be described on the same footing by our constituent-quark-pion ansatz.

\subsection{The magnetic form factors}
\label{sec:ff_mag}

For $G_{Mp}$ data are measured up to $Q^2$ = 30~(GeV/c)$^2$,
therefore in this respect the situation is more favourable here.
On the other side, the interpretation of the magnetic form
factor within our model is hampered by the 
additional degree of freedom of the vector coupling of the spins
and the magnetic moments:
While it is clear, that, e. g., the two $u$-quarks in the proton carry
the charge $2 \cdot 2/3$, the resulting magnetic moment depends
on the coupling of the quark spins.
Furthermore, it is not clear what magnetic moment one
has to attribute to the constituent quarks.
The same uncertainty holds for the contribution of the
pion. Though it should predominantly be in a $p$-state,
the related magnetic moment is not known, since
the pion is highly off-mass shell and therefore
its mass is not that of the free pion.
Furthermore, its contribution to the total magnetic
moment depends on the vector coupling.
Therefore, in the evaluation of the magnetic form factors
with the ansatz eqs.~(\ref{eq:gep_quark_pion},\ref{eq:gen_quark_pion}) 
with eqs.~(\ref{eq:quark_dipole},\ref{eq:pion_ff})
we have to take also the 
amplitudes $a^{qN}_0$ and $a^{\pi }_0$ as free parameters.

On the other side, one might think that
the parameters $a^{qN}_1$, which describe the spatial distributions,
might be the same for $G_M$ and $G_E$ such that they can
be taken from there. 
One could, however, only profit from this
for the sufficiently well determined bare proton part,
and here only for the dominating term from the $u$-quarks.
However, it is not clear whether the dipole parameter, determined at
relatively low $Q^2$, really should hold up to the highest $Q^2$
to which $G_{Mp}$ has been measured. Furthermore, the
magnetic operator does not weight the distribution in the same 
way as does the electric operator. Therefore, also the
parameters $a^{qN}_1$ have to be taken as free.

\begin{figure} [ht] 
\begin{center}
\vspace*{-0.0cm} \rotatebox{-90} {\resizebox{6.0cm}{!}
{\includegraphics[width=1.0\textwidth]{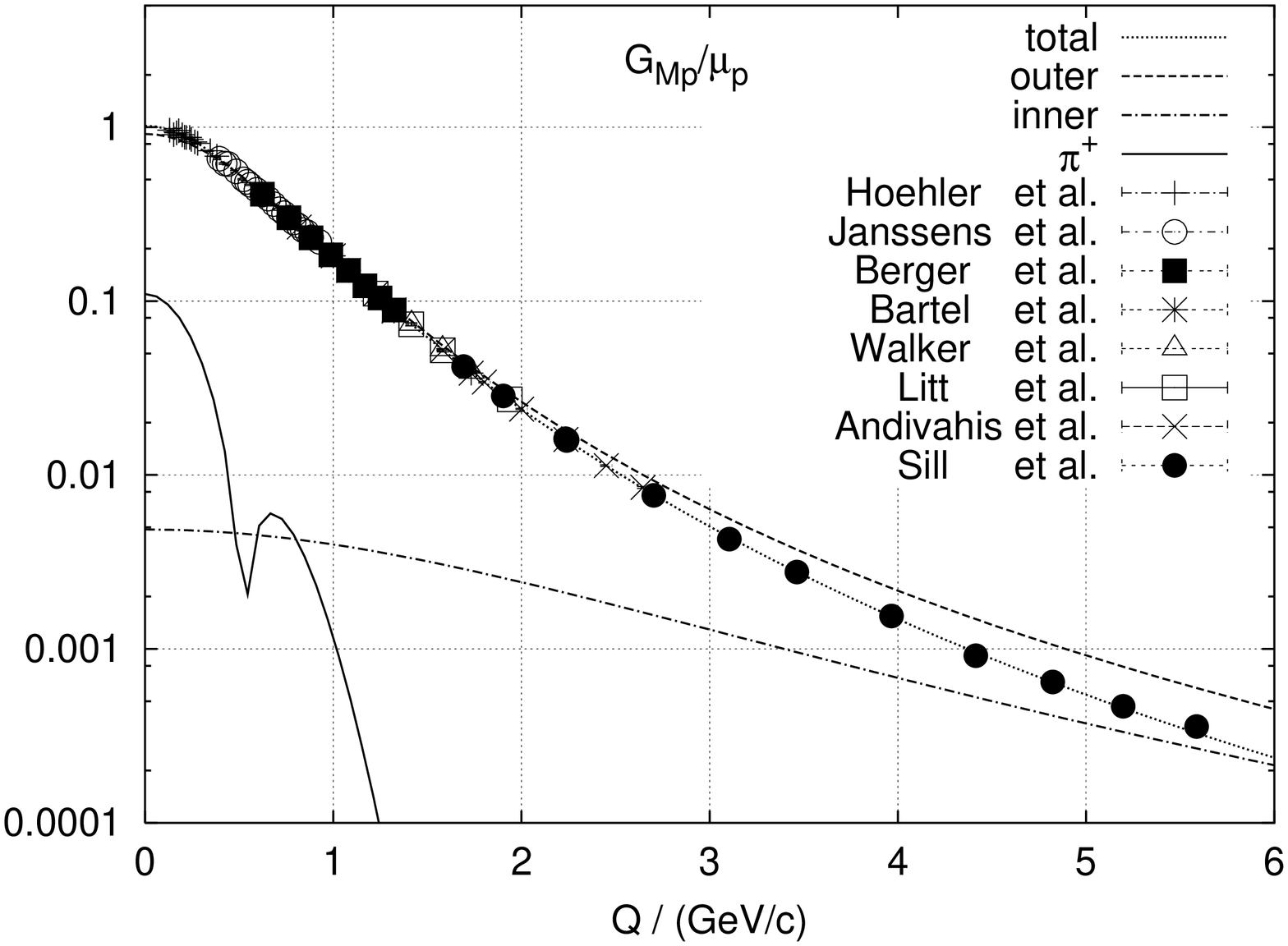}} } \\
\vspace*{-0.0cm} \rotatebox{-90} {\resizebox{6.0cm}{!}
{\includegraphics[width=1.0\textwidth]{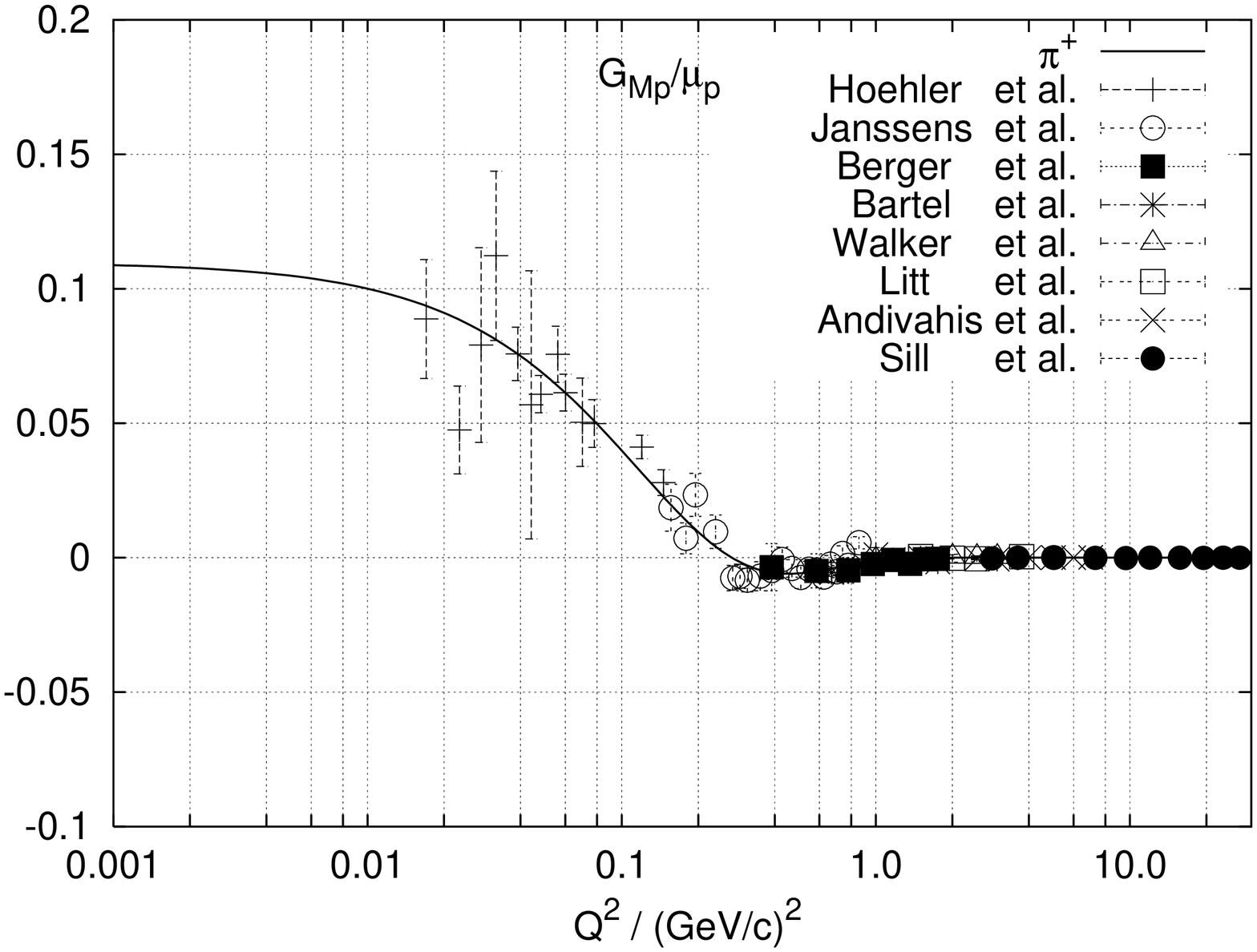}} } \\
\end{center}
\caption{The magnetic form factor of the proton (parameters from fit 1). 
Upper panel:
Comparison of the measured data with $G^{out} + G^{in} + G{\pi^+}$ (total)
and the three contributions to the fit separately. 
Lower panel: The data points
show the measurements$-[G^{out} + G^{in}]$, compared to the pion cloud 
$a_0^{\pi^+} \cdot G^{\pi^+}$.
Note that here the data are shown as function of $log(Q^2)$ in order to emphasise
the low-$Q^2$ region.}
\label{fig:ff_m_p_pion_cloud}
\end{figure}

\begin{figure} [ht] 
\begin{center}
\vspace*{-0.0cm} \rotatebox{-90} {\resizebox{6.0cm}{!}
{\includegraphics[width=1.0\textwidth]{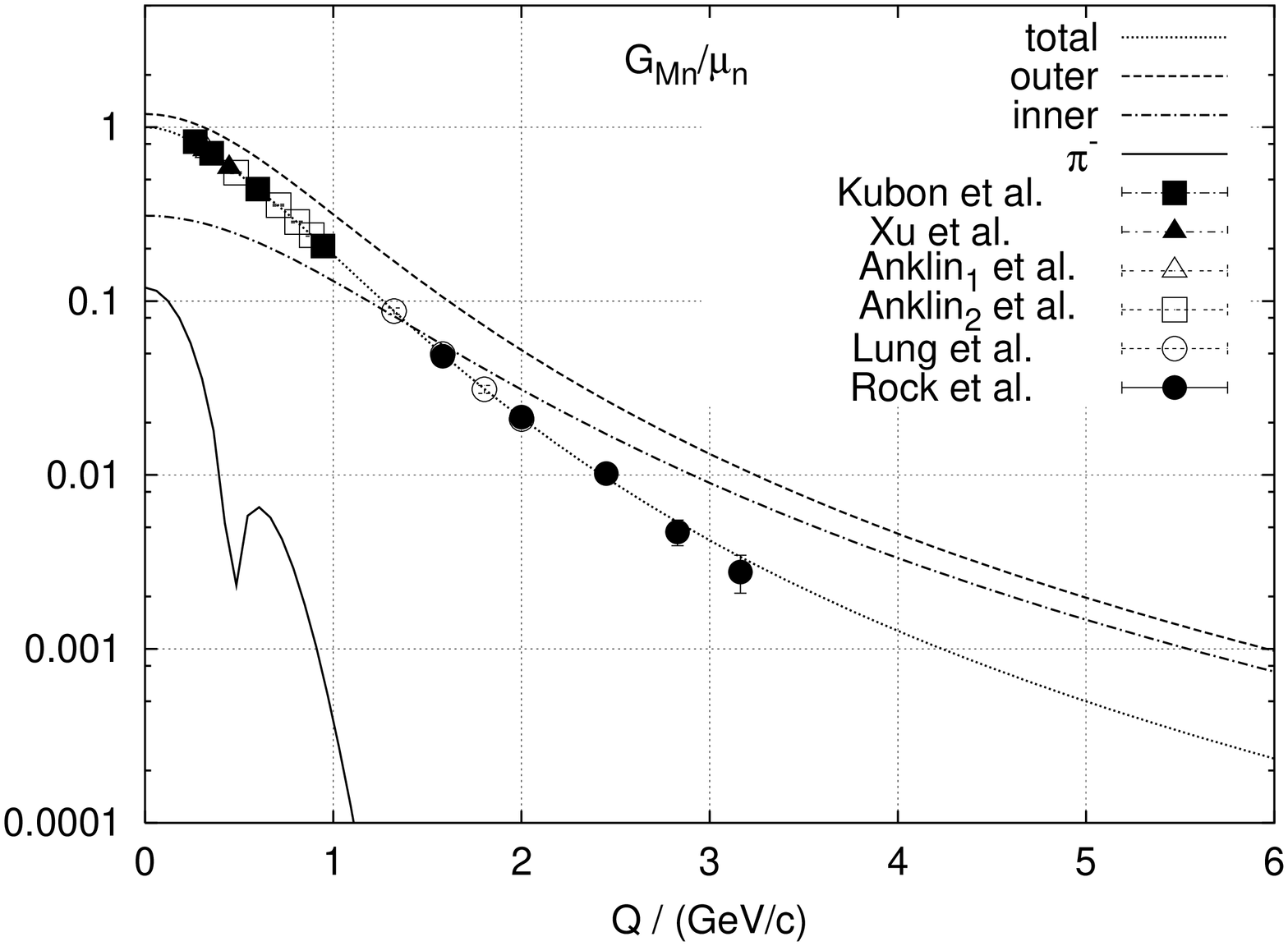}} } \\
\vspace*{-0.0cm} \rotatebox{-90} {\resizebox{6.0cm}{!}
{\includegraphics[width=1.0\textwidth]{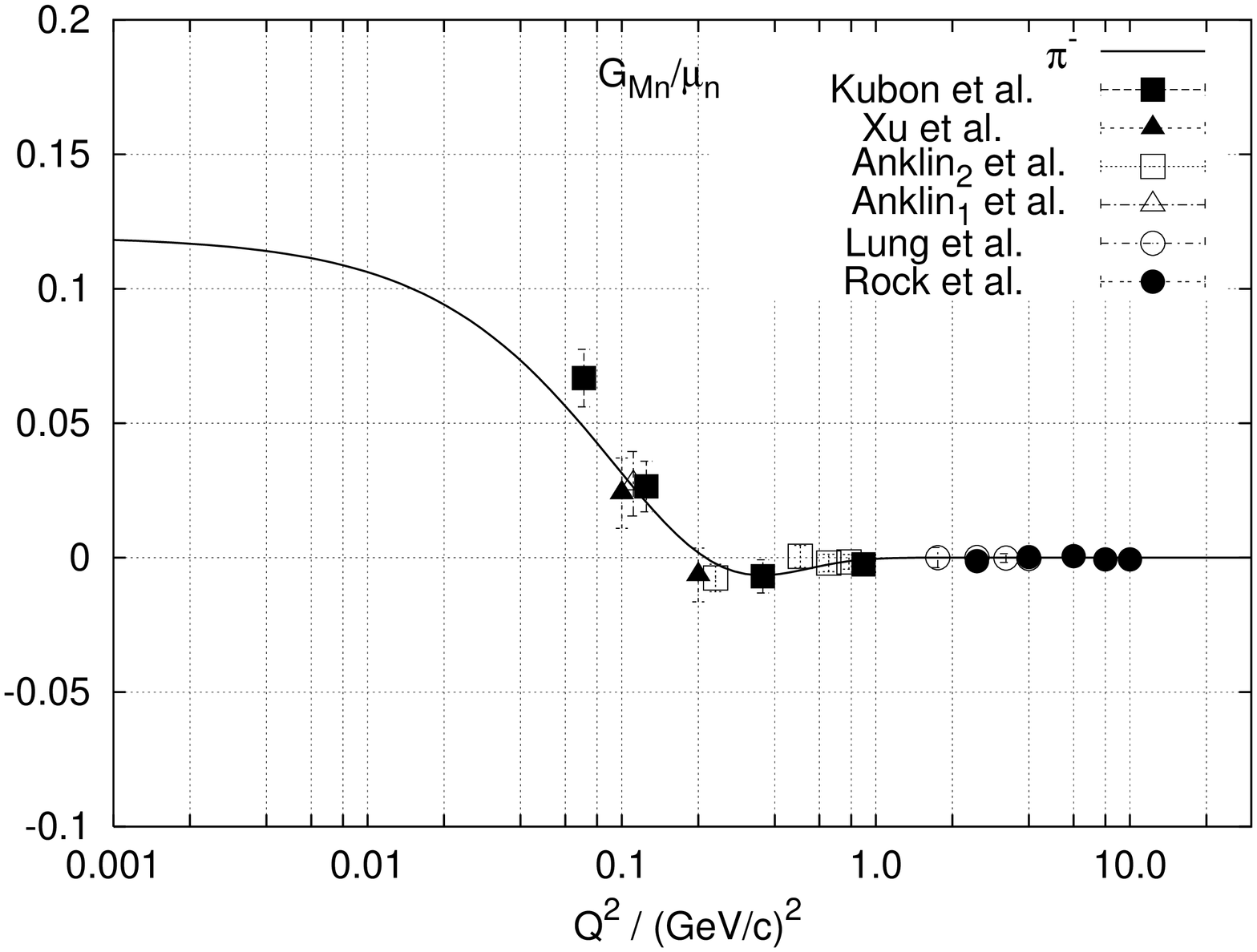}} } \\
\end{center}
\caption{Same as fig.\ref{fig:ff_m_p_pion_cloud}, now for $G_{Mn}$; 
fit 3 in table~\ref{tab:par_model_gm}.}
\label{fig:ff_m_n_pion_cloud}
\end{figure}

Isospin symmetry would suggest that there are only
two different distributions, that for $u$- and $d$-quark
in the proton, and in the neutron, respectively, the inverted
case. In this case, including the pion there are only 
three distributions left and we try the ansatz

\begin{equation}
G_{M} = a^{out}_0 \cdot G^{out} + a^{in}_0 \cdot G^{in} + a^{\pi }_0 \cdot G^{\pi } \, .
\label{eq:gm_quark_pion}
\end{equation}

Here, the nomenclature reminds on $inner$ and $outer$ quarks,
and we omit any discussion about the coupling of their magnetic
moments by just giving free amplitudes $a^{out,in}_0$ to their respective 
contributions to the magnetic form factors which, again,
are parametrised by the dipole form eq.~(\ref{eq:quark_dipole})
with the free parameters $a^{out,in}_1$.
In the same way we allow for a free amplitude for the pion cloud.
With the $a_0$ as free parameters, normalisation is not guaranteed. -
We only mention in passing, that we have checked fits with three
dipoles; even in the case of $G_{Mp}$, however, up to the 
highest $Q^2$ two dipoles are sufficient.

The parameters from the fits are tabulated in table~\ref{tab:par_model_gm}.
In the fits 1, all 6 parameters of the ansatz were free. 
Again $\chi ^2$ is comparable to the data evaluation with the
phenomenological ansatz. We show in the upper panels of
figs.~\ref{fig:ff_m_p_pion_cloud} and \ref{fig:ff_m_n_pion_cloud}
how well the data are described. Here, the three terms
are also shown separately.

For $G_{Mp}$ we find a surprisingly large value for $a^{in}_1$,
corresponding to a concentration of the respective distribution in 
r-space near the origin (see subsection~\ref{sec:charge_model} below),
however with very small amplitude.
For the sake of comparison with $G_{Mn}$, we have repeated the fit
with restricting the data to the $Q^2$-range for which there are
data for both magnetic form factors (fit 2). We find such large values 
for $a^{in}_1(G_{Mp})$ also from this restricted data base.
About 90\% of the (positive) magnetic moment of the proton is carried 
by the $outer$ distribution ($a^{out}_0(G_{Mp}) \approx 0.91$) and 10\%
by the pion cloud ($a^{\pi ^+}_0(G_{Mp}) \approx 0.11$). Note,
that the normalisation is violated by some 2\%. This should be acceptable
in view of the quality of the data at low $Q^2$.
The $inner$ distribution contributes only about -0.4\% to the magnetic moment,
while its contribution to the form factor becomes comparable at
large $Q^2$. Further, $a^{out}_1(G_{Mp})$ is 20\% smaller than $a^{up}_1(G_{Ep})$,
thus the related distribution in $r$-space extends further out
for the magnetism than for the charge.

In fit 1 of $G_{Mn}$, the dominant contribution to the magnetic moment
again comes from the $outer$ distribution (note that by referring to
$G_{Mn}/\mu _n$ the signs are inverted). 
The sign of the $inner$ distribution again is negative, in this case, however, 
its contribution to the magnetic moment is about 11\% and thus not negligible. 
The pion cloud contributes a factor of two more to $\mu _n$ than to $\mu _p$ in this fit.
It has to be admitted, however, that in this fit the normalisation is
off by about 10\%. In fact, the data do not extend sufficiently far
down in $Q^2$ to let the normalisation free in the fit,
and the pion cloud is particularly sensitive to the data at low $Q^2$.
Fit 2 shows the result of a fit with  $a^{\pi ^-}_1(G_{Mn})$ fixed to 
$a^{\pi ^+}_1(G_{Mp})$ = 0.213~GeV/c.
The fit now obeys normalisation to within a percent with the
amplitudes $a^{out}_0$ and $a^{in}_0$ having very large (correlated!) errors.
With fit 3 we went one step further by adding an additional data point
at $Q^2$ = 0 in order to fix the normalisation, while at the same time
letting the parameter $a^{\pi ^-}_1(G_{Mn})$ free. There is some redistribution
between $inner$ and $outer$ distribution, but all changes of the parameters
are within the errors. Thus, there is no problem with the normalisation
of the data. - We only mention in passing that
taking into account also the data by Markowitz et al.~\cite{Mar93}
and by Bruins et al.~\cite{Bru95} yield  $a^{\pi ^-}_0(G_{Mn})$ between 0.09 and 0.12
and $a^{\pi ^-}_1(G_{Mn})$ between 0.186 and 0.189~GeV/c.

It is beyond the scope of this analysis to try an explanation of these
findings.

\subsection{The distributions in r-space}
\label{sec:charge_model}

\begin{figure} 
\begin{center}
\vspace*{-0.0cm} \rotatebox{-90} {\resizebox{6.0cm}{!}
{\includegraphics[width=1.0\textwidth]{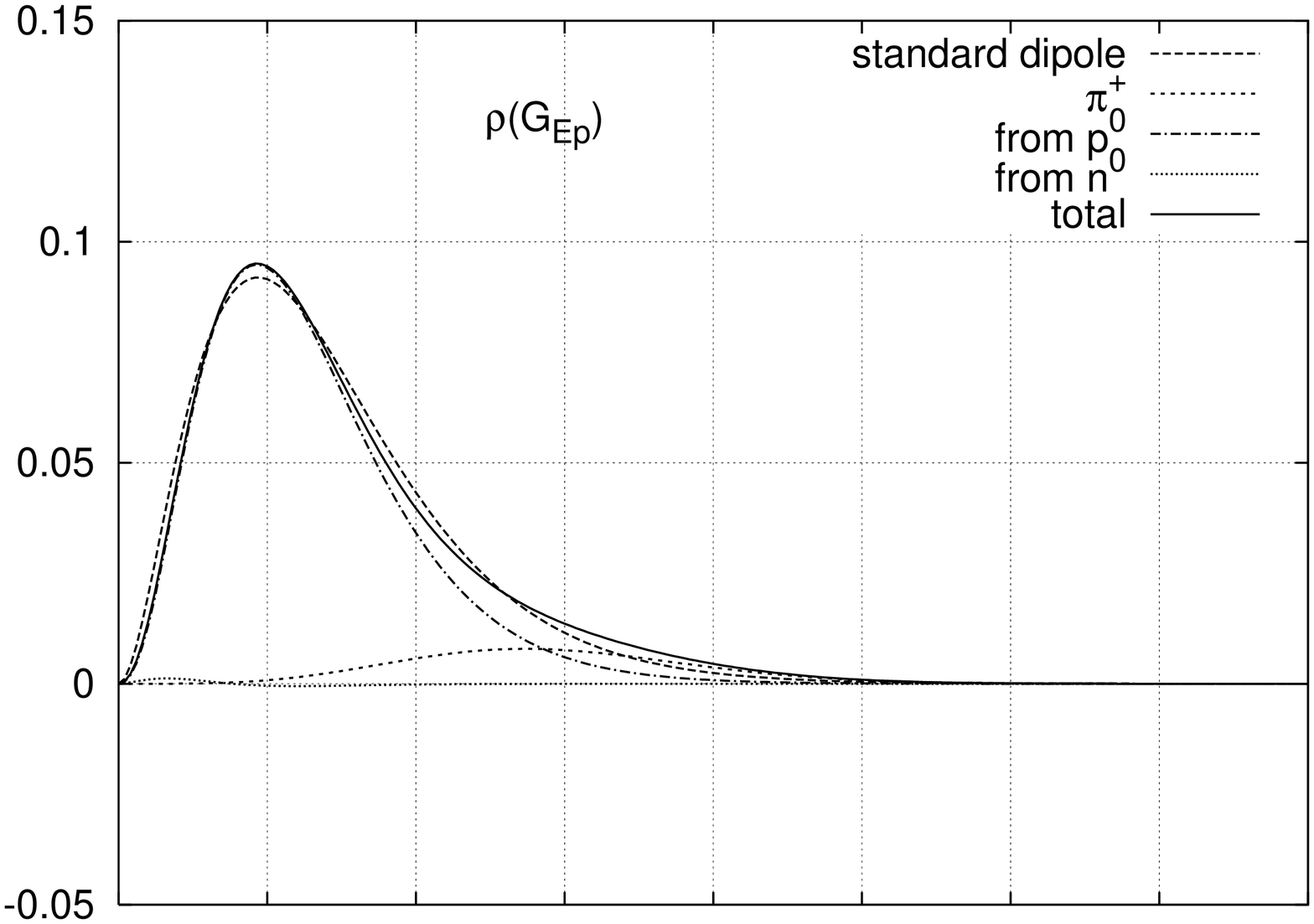}} } \\
\vspace*{-0.5cm} \rotatebox{-90} {\resizebox{6.0cm}{!}
{\includegraphics[width=1.0\textwidth]{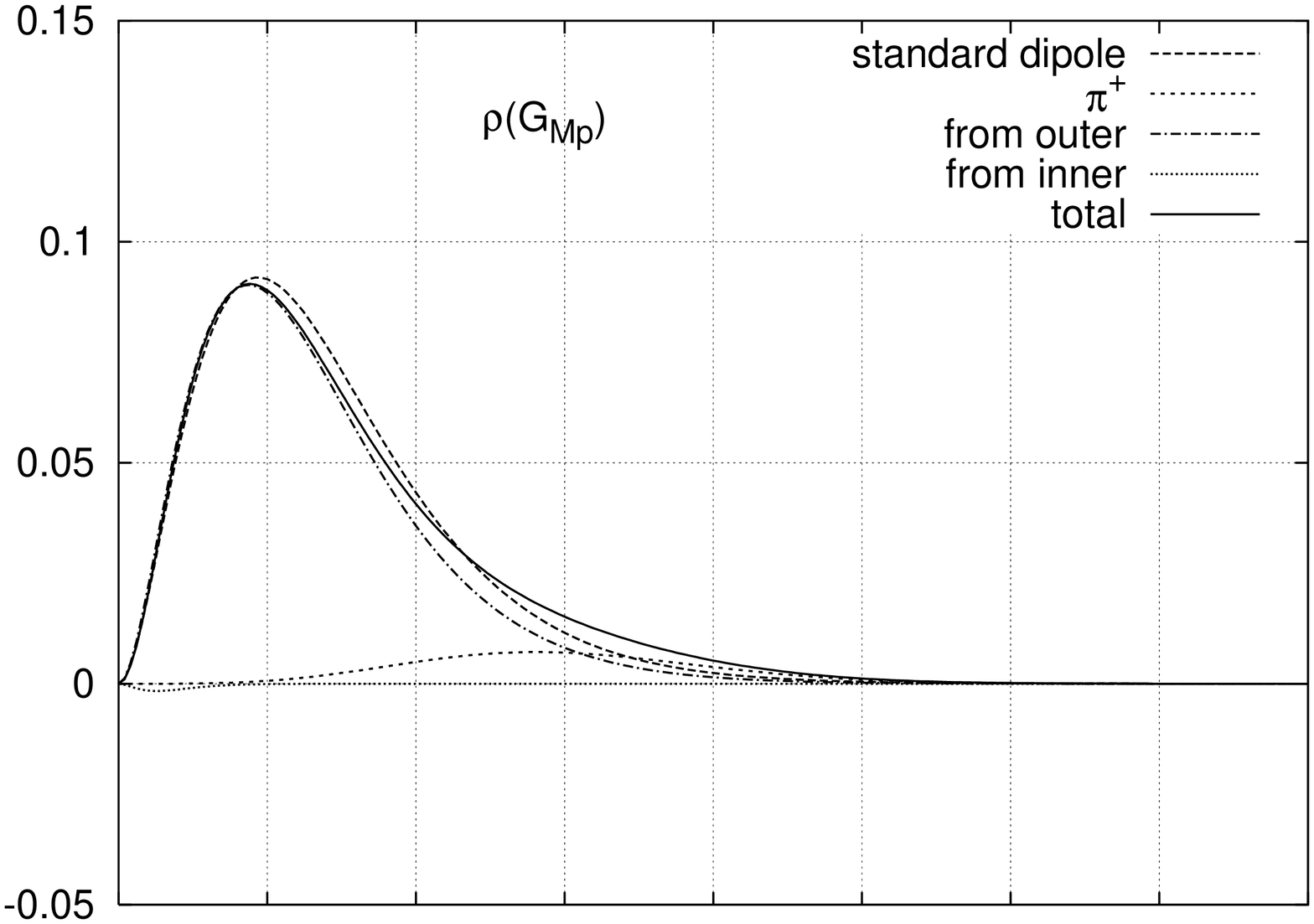}} } \\
\vspace*{-0.5cm} \rotatebox{-90} {\resizebox{6.0cm}{!}
{\includegraphics[width=1.0\textwidth]{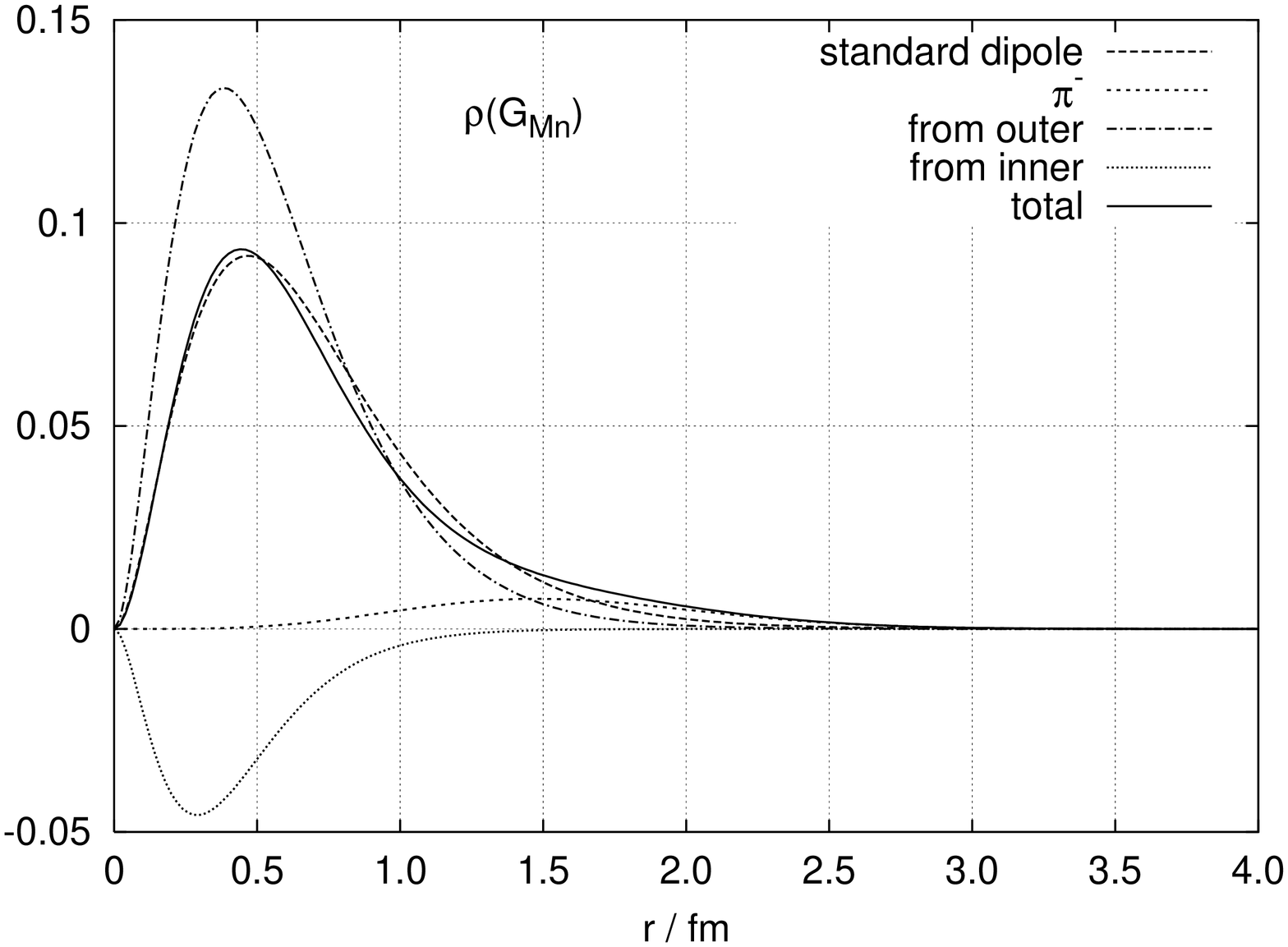}} } \\
\end{center}
\caption{$r^2 \cdot \rho (r)$ in the Breit frame, calculated with parameters
given in tables~\ref{tab:par_model} and \ref{tab:par_model_gm} 
($G_{Ep}$: Fit 3, $G_{Mp}$: Fit 2, $G_{Mn}$: Fit 3). }
\label{fig:rho_r2_mod}
\end{figure}

In fig.~\ref{fig:rho_r2_mod} we again show the distributions $r^2 \cdot \rho (r)$
in the Breit frame for the three standard form factors, 
now for the model evaluation calculated with parameters given
in tables~\ref{tab:par_model} and \ref{tab:par_model_gm}.
In the proton, the contribution from the $inner$ distribution
is practically invisible. This shows that, to the degree of precision 
visible in this plot, the proton form factors
are describable by one dipole plus the contribution from the $\pi ^+$,
which builds a shoulder on the distribution extending out beyond
2 fm. The magnetic distribution in the neutron has an appreciable 
contribution from the $inner$ distribution. Note that by evaluating
$G_{Mn}/\mu _n$ all signs are inverted such that, e. g., the
contribution from the $\pi ^-$ comes in with a positive sign.
It is worth to mention, however, that it is the
fit which yields the positive sign for the contribution
parametrised as form factor of a 1p wave function.

\begin{figure} 
\begin{center}
\vspace*{-0.0cm} \rotatebox{-90} {\resizebox{6.0cm}{!}
{\includegraphics[width=1.0\textwidth]{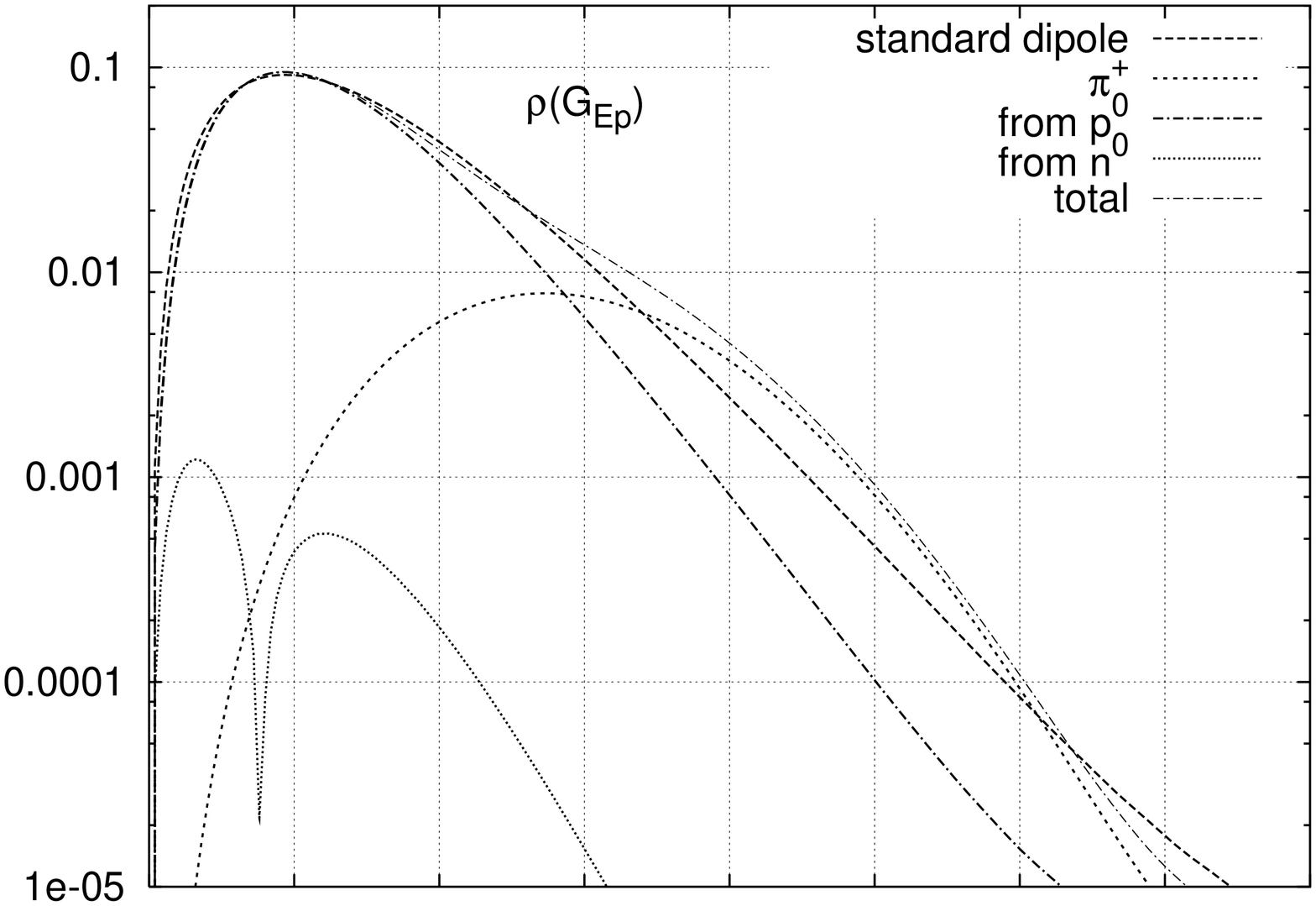}} } \\
\vspace*{-0.5cm} \rotatebox{-90} {\resizebox{6.0cm}{!}
{\includegraphics[width=1.0\textwidth]{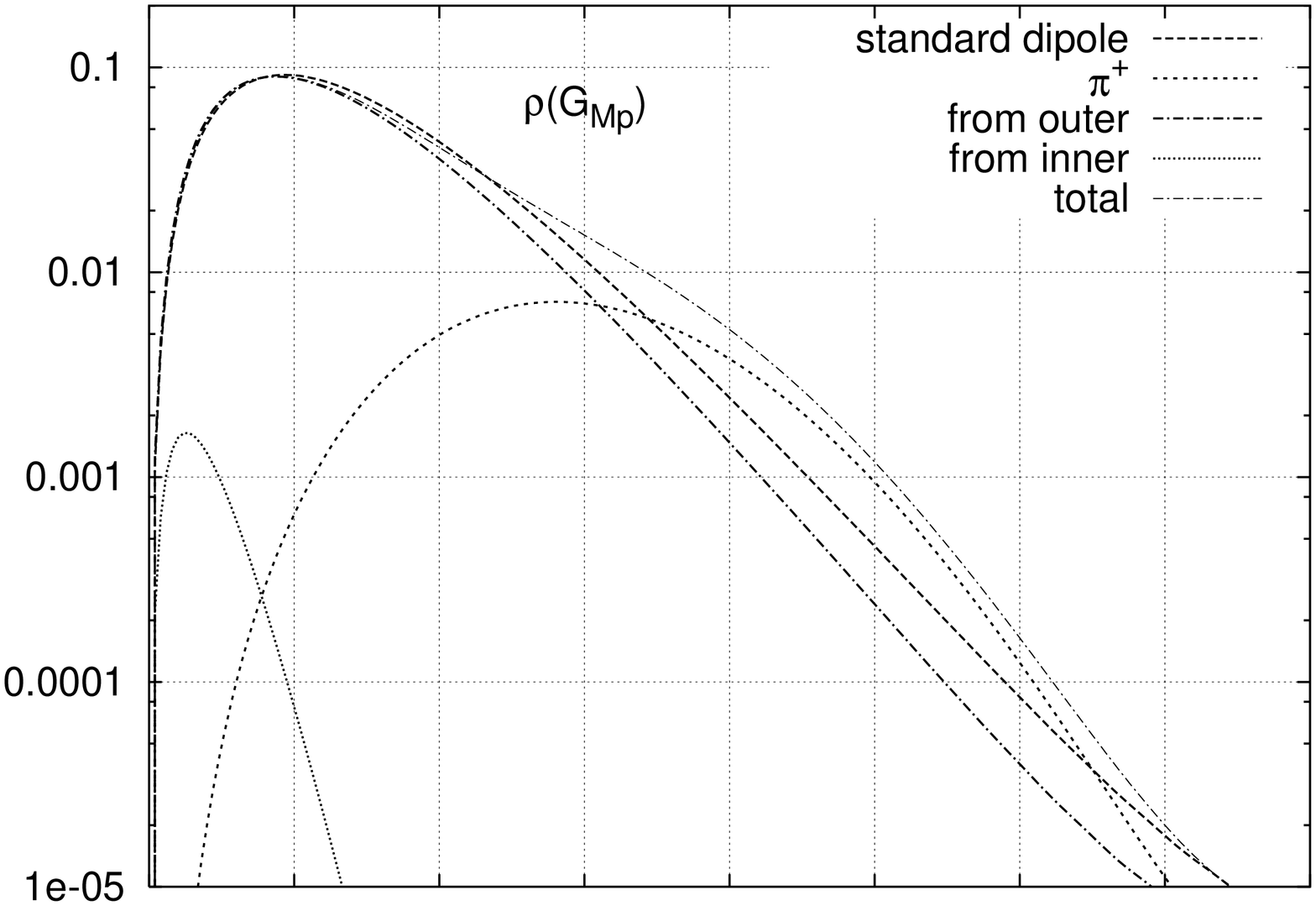}} } \\
\vspace*{-0.5cm} \rotatebox{-90} {\resizebox{6.0cm}{!}
{\includegraphics[width=1.0\textwidth]{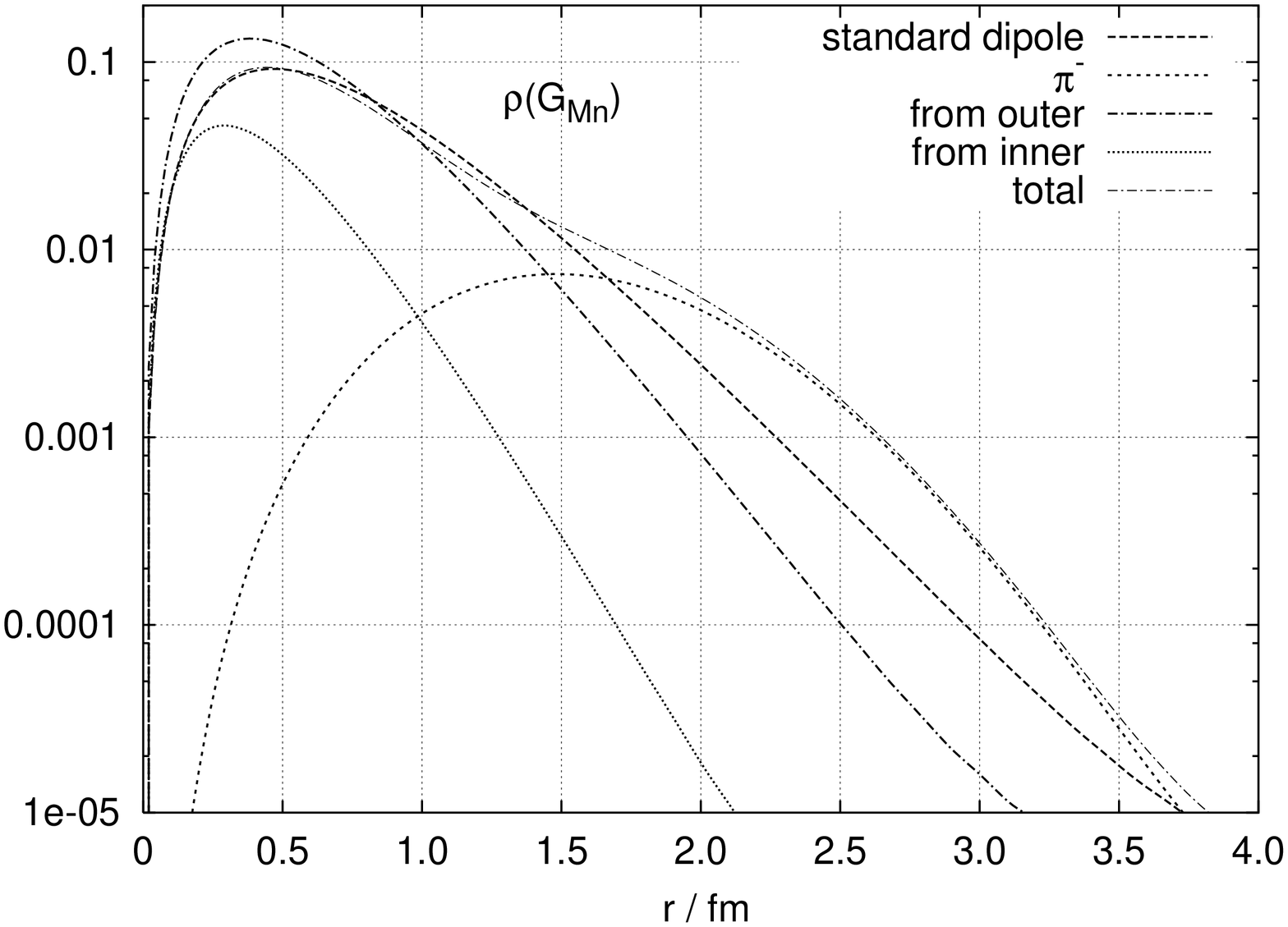}} } \\
\end{center}
\caption{Same as fig.~\ref{fig:rho_r2_mod}, now in logarithmic scale.}
\label{fig:rho_r2_mod_log}
\end{figure}

To emphasise again the smaller contributions and thus in particular
the outer region, fig.~\ref{fig:rho_r2_mod_log} shows $r^2 \cdot \rho (r)$
in logarithmic scale. By construction, the distinct structure at
the edge of the distribution now consists of only one bump,
which, according to the model, is due to the pion cloud.
This evaluation shows that the oscillations in the phenomenological
analysis in section~\ref{sec:phenomen} are not significantly
determined by the data, they result from the particular
phenomenological ansatz used there for the separation between
a ``smooth'' and a ``bump'' contribution to the form factor. 
The shoulders in all three standard form factors, however,
emerge in both evaluations, and we judge them as being
an unambiguous result from the data.

\begin{figure} 
\begin{center}
\vspace*{-0.0cm} \rotatebox{-90} {\resizebox{6.0cm}{!}
{\includegraphics[width=1.0\textwidth]{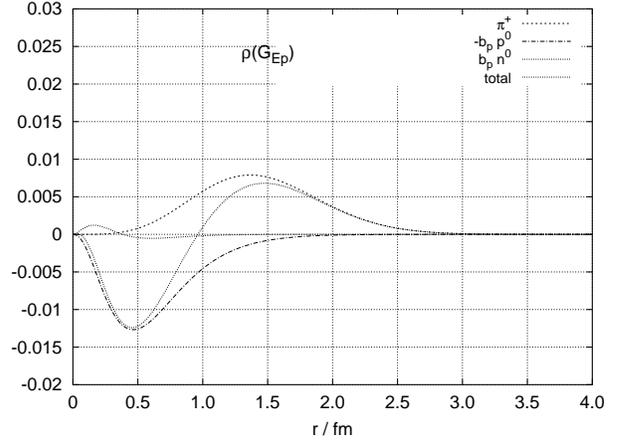}} } \\
\end{center}
\caption{The contribution of the polarisation term to $r^2 \cdot \rho (r)$ 
for the proton (in the Breit frame).}
\label{fig:rho_polarisation}
\end{figure}

\begin{figure} 
\begin{center}
\vspace*{-0.0cm} \rotatebox{-90} {\resizebox{6.0cm}{!}
{\includegraphics[width=1.0\textwidth]{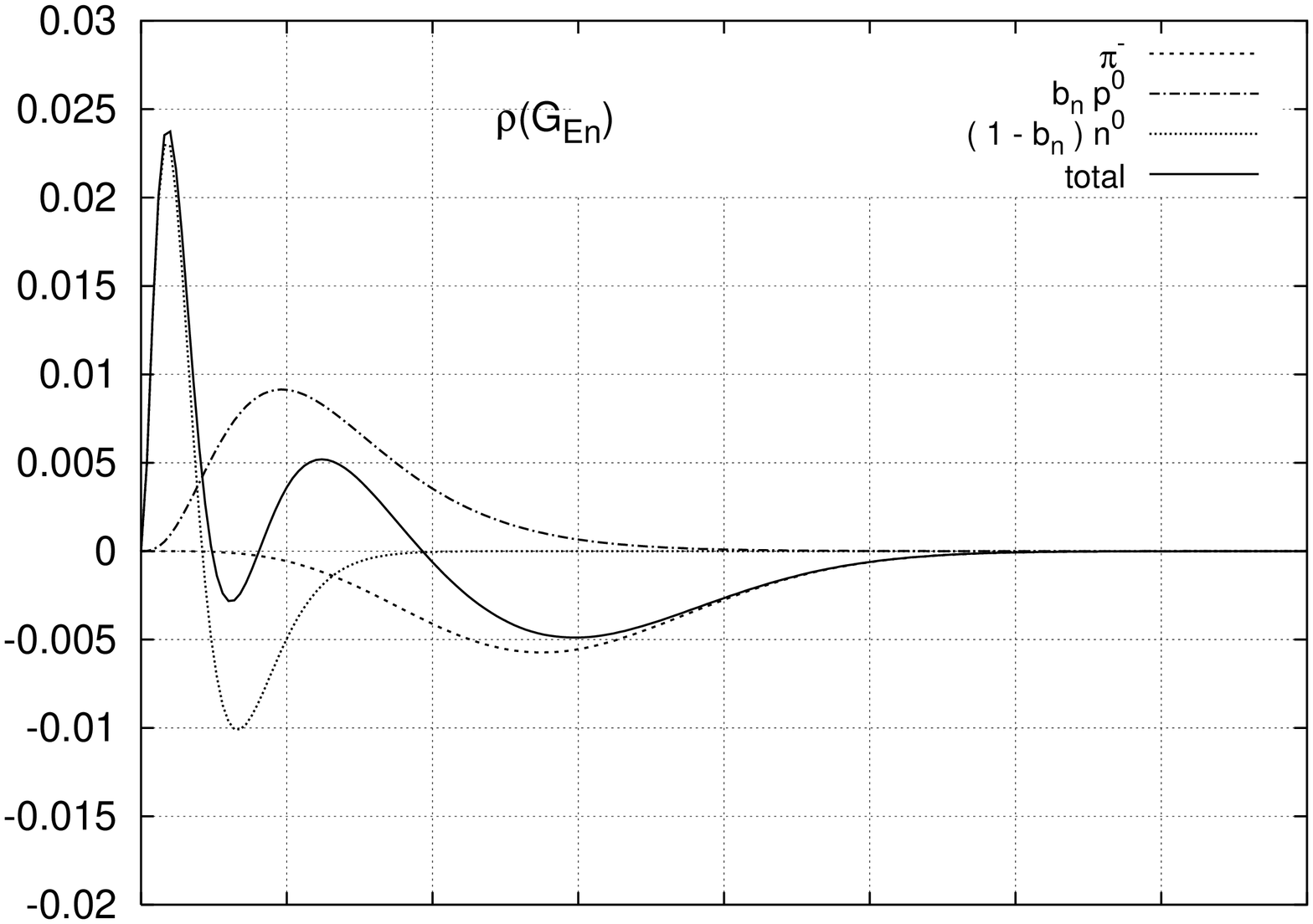}} } \\
\vspace*{-0.5cm} \rotatebox{-90} {\resizebox{6.0cm}{!}
{\includegraphics[width=1.0\textwidth]{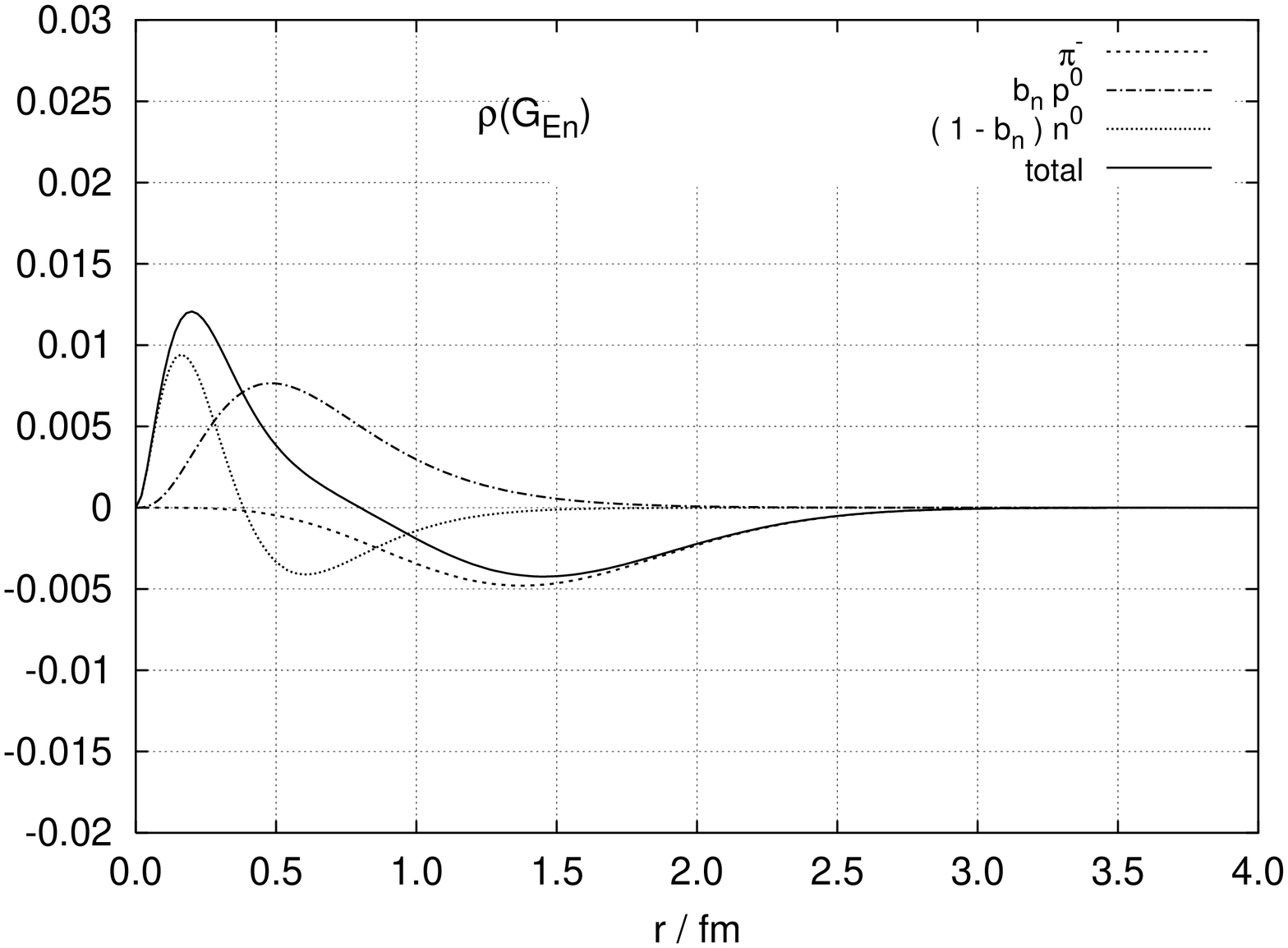}} } \\
\end{center}
\caption{$r^2 \cdot \rho (r)$ in the neutron (in the Breit frame). The contributions from
the bare neutron and proton and from the pion are shown separately. Upper panel: Fit 2,
lower panel: Fit 3 in table~\ref{tab:par_model}. }
\label{fig:rho_e_n_model}
\end{figure}

In fig.~\ref{fig:rho_polarisation} we show 
the polarisation contributions to the electric form factor of 
the proton in an enlarged scale for a closer comparison with
the situation in the neutron which is shown in fig.~\ref{fig:rho_e_n_model}.
The (tiny) neutron contribution to the polarisation part of the proton, 
$b_p \cdot n^0$, is situated in the inner region.
The superposition of $-b_p \cdot p^0$, i. e. from the reduction of $p^0$,
and $b_p \cdot \pi ^+$ yields just the two lobes which were also
seen in the phenomenological analysis and which are
emphasised in the logarithmic representation in fig.~\ref{fig:rho_r2_log}.
The small shift in the zero compared to fig.~\ref{fig:rho_r2_log} is due 
to the difference in what is regarded as the smooth part of the form factor. 
In fig.~\ref{fig:rho_r2_mod_log} the negative lobe of the ``bump'' in 
fig.~\ref{fig:rho_r2_log} in is not visible since $-b_p \cdot p^0$ is absorbed 
in the contribution from $p^0$ as a whole.

The charge distribution of the neutron, see
 fig.~\ref{fig:rho_e_n_model}, is dominated by the smooth polarisation oscillation,
i. e. by the positive lobe $b_n \cdot p^0$ and the negative lobe from the $\pi ^-$. 
These two contributions add up to the same form as the polarisation
in the proton, but with opposite sign.
Superimposed is now the charge distribution from the neutron, $(1-b_n) \cdot n^0$,
which modifies the smooth oscillation in a characteristic way.
In particular it reduces the positive lobe from $b_n \cdot p^0$ around
0.5 fm, possibly leading to a region with zero net charge.
The details depend on differences in the ansatz, the present data
do not contain sufficient information to discriminate between the
different solutions. It is, however, gratifying to note that this
feature is present throughout the different approaches in this paper:
It is also visible in figs.~\ref{fig:2} and, particularly clearly, in 
the lowest panel of fig.~\ref{fig:rho_r2}.


\section{Conclusion}
\label{sec:conclusion}

It is found as a common feature of all four nucleon form factors
that they exhibit a very similar structure at small momentum
transfer, which is related with some structure in $r$-space
at large $r$ around 2 fm. Such finding asks for a common explanation. 
We propose to interpret this as resulting from a pion cloud
around the bare nucleon. This is actually an old idea accounting for 
the chiral symmetry in quark bag models of the nucleon and used since many 
years \cite{Bro79,Tho84}. The phenomenologically successful ``cloudy
bag model'' (see e.g. \cite{Mil97} and references therein) was
recently used to describe the form factors of the nucleon
\cite{Lu97}. However, this description was still based on the old data
base and did not look for the effect of the pion cloud in the form
factor at low $Q^2$.

Also the other ingredient of our polarisation model has a deeper
theoretical basis. The division between the bare proton and neutron 
contributions into separate contributions from up and down quarks is
not only suggested by the naive constituent quark model but is well
justified through calculations in quenched lattice QCD
\cite{Hac00,Lei91}. 

A fit of the data with the ansatz of a p-wave for the pion
yields probabilities for the dissociation of the nucleons into
their counter part and a charged pion of 7 to 20\%, details
being dependent on the peculiarities of the ansatz. One could compare
these probabilities with similar results from high energy
experiments. But without a more thorough theoretical discussion such a
comparison is not very meaningful since our polarisation model is rather
crude. It is evident that next-to-leading order contributions as the
two pion continuum or the pion-Delta component in the nucleon wave
function have to be considered. In a dispersion-theoretical analysis 
such contributions can be included naturally and, in fact, have been
in the past \cite{Mer96}. Such analyses will be pursued again in the
frame work of chiral dynamics particularly suited to take higher
order effects to the pion cloud into account \cite{Mei03}.

From the experimental point of view our exciting result shows that
further studies at momentum transfers squared $Q^2 \lesssim$ 1~(GeV/c)$^2$  
down to the lowest reachable values are much needed with increased
precision. 

We have parametrised the smooth part of the form factors by
the superposition of dipoles, which lend themselves to an
interpretation in terms of the distribution of constituent
quarks. Data at high momentum transfers are needed to check 
this model assumption and constrain the distribution. \\ [0.2cm]

The authors are indebted to M. Seimetz and D. Glazier for making
available their $G_{En}$-measurements prior to publication.
A careful reading of the manuscript by Dr. U. M\"uller is 
gratefully acknowledged. \\

This work was supported by the state of Rhine\-land-Pala\-tin\-ate and by
the Deutsche Forschungsgemeinschaft (SFB~443).

\end{document}